\documentclass[longauth]{aa}  

\usepackage{tabularx}
\usepackage{color,graphicx}
\usepackage{txfonts}
\begin{document}

\title{Shape and spin determination of Barbarian asteroids}

\author{M. Devog\`ele \inst{\ref{ULg},\ref{OCA}}
\and    P. Tanga                        \inst{\ref{OCA}}
\and    P. Bendjoya             \inst{\ref{OCA}}
\and    J.P. Rivet                      \inst{\ref{OCA}}
\and    J. Surdej                       \inst{\ref{ULg}}
\and    J. Hanu{\v s}           \inst{\ref{OCA},\ref{Prague}}
\and    L. Abe                          \inst{\ref{OCA}}
\and    P. Antonini             \inst{\ref{CdR-CdL}}
\and    R.A. Artola             \inst{\ref{Arg}}
\and    M. Audejean             \inst{\ref{CdR-CdL},\ref{Chinon}}
\and    R. Behrend              \inst{\ref{CdR-CdL},\ref{Geneva}}
\and    F. Berski                       \inst{\ref{Poznan}}
\and    J.G. Bosch              \inst{\ref{CdR-CdL}}
\and    M. Bronikowska\inst{\ref{Krygowskiego}}
\and    A. Carbognani   \inst{\ref{Aosta}}
\and    F. Char                         \inst{\ref{Antofagasta}}
\and    M.-J. Kim                       \inst{\ref{Korea}}
\and    Y.-J. Choi                      \inst{\ref{Korea}}
\and    C.A. Colazo             \inst{\ref{Arg}}
\and    J. Coloma                       \inst{\ref{CdR-CdL}}
\and    D. Coward               \inst{\ref{Zadko}}
\and    R. Durkee                       \inst{\ref{Shed}}
\and O. Erece                   \inst{\ref{Tur}, \ref{Tur2}}
\and    E. Forne                        \inst{\ref{CdR-CdL}}
\and P. Hickson                 \inst{\ref{Canada}}
\and    R. Hirsch                       \inst{\ref{Poznan}}
\and    J. Horbowicz    \inst{\ref{Poznan}}
\and    K. Kami{\'n}ski \inst{\ref{Poznan}}
\and    P. Kankiewicz   \inst{\ref{Kielce}}
\and M. Kaplan                  \inst{\ref{Tur}}
\and T. Kwiatkowski   \inst{\ref{Poznan}}
\and    I. Konstanciak  \inst{\ref{Poznan}}
\and    A. Kruszewki            \inst{\ref{Poznan}}
\and V. Kudak                   \inst{\ref{Uzghorod},\ref{Pavol}}
\and    F. Manzini                      \inst{\ref{CdR-CdL},\ref{Sozzago}}
\and    H.-K. Moon              \inst{\ref{Korea}}
\and    A. Marciniak            \inst{\ref{Poznan}}
\and    M. Murawiecka   \inst{\ref{Namur}}
\and    J. Nadolny              \inst{\ref{IAC},\ref{Tenerife}}
\and    W. Og{\l}oza            \inst{\ref{Cracow}}
\and    J.L Ortiz                       \inst{\ref{Granada}}
\and    D. Oszkiewicz   \inst{\ref{Poznan}}
\and    H. Pallares             \inst{\ref{CdR-CdL}}
\and N. Peixinho                \inst{\ref{Antofagasta},\ref{Coimbra}}
\and    R. Poncy                        \inst{\ref{CdR-CdL}}
\and    F. Reyes                        \inst{\ref{Orihuela}}
\and    J.A. de los Reyes       \inst{\ref{Arroyo}}
\and    T. Santana--Ros\inst{\ref{Poznan}}
\and    K. Sobkowiak    \inst{\ref{Poznan}}
\and    S. Pastor       \inst{\ref{Arroyo}}
\and    F. Pilcher                      \inst{\ref{OrganMesa}}
\and    M.C. Qui{\~n}ones               \inst{\ref{Arg}}
\and    P. Trela                                \inst{\ref{Poznan}}
\and D. Vernet                  \inst{\ref{OCA}}
}  
 \institute{Universit\'{e} de Li\`{e}ge, Space sciences, Technologies and Astrophysics Research (STAR) Institute, All\'{e}e du 6 Ao\^{u}t 19c, Sart Tilman, 4000 Li\`{e}ge, Belgium \\ \label{ULg}
 \email{devogele@astro.ulg.ac.be} 
\and
Universit\'{e} C{\^o}te d'Azur, Observatoire de la C{\^o}te d'Azur, CNRS, Laboratoire Lagrange\label{OCA}
\and
Astronomical Institute, Faculty of Mathematics and Physics, Charles University in Prague, Czech Republic  \label{Prague}
\and
CdR \& CdL Group: Light-curves of Minor Planets and Variable Stars, Switzerland \label{CdR-CdL}
\and
Estaci{\'o}n Astrof{\'{\i}}sica Bosque Alegre, Observatorio Astron{\'o}mico C{\'o}rdoba, Argentina \label{Arg}
\and
Institute of Geology, Adam Mickiewicz University, Krygowskiego 12, 61-606 Pozna{\'n}  \label{Krygowskiego}
\and
Observatoire de Chinon, Chinon, France \label{Chinon}
\and 
Geneva Observatory, Switzerland \label{Geneva} 
\and
Astronomical Observatory Institute, Faculty of Physics, A. Mickiewicz University, S{\l}oneczna 36, 60-286 Pozna{\'n}, Poland \label{Poznan} 
\and 
Unidad de Astronom{\'{\i}}a, Fac. de Ciencias B{\'a}sicas, Universidad de Antofagasta, Avda. U. de Antofagasta 02800, Antofagasta, Chile \label{Antofagasta} 
\and
Korea Astronomy and Space Science Institute, 776 Daedeokdae-ro, Yuseong-gu, Daejeon, Republic of Korea \label{Korea} 
\and
Osservatorio Astronomico della regione autonoma Valle d'Aosta,Italy \label{Aosta} 
\and
School of Physics, University of Western Australia, M013, Crawley WA 6009, Australia \label{Zadko} 
\and
Shed of Science Observatory, 5213 Washburn Ave. S, Minneapolis,MN 55410, USA \label{Shed} 
\and
Akdeniz University, Department of Space Sciences and Technologies, Antalya, Turkey \label{Tur}
\and
TUBITAK National Observatory (TUG), 07058, Akdeniz University Campus, Antalya, Turkey \label{Tur2}
\and
Department of Physics \& Astronomy, University of British Columbia, 6224 Agricultural Road, Vancouver, B.C. V6T 1Z1, Canada \label{Canada} 
\and
Astrophysics Division, Institute of Physics, Jan Kochanowski University, {\'S}wi{\,e}tokrzyska 15,25-406 Kielce Poland \label{Kielce} 
\and 
Laboratory of Space Researches, Uzhhorod National University, Daleka st., 2a, 88000, Uzhhorod, Ukraine \label{Uzghorod}
\and
Institute of Physics, Faculty of Natural Sciences, University of P.J. Safarik, Park Angelinum 9, 040 01 Kosice, Slovakia \label{Pavol}
\and 
Stazione Astronomica di Sozzago, Italy \label{Sozzago}
 \and
NaXys, Department of Mathematics, University of Namur, 8 Rempart de la Vierge, 5000 Namur, Belgium \label{Namur}
\and
Instituto de Astrof{\'i}sica de Canarias (IAC), E-38205 La Laguna, Tenerife, Spain \label{IAC}
\and
Departamento de Astrof{\'i}sica, Universidad de La Laguna (ULL), E-38206 La Laguna, Tenerife, Spain \label{Tenerife}
\and
Mt. Suhora Observatory, Pedagogical University, Podchor\c{a}{\.z}ych 2, 30-084, Cracow, Poland \label{Cracow} 
\and
Instituto de Astrof{\'i}sica de Andaluc{\'i}a, CSIC, Apt 3004, 18080 Granada, Spain \label{Granada} 
\and
CITEUC -- Centre for Earth and Space Science Research of the University of Coimbra, Observat\'orio Astron\'omico da Universidade de Coimbra, 3030-004 Coimbra, Portugal \label{Coimbra} 
\and
Agrupaci{\'o}n Astron{\'o}mica Regi{\'o}n de Murcia, Orihuela, Spain \label{Orihuela} 
\and
Arroyo Observatory, Arroyo Hurtado, Murcia, Spain \label{Arroyo} 
\and
4438 Organ Mesa Loop, Las Cruces, New Mexico 88011 USA \label{OrganMesa}
 }

        \titlerunning{Shape and spin determination of Barbarian asteroids}
        \authorrunning{Devog{\`e}le et al.}

\newcommand{\subf}[2]{%
  {\small\begin{tabular}[t]{@{}c@{}}
  #1\\#2
  \end{tabular}}%
}

   \date{Received \ldots; accepted \ldots}


   \abstract
   {The so-called Barbarian asteroids share peculiar, but common polarimetric properties, probably related to both their shape and composition. They are named after (234)~Barbara, the first on which such properties were identified.
As has been suggested, large scale topographic features could play a role in the polarimetric response, if the shapes of Barbarians are particularly irregular and present a variety of scattering/incidence angles. This idea is supported by the shape of (234)~Barbara, that appears to be deeply excavated by wide concave areas revealed by photometry and stellar occultations.}
   {With these motivations, we started an observation campaign to characterise the shape and rotation properties of Small Main-Belt Asteroid Spectroscopic Survey (SMASS) type L and Ld asteroids. As many of them show long rotation periods, we activated a worldwide network of observers to obtain a dense temporal coverage.}
   { We used light-curve inversion technique in order to determine the sidereal rotation periods of 15 asteroids and the convergence to a stable shape and pole coordinates for 8 of them. By using available data from occultations, we are able to scale some shapes to an absolute size. We also study the rotation periods of our sample looking for confirmation of the suspected abundance of asteroids with long rotation periods.     }
  {Our results show that the shape models of our sample do not seem to have peculiar properties with respect to asteroids with similar size, while an excess of slow rotators is most probably confirmed. 
    }
   { }

   \maketitle

\section{Introduction}

Shape modeling is of primary importance in the study of asteroid properties. In the last decades, $\sim1000$ asteroids have had a shape model determined by using inversion techniques\footnote{See the Database of Asteroid Models from Inversion Techniques (DAMIT) data base for an up-to-date list of asteroid shape models: http://astro.troja.mff.cuni.cz/projects/asteroids3D/}; most of them are represented by convex shape models. This is due to the fact that the most used technique when inverting the observed light-curves of an asteroid -- the so-called light-curve inversion -- is proven to mathematically converge to a unique solution only if the convex hypothesis is enforced \citep{b11,b12}. At the typical phase angles observed for Main Belt asteroids, the light-curve is almost insensitive to the presence of concavities according to \citet{Dure_2003}. Although the classical light-curve inversion process cannot model concavities, \citet{b12} point out that the convex shape model obtained by the inversion is actually very close to the convex hull of the asteroid shape. Since the convex hull corresponds to the minimal envelope that contains the non-convex shape, the location of concavities corresponds to flat areas. \citet{Dev_2015} developed the flat surfaces
derivation technique (FSDT) which considers flat surfaces on such models to obtain indications about the possible presence of large concavities.

\citet{b5} reported the discovery of the anomalous polarimetric behaviour of (234)~Barbara. Polarisation measurements are commonly used to investigate asteroid surface properties and albedos. Usually, the polarisation rate is defined as the difference of the photometric intensity in the directions parallel and perpendicular to the scattering plane (normalised to their sum): $P_{\rm r} = \frac{I_\perp-I_{\parallel}}{I_\perp+I_ {\parallel}}$. It is well known that the intensity of polarisation is related to the phase angle, that is, to the angle between the light-source direction and the observer, as seen from the object. The morphology of the phase-polarisation curve has some general properties that are mainly dependent on the albedo of the surface. A feature common to all asteroids is a ``negative polarisation branch'' for small phase angles, corresponding to a higher polarisation parallel to the scattering plane. For most asteroids, the transition to positive polarisation occurs at an ``inversion angle'' of $\sim20^\circ$. In the case of (234)~Barbara, the negative polarisation branch is much stronger and exhibits an uncommonly large value for the inversion angle of approximately $30^{\circ}$. Other Barbarians were found later on \citep{b6, b31, b26, b30,Bag_2015,Dev_2017a}, and very recently, it was discovered that the family of Watsonia is composed of Barbarians \citep{b26}.

Several hypotheses have been formulated in the past to explain this anomaly. Strong backscattering and single-particle scattering on high-albedo inclusions were invoked. Near-infrared (NIR) spectra exhibit an absorption feature that has been related to the presence of spinel absorption from fluffy-type Calcium-Aluminium rich Inclusions (CAIs) \citep{b15, b16}. The meteorite analogue of these asteroids would be similar to CO3/CV3 meteorites, but with a surprisingly high CAI abundance ($\sim30$\%), never found in Earth samples. If this interpretation is confirmed, the Barbarians should have formed in an environment very rich in refractory materials, and would contain the most ancient mineral assemblages of the Solar System. This fact link, the explanation of the polarisation to the presence of high-albedo CAIs, is the main motivation for the study of Barbarians, whose composition challenges our knowledge on meteorites and on the mechanisms of the formation of the early Solar System.

The compositional link is confirmed by the evidence that all Barbarians belong to the L and Ld classes of the Small Main-Belt Asteroid Spectroscopic Survey (SMASS) taxonomy \citep{Bus_Bin}, with a few exceptions of the K class. While not all of them have a near-infrared spectrum, all Barbarians are found to be L-type in the \citet{b14} NIR-inclusive classification \citep{Dev_2017b}.

However, the variety of polarimetric and spectroscopic properties among the known Barbarians, also suggests that composition cannot be the only reason for the anomalous scattering properties. In particular, \citet{b5} suggested the possible presence of large concavities on the asteroid surfaces, resulting in a variety of (large) scattering and incidence angles. This could have a non-negligible influence on the detected polarisation, but such a possibility has remained without an observational verification up to now. 

Recent results concerning the prototype of the class, (234)~Barbara, show the presence of large-scale concavities spanning a significant fraction of the object size. In particular, interferometric measurements \citep{b27} and well-sampled profiles of the asteroid obtained during two stellar occultations indicate the presence of large-scale concave features \citep{b13}. Testing the shape of other Barbarians for the presence of concavities is one of the main goals of the present study, summarising the results of several years of observation. To achieve this goal, we mainly exploit photometry to determine the shape, to be analysed by the FSDT. We compare our results with the available occultation data \citep{Dun_2016} in order to check the reliability of the inverted shape model, scale in size, and constrain the spin axis orientation. We also introduce a method capable of indicating if the number of available light-curves is sufficient to adequately constrain the shape model solution.

As we derive rotation periods in the process, we also test another evidence, rather weak up to now: the possibility that Barbarians contain an abnormally large fraction of ``slow'' rotators, with respect to a population of similar size. This peculiarity -- if confirmed -- could suggest a peculiar past history of the rotation properties of the Barbarians. 

The article is organised as follows. In Sect. 2, we describe the observation campaign and the obtained data. Sect. 3 presents the approach that was followed to derive the shape models. A new method for analysing the shape model of asteroids is presented and tested on well-known asteroids. In Sect. 4, the characteristics of some individual asteroids are presented as well as a general discussion about the main results of the present work. The incidence of concavities and the distribution of the rotation periods and\ pole orientations of L-type asteroids are presented. Finally Sect. 5 contains the conclusions of this work.

\section{The observation campaigns}

\subsection{Instruments and sites}

Our photometric campaigns involve 15 different telescopes and sites distributed on a wide range in Earth longitudes in order to optimise the time coverage of slowly rotating asteroids, with minimum gaps. This is the only strategy to complete light-curves when the rotation periods are close to 1 day, or longer, for ground-based observers.

The participating observatories (Minor Planet Center (MPC) code in parentheses) and instruments, ordered according to their geographic longitudes (also Tab.~\ref{Obs_Table}), are:
\begin{itemize}
\item the 0.8m telescope of Observatori Astron\`omic del Montsec (OAdM), Spain (C65),
\item the 1.04m ``Omicron'' telescope at the Centre P{\'e}dagogique Plan{\`e}te et Univers (C2PU) facility, Calern observatory (Observatoire de la C\^ote d'Azur), France (010),
\item the 0.81m telescope from the Astronomical Observatory of the Autonomous Region of the Aosta Valley (OAVdA), Italy (B04),
\item the 0.4m telescope of the Borowiec Observatory, Poland (187),
\item the 0.6m telescope from the Mt. Suhora Observatory, Cracow, Poland,
\item the 0.35m telescope from Jan Kochanowski University in Kielce, Poland (B02),
\item the 1m Zadko telescope near Perth, Australia (D20),
\item the 0.61m telescope of the Sobaeksan Optical Astronomy Observatory (SOAO), South Korea (345),
\item the 0.7m telescope of Winer Observatory (RBT), Arizona, USA (648),
\item the 0.35m telescope from the Organ Mesa Observaory (OMO), NM, USA (G50),
\item the 0.6m telescope of the Southern Association for Research in Astrophysics (SARA), La Serena Observatory, Chile (807),
\item the 0.35m telescope from the UBC Southern Observatory, La Serena Observatory, Chile (807),
\item the 1.54m telescope of the Estaci{\`o}n Astrof{\`i}sica de Bosque Alegre (EABA), Argentina  (821),
\item the 1.5m telescope of the Instituto Astrof{\`i}sica de Andaluc{\`i}a (IAA) in Sierra Nevada, Spain (J86),
\item the 0.77m telescope from the Complejo Astron{\o}mico de La Hita, Spain (I95).

\end{itemize} 

All observations were made in the standard V or R band as well as sometimes without any filter. The CCD images were reduced following standard procedures for flat-field correction and dark/bias subtraction. Aperture photometry was performed by the experienced observers operating the telescopes, and the auxiliary quantities needed to apply the photometric inversion (light-time delay correction, asterocentric coordinates of the Sun and the Earth) were computed for all the data.

The whole campaign is composed of $244$ individual light-curves resulting from approximatively $1400$ hours of observation and $25000$ individual photometric measurements of asteroids. All the new data presented in this work are listed in Table A1 of Appendix A.

In addition to these observations, we exploited light-curves published in the literature. For the most ancient light-curves, the APC (Asteroid Photometric Catalogue \citep{Lag_2001}) was used. The most recent ones are publicly available in the DAMIT database \citep{Dure_2010}. 

As explained in the following section, we also used sparse photometry following the approach of \citet{Han_2011,b23}. The selected data sets come from observations from the United States Naval Observatory (USNO)-Flagstaff station (International Astronomical Union (IAU) code 689),  Catalina Sky Survey Observatory (CSS, IAU code 703 \citep{Lar_2003}), and Lowell survey \citep{bowe_2014}.

The already published data used in this work are listed in Table B1 of Appendix B.

\begin{table}
\begin{tabular}{|lcllcl|}
\hline
Observat. & D  & Longitude & Latitude & MPC  & N$_{lc}$ \\
                                &       [m]             &                                       &                               &       code    &                               \\
\hline
OAdM & 0.8                              & $000^{\circ}44'13"$   & $42^{\circ}01'28"$& C65 & 16 \\
C2PU & 1.04                     & $006^{\circ}56'00"$   & $43^{\circ}45'00"$& 010 & 89 \\
OAVdA & 0.81                    & $007^{\circ}28'42"$   & $45^{\circ}47'23"$ & B04 & 16 \\ 
Borowiec & 0.4                  & $016^{\circ}52'48"$   & $52^{\circ}24'00"$& 187 & 41 \\
Suhora & 0.60                   & $020^{\circ}04'01"$   & $49^{\circ}34'09"$& - & 2 \\
Kielce & 0.35                   & $020^{\circ}39'24"$   & $50^{\circ}52'52" $& B02 & 2 \\
Zadko & 1                               & $115^{\circ}42'49"$   & $-31^{\circ}21'24"$& D20 & 9 \\
SOAO & 0.61                     & $128^{\circ}27'27"$   & $36^{\circ}56'04"$ & 345 & 4 \\
RBT & 0.7                               & $248^{\circ}01'15"$   & $31^{\circ}42'00"$& 648 & 21 \\
OMO & 0.35                              & $253^{\circ}19'48"$   & $32^{\circ}17'46"$ & G50 & 11\\
SARA & 0.6                              & $289^{\circ}11'47"$   & $-30^{\circ}10'10"$& 807 & 11 \\
UBC & 0.35                              & $289^{\circ}11'37"$   & $-30^{\circ}10'07"$& 807& 15 \\
EABA & 1.54                     & $295^{\circ}27'25"$   & $-31^{\circ}35'46"$& 821 & 12\\
IAA & 1.5                                       & $356^{\circ}36'51"$   & $37^{\circ}03'47"$& J86 & 5 \\
LA HITA & 0.77          & $356^{\circ}48'50''$  & $39^{\circ}34'07'' $& I95 & 1 \\
\hline

\end{tabular}
\caption{List of the observatories participating in the observation campaigns, with telescope aperture (D), position, and number of light-curves. A rather good longitude coverage was ensured.}
\label{Obs_Table}
\end{table} 

\subsection{The target sample}

We coordinated the observation of 15 asteroids, selected on the basis of the following criteria:
\begin{itemize}
\item Known Barbarians, with a measured polarimetric anomaly;
\item SMASS L or Ld type asteroids, exhibiting the 2 $\mu m$ spinel absorption band;
\item members of dynamical families containing known Barbarians and/or L-type asteroids: Watsonia \citep{b26}, the dispersed Henan \citep{Nes_2015}; Tirela \citep{Mot_2008}, renamed Klumpkea in \citet{Mil_2014}.
\end{itemize} 

\begin{figure*}
\centering
\includegraphics[width=18cm]{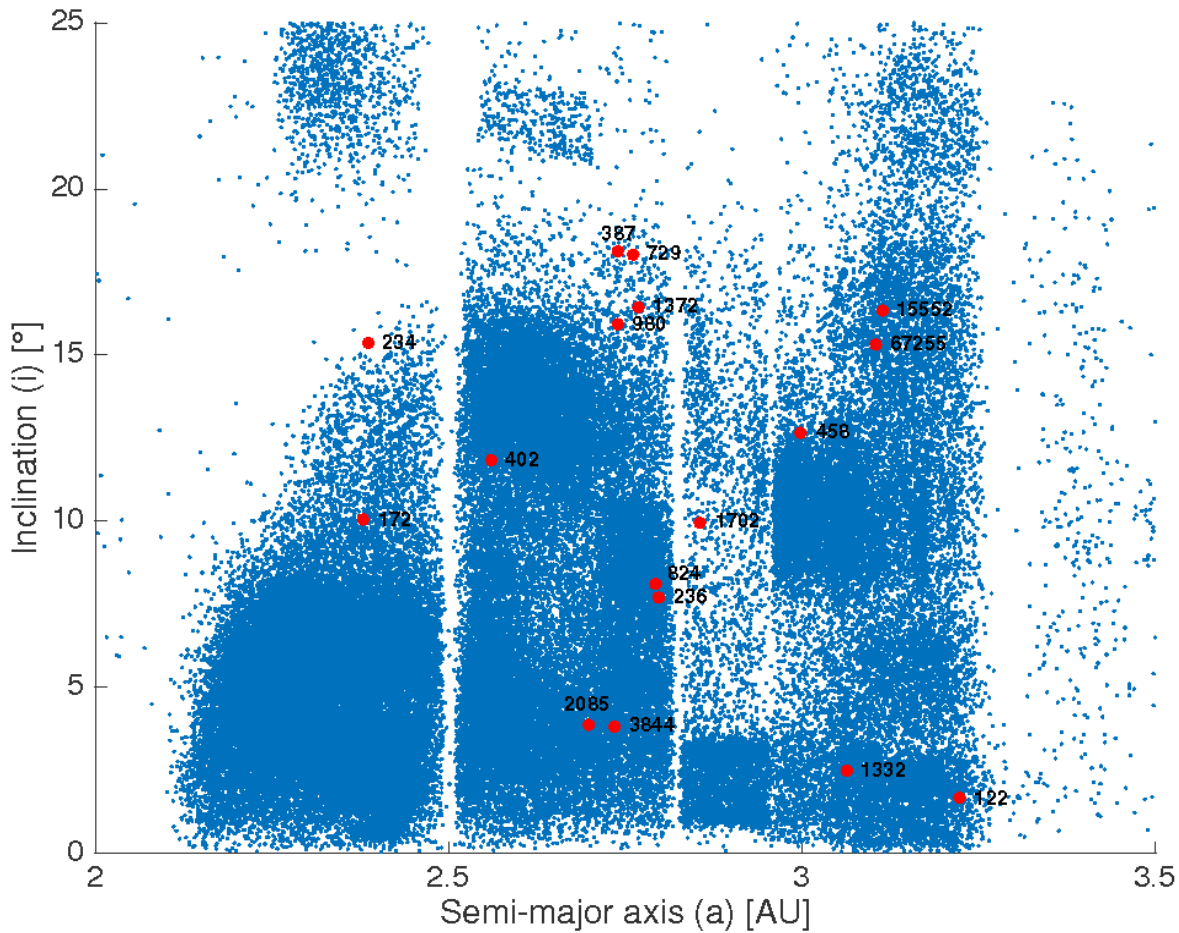}
\caption{Inclination versus semi-major axis plot of main-belt asteroids. The asteroids studied in this work are plotted as red circular dots, while the thinner blue dots represent the asteroid population as a whole.}
\label{fig:aVSi}
\end{figure*}

Fig. \ref{fig:aVSi} shows the location (red dots) in semi-major axis and inclination of the asteroids studied in this work. In general, Barbarians are distributed all over the main belt. The concentration in certain regions is a direct consequence of the identification of the Watsonia, Henan, and Tirela families. Future spectroscopic surveys including objects with a smaller diameter could better portray the global distribution.

For each of the 15 asteroids that have been targeted in this work, our observations have led to the determination of a new rotation period, or increase the accuracy of the previously known one. For eight of them, the light-curves dataset was sufficiently large to determine the orientation of the spin axis and a reliable convex shape model.

Table~\ref{tab:Param} lists the class in the Tholen \citep{Tholen84}, SMASS \citep{Bus_Bin,Mot_2008} and Bus-Demeo \citep{b14,Bus_2009} taxonomies of the 15 asteroids observed by our network. Each known Barbarian from polarimetry is labelled as ``Y'' in the corresponding column. The number of the family parent member is given in the corresponding column. The bottom part of Table~\ref{tab:Param} gives the same information for asteroids not observed by our network, but included in this work to increase the sample.

\begin{table*}
\centering
\begin{tabular}{| l | lll | c | l |}
\hline
Asteroid                                &       Tholen  &       SMASS           &       DM      &       Barbarian               &        Family  \\
                                                &                               &                                               &                                               &                                                                                               &                       \\
\hline 
(122)~Gerda                     &       ST              &       L                               &                       &                                               &       \\
(172)~Baucis                    &       S                       &       L                               &                       &       Y                               &               \\
(234)~Barbara                   &       S                       &       Ld                              &       L                                       &       Y                                               &                       \\
(236)~Honoria                   &       S                       &       L                               &       L                                       &       Y                                                               &                       \\
(387)   ~Aquitania                      &       S                       &       L                               &       L                                       &       Y                                               &                       \\          
(402)   ~Chloe          &       S                       &       K                               &       L                                       &       Y                                                               &               \\ 
(458)   ~Hercynia                       &       S                       &       L                               &               &       Y                                                                               &       \\
(729)~Watsonia                          &       STGD    &       L                               &       L                                       &       Y                                               & 729     \\
(824)~Anastasia                         &       S                       &       L                               &       L                                       &                                               &                       \\
(980)~Anacostia                 &       SU              &       L                               &       L                                       &       Y                                               &                       \\
(1332)~Marconia                 &                               &       Ld                              &       L                                       &                                               &                       \\
(1372)~Haremari                 &                               &       L                               &               &                       &       729     \\ 
(1702)~Kalahari                         &       D                       &       L                               &                                               &               &                       \\
(2085)~Henan                            &                               &       L                               &       L                                       &                & 2085 \\
(3844)~Lujiaxi                                  &               &       L                       &       L               &                                                       &                                                       2085 \\
(15552)~Sandashounkan                                   &                               &                                       &                                               &                                               &1400   \\
\hline \hline
(234)~Barbara                   &       S                       &       Ld                              &       L                                       &       Y                                                       &                       \\
(599)   ~Luisa                  &       S                       &       K                               &       L                                       &       Y                                               &               \\
(606)   ~Brangane                       &       TSD             &       K                               &       L                                       &                       &       606             \\
(642)~Clara                     &       S                       &       L                               &                                               &                                                               &               \\
(673)~Edda                                              &       S       &       S                       &       L               &                                                                                                       &                               \\
(679)~Pax                       &       I                       &       K                               &       L                                       &       Y                                                       &                       \\
(1284)~Latvia                   &       T                       &       L                               &                                               &                       &                       \\
(2448)~Sholokhov                                                &                               &       L                               &       L                                       &               &                       \\
\hline
\end{tabular}
\caption{List of the targets observed by our network (upper part). Some targets that were not observed by us but discussed in this work were added in the lower part. The first column corresponds to the number and name of the considered asteroid. The columns Tholen \citep{Tholen84}, SMASS \citep{Bus_Bin,Mot_2008} and Bus-Demeo (DM) \citep{b14,Bus_2009} stand for the taxonomic class in these three types of taxonomy. The Barbarian column indicates whether or not the asteroid is considered as a Barbarian \citep{b5,b6,b31,Bag_2015}. Finally, the Family column indicates the number of the parent member of the family in which the asteroid is classified (606 for the Brangane, 729 for the Watsonia, 1400 for the Tirela/Klumpkea and 2085 for the Henan family). }
\label{tab:Param}
\end{table*}

\section{Modelling method and validation of the results} 
\label{Red_pro}

The sidereal rotation period was searched for by the light-curve inversion code described in \citet{b11} and \citet{b12} over a wide range. The final period is the one that best fits the observations. It is considered as unique if the chi-square of all the other tested periods is $> 10 \%$ higher. In order to obtain a more precise solution, sparse data are also included in the procedure, as described by \citet{b23}. When the determination of a unique rotation period is possible, the pole solution(s) can be obtained. As is well known, paired solutions with opposite pole longitudes (i.e. differing by $180^{\circ}$) can fit the data equally well. If four or less pole solutions (usually two sets of two mirror solutions) are found, the shape model is computed using the best fitting pole solution.

The accuracy of the sidereal rotation period is usually of the order of $0.1$ to $0.01$ times the temporal resolution interval $P^2/(2T)$ (where P is the rotation period and T is the length of the total observation time span). For the typical amount of optical photometry we have for our targets the spin axis orientation accuracy for the light-curve inversion method is, in general, $\sim 5^{\circ}/\cos{\beta}$ for $\lambda$ and $\sim 10^{\circ}$ for $\beta$ \citep{Han_2011}.

All possible shapes were compared with the results derived from stellar occultations whenever they were available. The projected shape model on the sky plane is computed at the occultation epoch using the \citet{Dur_2011} approach. Its profile is then adjusted, using a Monte-Carlo-Markov-Chain algorithm, to the extremes of the occultation chords. 

With this procedure, the best absolute scale of the model is found, thus allowing us to determine the size of the target. For ease of interpretation, the absolute dimensions of the ellipsoid best fitting- the scaled shape model and the volume-equivalent sphere $R_{\rm eq}$ are computed.

In some cases, the result of stellar occultations also allows one to discriminate different spin axis solutions.

Eventually, the FSDT is used to analyse all the derived convex shape models. As this procedure depends on the presence of flat surfaces, we must ensure that the shape models are reliable. In fact, portions of the asteroid surface not sufficiently sampled by photometry, can also produce flat surfaces. We thus established the approach explained in the following section, allowing us to estimate the reliability of a shape derived by the light-curve inversion.

\subsection{Shape validation by the bootstrap method} 
\label{Bootstrap}

Our goal is to evaluate the reliability of shape model details, and to check if a set of light-curves provides a good determination of the spin parameters.

As mentioned above, the light-curve inversion process often provides several sets of spin parameters (rotation period and spin axis orientation) that fit the optical data equally well. In other cases, when only a few light-curves are available, the parameter set may converge with a single, but incorrect solution, often associated with an incorrect rotation period. We see below that the bootstrapping method detects such cases.  

Our approach consists in computing the shape model that best fits a subset of light-curves, randomly selected among all those that are available for a given asteroid. By studying how the fraction of flat surfaces $\eta_s$ evolves with the number of selected light-curves, we obtain a solid indication on the completeness of the data, that is, the need for new observations or not. 

In more detail, we proceed as follows: 
\begin{itemize}
\item First, a large number of light-curve samples is extracted from the whole data set. Each of these subsets may contain a number of light-curves $n_l$ ranging from $n_l = 1$ to all those available ($n_l = N_{dl}$). Sets of sparse data are also included in the process and equal for all samples. 
In fact, the availability of a sparse data set is a necessary condition to apply the bootstrapping technique. Since shape models can sometimes be derived using sparse data only, they allow the inversion process to converge to a solution even for low values of $n_l$.
\item The second step consists in deriving the shape model corresponding to all the light-curve samples generated in the previous step. The spin axis parameters found by using all the light-curves are exploited as initial conditions for each inversion computation. If different sets of spin parameters are found with equal probability, the whole procedure is executed for each of them. 
\item Once all the shape models are determined, we apply the FSDT in order to find the best value of $\eta_s$ for each shape model.
\item The last step consists in determining the mean value and the dispersion of $\eta_s$ corresponding to each subset group characterised by $n_l=1,\ 2,\ 3,...,N_{dl}$ light-curves.
\end{itemize}

An excess of flat surfaces is often interpreted as the qualitative indication that the available photometry does not constrain the shape model adequately. The addition of supplementary light-curves is then a necessary condition to improve the shape. In our procedure, we make the hypothesis that, as the number of light-curves increases, the $\eta_s$ parameter decreases. In the process, the model converges to the convex hull of the real shape. Ideally, the only flat surfaces remaining should then correspond to concave (or really flat) topological features. Such a convergence should ideally show up as a monotonic decrease of $\eta_s$ towards an asymptotic value for a large $n_l$. Here, we apply our procedure and show that our quantitative analysis is consistent with this assumption.

We stress here that the number of possible subsets that can be extracted is potentially large, as it is given by the binomial coefficient ${{N}\choose{k}}$.

If the number of light-curves is below 14, all subsets are exploited. If there are additional light-curves, the number of possible combinations increases, but we do not generate more than $10,000$ random subsets to limit the computation time. Our results show that this is not a limitation when the light-curve sample is very rich.

\subsubsection{Test on (433)~Eros}
\label{Eros}
The bootstrap method was first tested on (433)~Eros for which the shape is well known thanks to the images taken by the NEAR mission \citep{Gas_2008}. A large number of dense photometric light-curves ($N = 134$) is also available for this asteroid which allows us to derive an accurate shape model using light-curve inversion. 

\begin{figure}
\includegraphics[width=8.8cm]{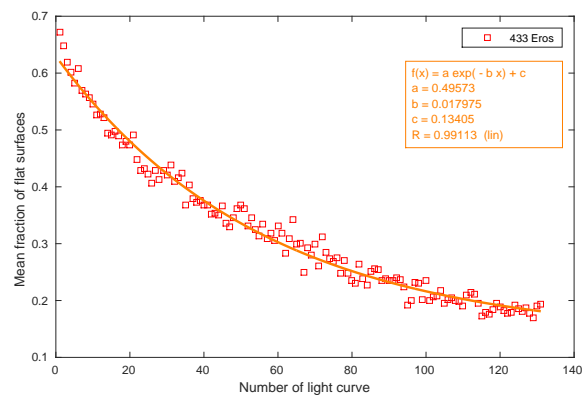}
\caption{ Bootstrap curves for the asteroid (433)~Eros}
\label{Eros_BS}
\end{figure}

The data set of (433)~Eros contains $134$ dense light-curves and $2$ sets of sparse data. Fig. \ref{Eros_BS} shows the average of the flat surface fraction ($\eta_s$) as a function of the number of dense light-curves ($N_{\rm dl}$) used for the shape modelling.

As expected  (Fig. \ref{Eros_BS}), the $\eta_s$ versus $N_{dl}$ curve (hereafter called the bootstrap curve) is monotonically decreasing. It is straightforward to verify that an exponential function $y = a\exp{(-bx)}+c$ is well suited to fit the bootstrap curve. In this expression the $c$ parameter corresponds to the asymptotic value $\eta_a$, that is, the $\eta_s$ that would be detected if a very large number of light-curves was used. In the case of (433)~Eros we find $\eta_a=0.134$.

For a comparison to the ``real'' shape, we can take advantage of the detailed shape model of (433)~Eros obtained by the analysis of the NEAR space probe images, obtained during the close encounter. For our goal, a low-resolution version \citep{Gas_2008} is sufficient, and was used to check the result described above. 

By taking the convex-hull (i.e. the smallest convex volume that encloses a non-convex shape), we obtain the shape model that, in an ideal case, the light-curve inversion should provide. In that specific case, the FSDT yields $\eta_a = 0.12$. This value is within $6 \%$ of the value obtained by the bootstrapping method. This is a first indication that our approach provides a good approximation and can be used to evaluate the ``completeness'' of a light-curve set.

We may compare our result to the value corresponding to the complete set of $N_{dl}$ available light-curves. In this case, $\eta_s = 0.18$, a much larger discrepancy ($20 \%$) with respect to the correct value. This result means that the convex shape model of (433)~Eros (derived by photometry only) could probably be somewhat improved by adding new dense light-curves. 

In the following sections, we apply this analysis to other objects for the evaluation of their shape as derived from the photometric inversion.

\subsubsection{Sensitivity of the bootstrap method to the spin axis coordinates}

The behaviour of the bootstrap curve (exponentially decreasing) is related to the choice of the correct spin axis parameters. 

This property is illustrated on the asteroid (2)~Pallas. Even though Pallas has not been visited by a space probe, its spin axis orientation and shape are well known thanks to stellar occultations and adaptive optics observations \citep{b24}.

The other difference with the case of (433)~Eros is that there are less observed dense light-curves. As a matter of fact, the light-curve inversion provides two ambiguous solutions for the spin axis coordinates, at ($37.9^\circ$,$ -15.1^\circ$) and ($199.6^\circ$, $ 39.4^\circ$). Both poles yield the same RMS residuals for the fit to the light-curves.

\begin{figure*}
\centering
\includegraphics[width=18cm]{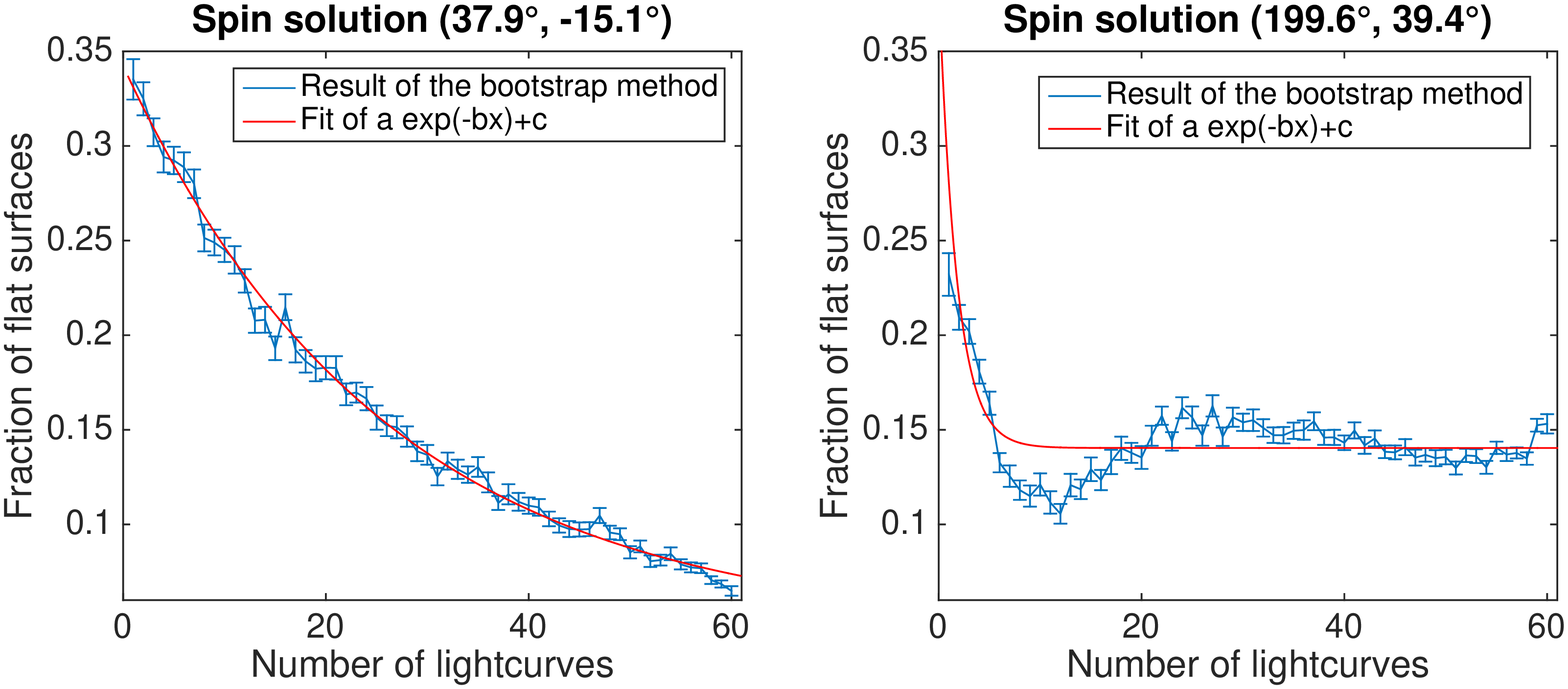}
\caption{ Bootstrap curves for the asteroid (2)~Pallas. The panels illustrate the curve obtained for different coordinates of the spin axis. }
\label{Pallas_BC}
\end{figure*}

A priori, there is no way to know which solution is the best, without additional constraint such as disk-resolved observations, since both solutions reproduce equally well the observed photometry. However, there is a clear difference between the two when analysing the bootstrap curve (Fig. \ref{Pallas_BC}). The first solution follows the well-defined exponential convergence providing $\eta_a=0.045$, which seems to be consistent with the shape model as it is known so far. For the second spin axis orientation solution, we see that the bootstrap curve does not seem to follow the exponential trend. The corresponding $\eta_a\sim 0.15$ is in contradiction with the fact that the shape of (2)~Pallas does not show any sign of large concave topological features. The rejection of the second pole solution is consistent with the result of stellar occultations and adaptive optics \citep{b24}.

\subsubsection{Distribution of flat surfaces across the available shape-model population.}
\label{Sec:flatVSDiam}

The bootstrap method gives us information on the possible presence of concave topological features. Here, a large number of asteroid shape models are analysed. These shape models are used as a reference population with which the derived shape models of Barbarians can be compared.

This reference was constructed using the shape models available on the DAMIT database \citep{Dure_2010}. More than $200$ asteroid shape models were analysed. For many of them, the result of the bootstrap method clearly shows that more data would be needed to obtain a stable shape. In a minority of cases, we find a behaviour suggesting incorrect pole coordinates. We selected a sample of $130$ shape models, for which the bootstrap curve had the expected, regular behaviour towards a convergence, and for them we computed $\eta_a$.

\section{Results and interpretation}

\begin{figure*}[!h]
\centering
\begin{tabular}{|c|c|}
\hline
\includegraphics[width=8.4cm]{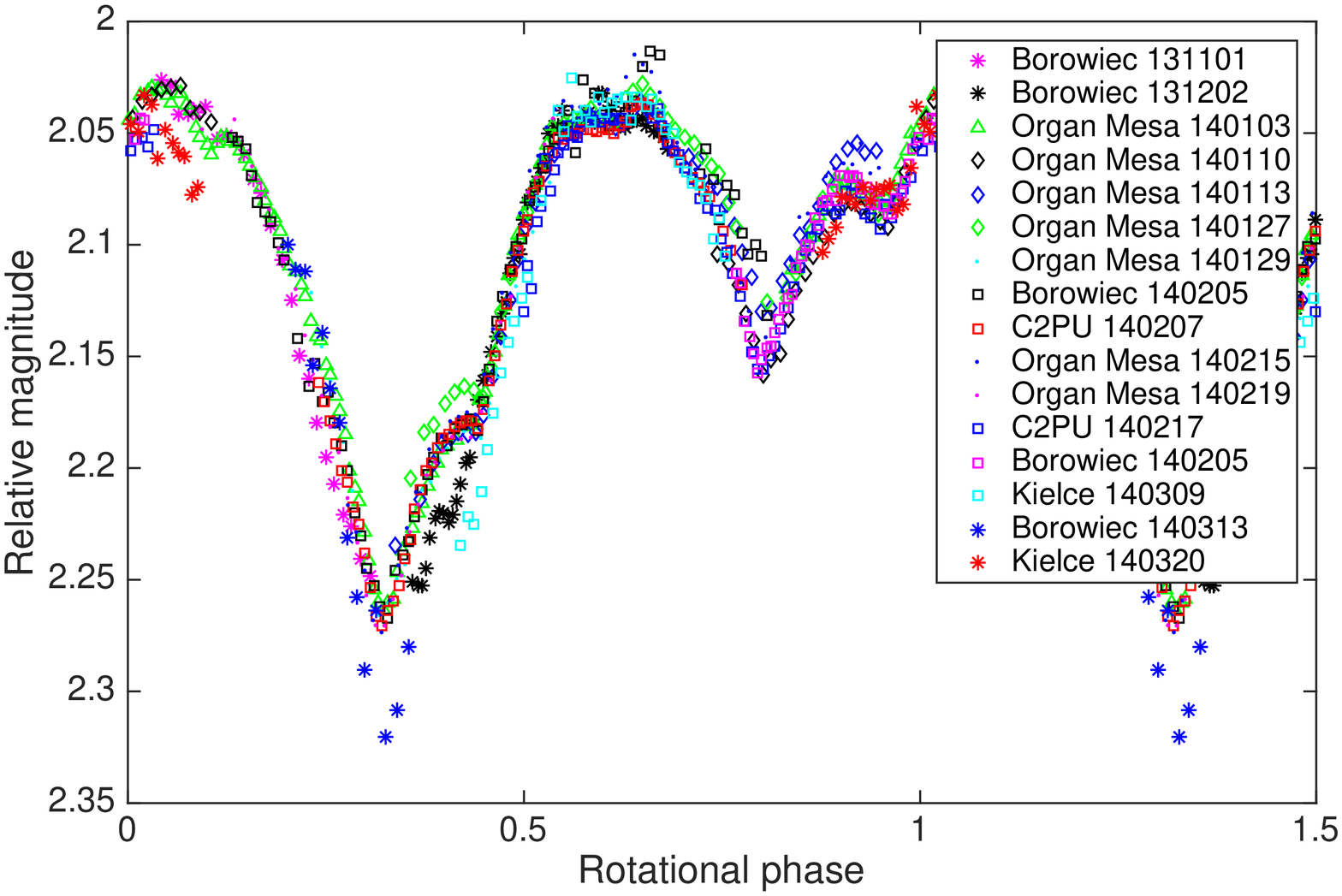}
&
\includegraphics[width=8.4cm]{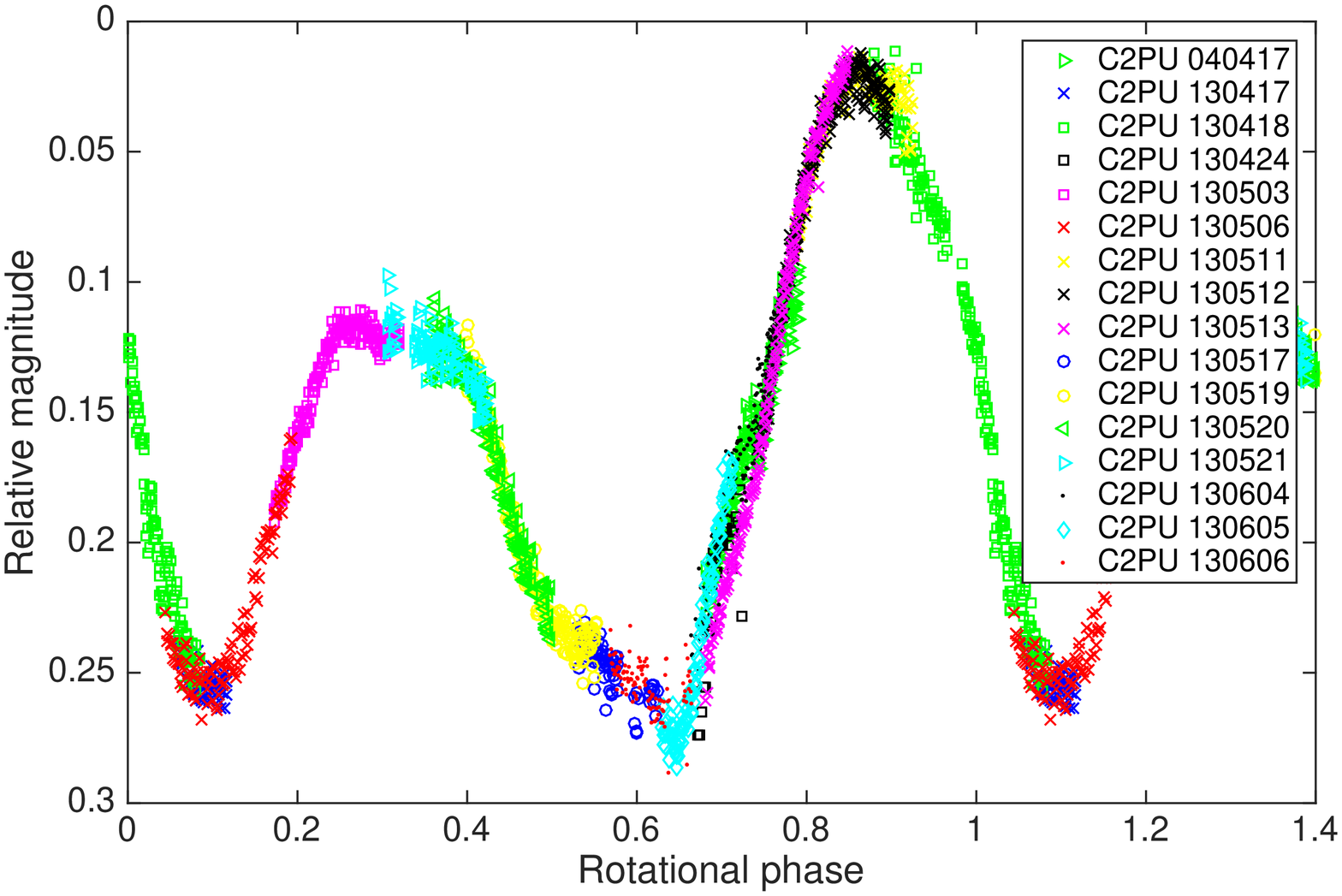}

\\
     {(236)~Honoria, $P_{\rm syn} = 12.3373 \pm 0.0002$~h}
&
{(729)~Watsonia, $P_{\rm syn} = 25.19 \pm 0.03$~h}

\\
\hline

\end{tabular}
\caption{Composite light-curves of two asteroids in our sample. Each light-curve is folded with respect to the synodic period of the object, indicated below the plot. 
}
\label{LC_Ex}
\end{figure*}

\begin{figure*}
\centering
\begin{tabular}{|c|}
\hline

\subf{\includegraphics[width=17cm]{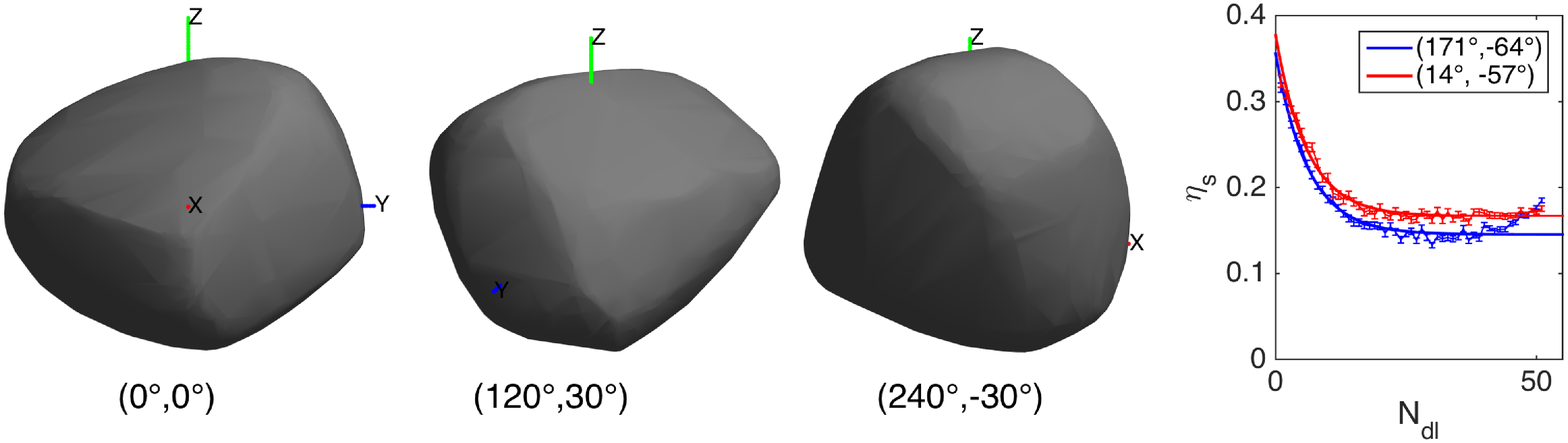}}
     {(172)~Baucis, pole solution ($14^{\circ}$, $-57^{\circ}$).}

\\
\hline
\subf{\includegraphics[width=17cm]{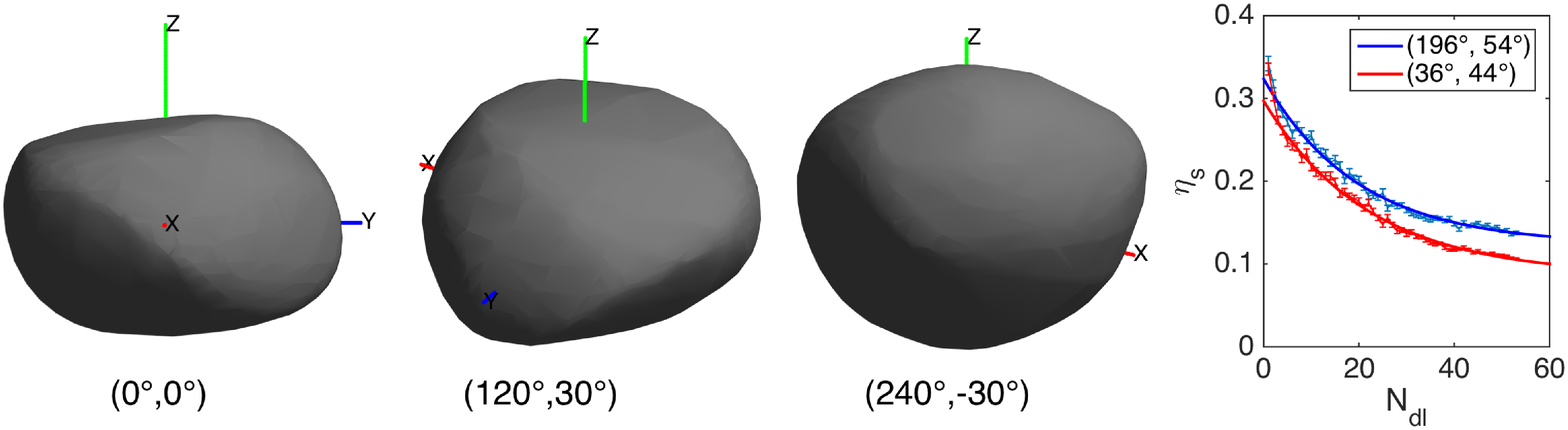}}
     {(236)~Honoria, pole solution ($196^{\circ}$, $54^{\circ}$).}

\\
\hline
\end{tabular}
\caption{Example of two asteroid shape models derived in this work. For both shape model, the reference system in which the shape is described by the inversion procedure, is also displayed. The $z$ axis corresponds to the rotation axis. The $y$ axis is oriented to correspond to the longest direction of the shape model on the plane perpendicular to $z$. Each shape is projected along three different viewing geometries to provide an overall view. The first one (left-most part of the figures) corresponds to a viewing geometry of $0^{\circ}$ and $0^{\circ}$ for the longitude and latitude respectively (the $x$ axis is facing the observer). The second orientation corresponds to ($120^{\circ}$, $30^{\circ}$) and the third one to ($240^{\circ}$, $-30^{\circ}$). The inset plot shows the result of the Bootstrap method. The $x$ axis corresponds to the number of light-curves used and the $y$ axis is $\eta_a$.}
\label{fig:Shape_Ex}
\end{figure*}

In Fig.~\ref{LC_Ex}, one example of a composite light-curve for two asteroids observed by our network is shown. For each target, the synodic period ($P_{\rm syn}$) associated to the composite light-curve is provided. Additional light-curves are displayed in Appendix~\ref{App:LC}. For the cases where too few light-curves have been obtained by our survey to derive a reliable synodic period, the sidereal period ($P_{\rm sid}$) determined based on multi-opposition observations and obtained by the light-curve inversion method is used as an initial guess. In the case of (824)~Anastasia, no error bars are provided since our observations were obtained during only one (very long) revolution. In Fig.~\ref{fig:Shape_Ex} is displayed two examples of shape model derived in this work. Additional shape models are displayed in Appendix~\ref{App:Shape}. The bootstrap curves (see Sec.~\ref{Eros}) are also shown for the different solutions of the pole orientation (except for (387)~Aquitania for which no sparse data were used). Table~\ref{tab:Obs_Table} summarises the information about these asteroids, whose physical properties have been improved by our observations. Table~\ref{tab:Obs_Tab2} is the same as Table~\ref{tab:Obs_Table}, except that it lists asteroids that we did not directly observe ourselves, but that are relevant to our discussion.

\begin{table*}
\hspace{-0.7 cm}
\small
\begin{tabular}{| l | ll | l | llll | lllll | l | l|}
\hline
Aster. & $b/a$   & $c/b$ & $P_{\rm sid}$                 & $\lambda_1$ & $\beta_1$ & $\lambda_2$ & $\beta_2$ & N$_{\rm dl} $ & N$_{\rm opp} $& N$_{689}$& N$_{703}$& N$_{\rm LO}$&$\eta_a$  & D  \\
                        &             &             &   [hours]              &     [deg]          &   [deg]       &  [deg]             & [deg]         &                 &  &   &           &  &      & [km] \\
\hline  
$122$   & $0.85$ & $0.96$ & $10.6872 \pm 1$ &$ 30 \pm 5$          &  $20 \pm 10$     & $209 \pm 5 $         & $22 \pm 10$      &  $24$  & $6$    &$181$  & $108$ & & $0$       & \textbf{($70.7 \pm 0.9$)}  \\
$172$   & $0.93 $& $0.90 $&$27.4097 \pm 4$  & $171 \pm 11$        & $-64 \pm 10$    & $14 \pm 9$           & $-57 \pm 10$     &   $61$&  $6$   & $159$ &  $75$  & &$0.17$    & $77.4 \pm 3.9$ \\
$236$   & $0.88$ & $0.86 $&$12.3375 \pm 1 $ &  $196 \pm 9$       & $54 \pm 10$      & $36 \pm 7 $             & $44 \pm 10$         &   $57$   &  $5$    & $187$ & $120$ &  &$0.16$  & $86.0 \pm 4.3 $ \\
$387$   &  $0.93$& $0.88$ &$24.14 \pm 2$     & $142 \pm 8$    &  $51 \pm 10$     & & &                                $26$     &  $6$   &   &  & & & $100.7 \pm 5.3$ \\          
$402$   & $0.88 $& $0.70 $&$10.6684 \pm 1$     & $306 \pm 10 $         & $-61 \pm 10$     &$162 \pm 7$         &  $ -41 \pm 10 $   &    $15$     & $4$     & $169$  & $65$ & & $0$        &  $64.6 \pm 3.2$ \\ 
$458$   & $ 0.86$& $0.80 $&$21.81 \pm 5$             & $274 \pm 6$         & $33 \pm 10$      &$86 \pm 5$          &  $ 14 \pm 10 $   &     $14$       & $3$    & $197$  & $103$ & &$0.38$  & \textbf{($36.7 \pm 0.4$)}\\
$729$   &       $0.88$ &$0.86$  & $25.195 \pm 1$  & $ 88 \pm 26$           & $-79 \pm 10$   &                      &                    &                $60 $            &    $3$ & $182$ & $104$ &  & $0.14$  & \textbf{($50.0 \pm 0.4$)} \\
$824$   &       &  &$252.0 \pm 7$    &             &   &  &  &     $25$          & $2$        & $133$       & $149$ & &  & \textbf{$(32.5 \pm 0.3)$} \\
$980$   & $ 0.93$& $ 0.83 $&$20.114 \pm 4$           & $24.2 \pm 6$             & $ 35 \pm 10 $      & $203 \pm 5 $        & $-5 \pm 10$     &  $48$     & $6$   &  $167$ & $68$   &    & $0.28$  & \textbf{($74.7 \pm 0.6$)}\\
$1332$ &  &  & $32.120 \pm 1$             &           &     &          &        & $16$      & $2$  &   &    & $408$   &    & ($46.8 \pm 0.1$)\\
$1372$ & $ $       & $ $        & $15.23 \pm 4$              & $ $                 & $ $           & & &  $ 18 $         &   $2$      &  $84$    & $60$  & &  &\textbf{($26.5 \pm 0.3 $ ) }  \\ 
$1702$ & $ $       & $ $        & $21.15 \pm 2$            & $ $                 & $ $           & & &$ 21$           &   $2$      & $90$        & $155$  & &  & \textbf{($34.6 \pm 0.1$)}\\
$2085$ & $ $       & $ $        & $111 \pm 1 $    & $ $                  & $ $          & & & $12 $           &   $1$     &  $27$      & $93$   & & & \textbf{($13.35 \pm 0.04$)}    \\
$3844$ & $ $       & $ $        & $13.33 \pm 2$  & $ $                 & $ $           & & &  $10$             &  $1$        &       &$87$   & & &($15.5 \pm 0.7 $)      \\
$15552 $ & $ $  & $ $        & $ 33.63 \pm 7 $            & $ $                 & $ $           & & & $19 $         &  $2$        &       &  $58$  & & & \textbf{($ 6.2 \pm 0.2$ )}     \\
\hline
\end{tabular}
\caption{Summary of the results pertaining to our survey. The $b/a$ and $c/b$ columns represent the relative axis dimensions of the ellipsoid best fitting the shape model. The $P_{\rm sid}$ column indicates the sidereal rotation period of the asteroid in hours. The uncertainties are given with respect to the last digit. Columns $\lambda_n$ and $\beta_n$, with $n=1$ and 2, represent the two or the unique pole solution(s). The N$_x$ columns represent respectively the number of dense light-curves, the number of oppositions and the number of sparse points from the USNO (MPC code 689), Catalina (MPC code 703) and Lowell surveys. $\eta_a$ represents the fraction of flat surfaces present on the shape model as inferred by the bootstrap method. Finally, the $D$ column represents the equivalent diameter of the sphere having the same volume as the asteroid shape model. When we were not able to scale the shape model of the asteroid, the NEOWISE diameter \citep{Mai_2016} (or WISE diameter \citep{Mas_2011} when NEOWISE data are unavailable)  is given in parentheses.}
\label{tab:Obs_Table}
\end{table*} 

\begin{table*}
\centering
\begin{tabular}{|l|ll|l|llll|l|l|l|}
\hline
Asteroid& $b/a$   & $c/b$   & $P_{\rm sid}$                & $\lambda_1$ & $\beta_1$ & $\lambda_2$ & $\beta_2$ &$\eta_a$         & D  & Ref.  \\
                        &             &             &   [hours]              &     [deg]          &   [deg]       &  [deg]             & [deg]         &               & [km]  & \\
\hline
$234$ & $0.90 $     & $0.88 $  &$26.474$ & $144$     & $-38$& & & $0.33$ & $43.7 $ & [1] \\
$599$ &                                 &                               & $9.3240$ &                                      &                       &       &                       &       &$71$         & [2] \\
$606$ & $0.82 $   & $0.92  $   &$12.2907$ & $183$  &$20$   &    $354$  & $26$ & &  $35.5$   & [3] \\
$642$ &                                 &                                       & $8.19$  &                               &                       &                               &                         & &     $38.2$  & [4] \\
$673$ &                                 &                                       &$22.340$ &            &                  &                               &                       & &       $41.6$  &  [5] \\
$679$ & $0.72 $   & $0.92 $    &$8.45602$ & $220$ & $32$ &$42$ & $-5$ &$0.38$  & $51.4 $ & [6]\\
$1284$ &                                &                                       & $ 9.55$  &                              &                       &               &                         &                               & $46  $  & [7] \\
 $2448$ &                               &                               & $10.061$&                       &                       &                               &                         &       & $43$ & [8] \\
 \hline
\end{tabular}
\caption{Same as Table \ref{tab:Obs_Table}, but for asteroids which were not observed during this campaign. The bootstrap method was not applied to (606) because the number of dense light-curves was not sufficient. References : [1] \citet{b13}, [2] \citet{Deb_1977}, [3] \citet{Han_2011}, [4] \citet{Lec_2005}, [5] \citet{Mar_2016b}, [6] \citet{Mar_2011}, [7] \citet{Car_2016}, [8] \citet{Str_2013} }

\label{tab:Obs_Tab2}
\end{table*}

\subsection{Individual asteroids}
In the following we compare our results for each asteroid to some data available in the literature; asteroid occultation results \citep{Dun_2016} in particular.

\begin{figure*}
\centering
\begin{tabular}{|cc|}
\hline
\subf{\includegraphics[width=8cm]{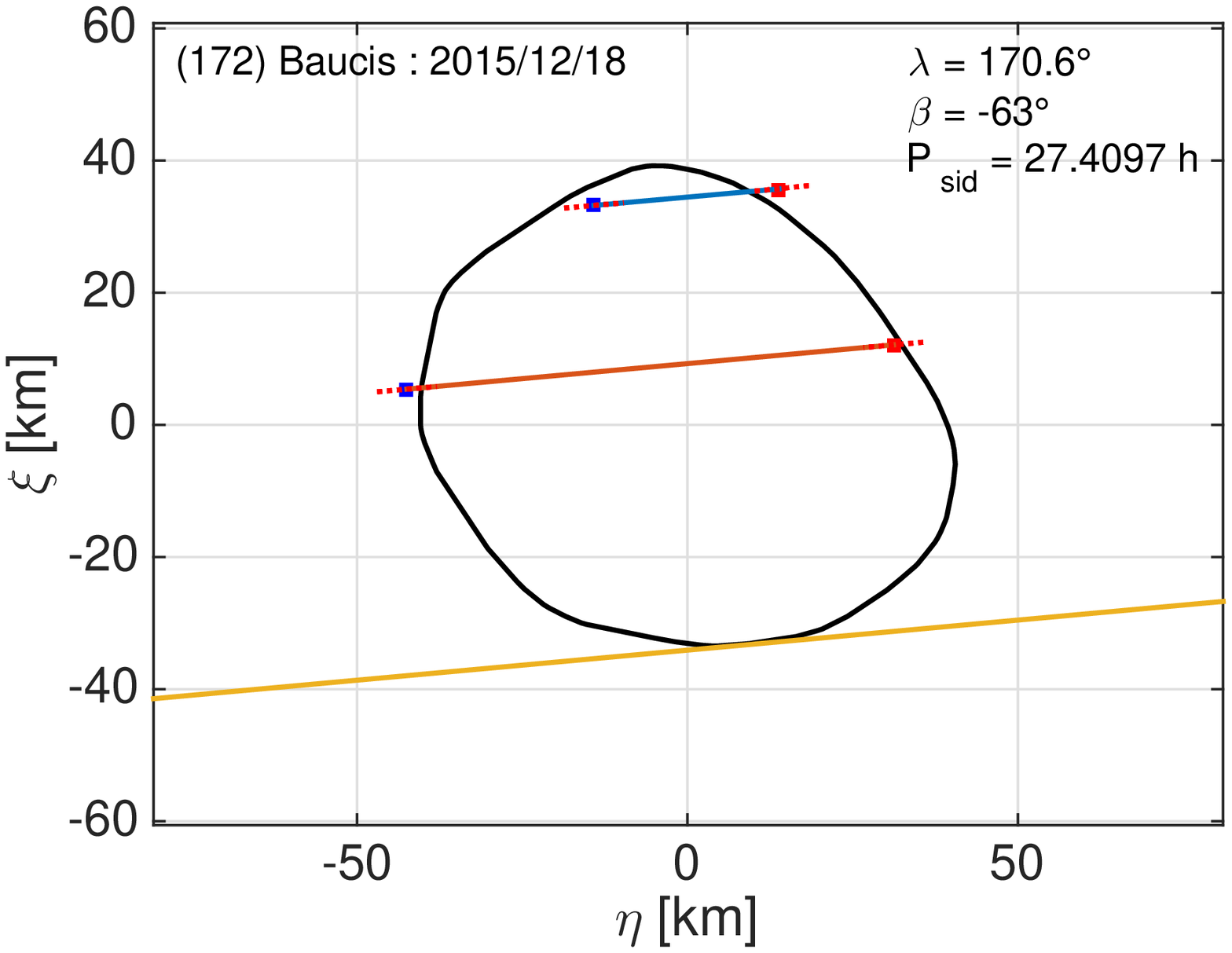}}
     {}
     &
     \subf{\includegraphics[width=8cm]{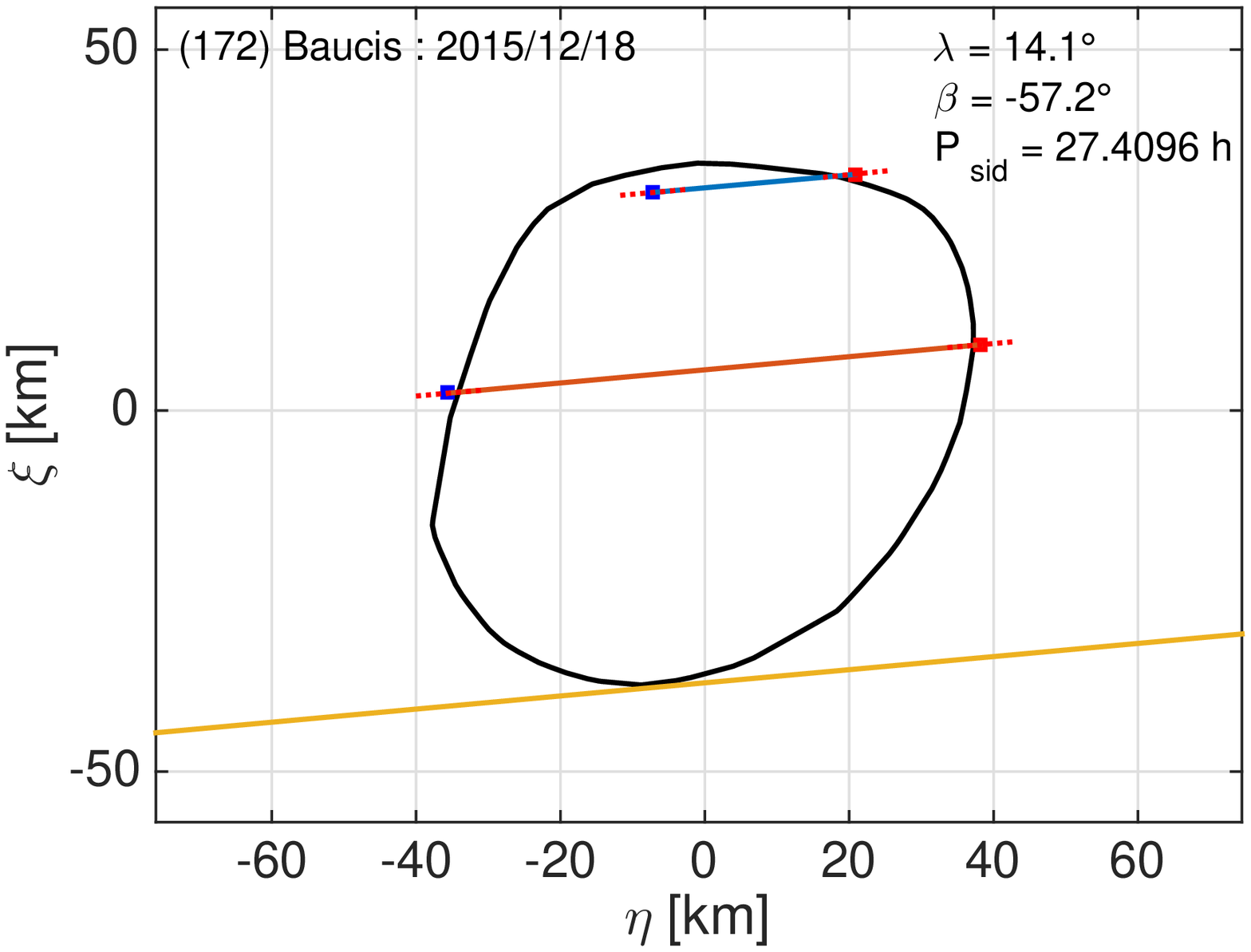}}
     {}
\\
\hline
\end{tabular}
\caption{Fit of the derived shape models for (172)~Baucis on the observed occultation chords from the 18 December 2015 event. The left panel represents the shape model corresponding to the first pole solution ($\lambda_1$, $\beta_1$). The right panel represents the shape model obtained using the second pole solution.}
\label{fig:Occ_Baucis}
\end{figure*}

\begin{figure*}
\centering
\begin{tabular}{|cc|}
\hline
\subf{\includegraphics[width=8cm]{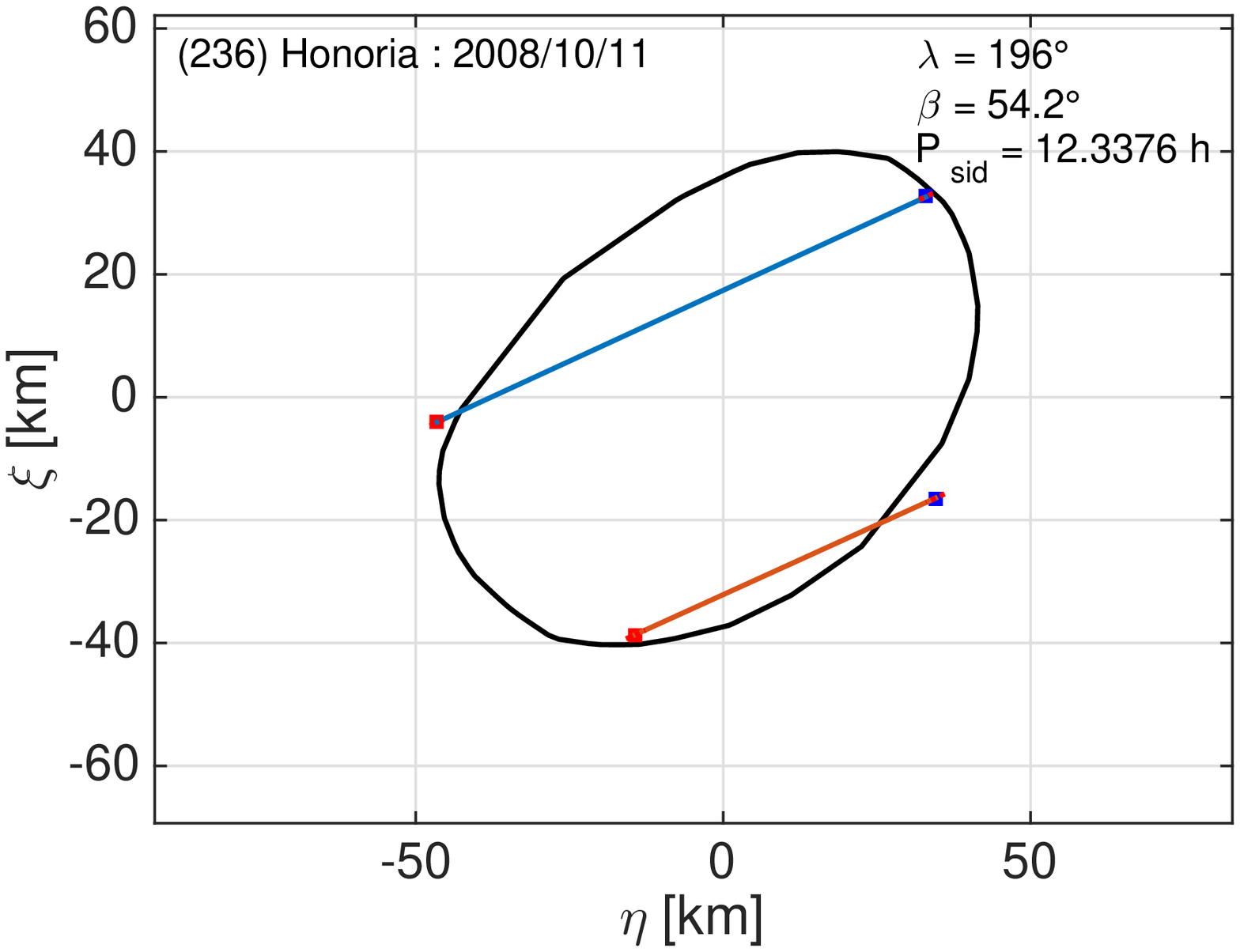}}
     {}
     &
     \subf{\includegraphics[width=8cm]{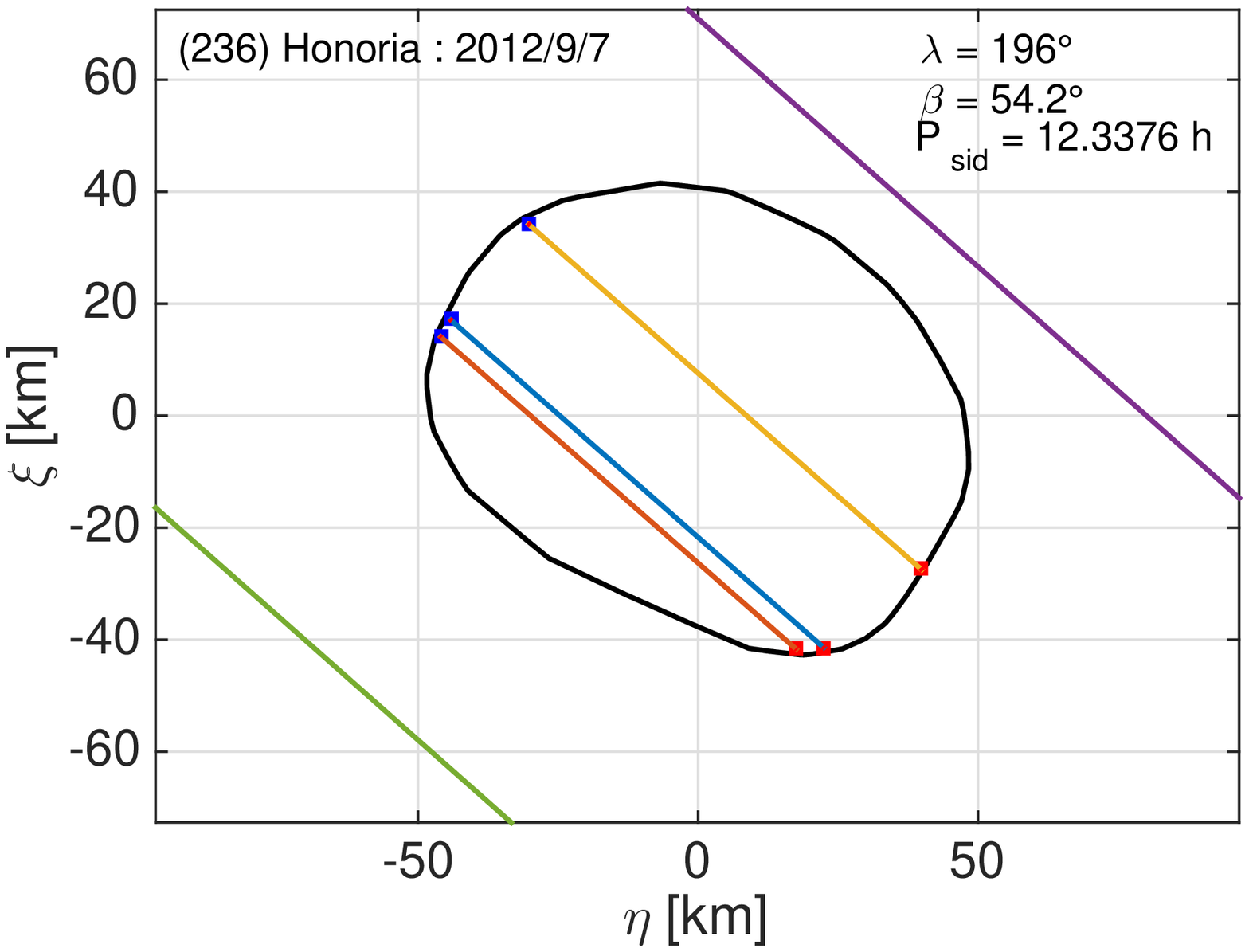}}
     {}
\\
\hline

\subf{\includegraphics[width=8cm]{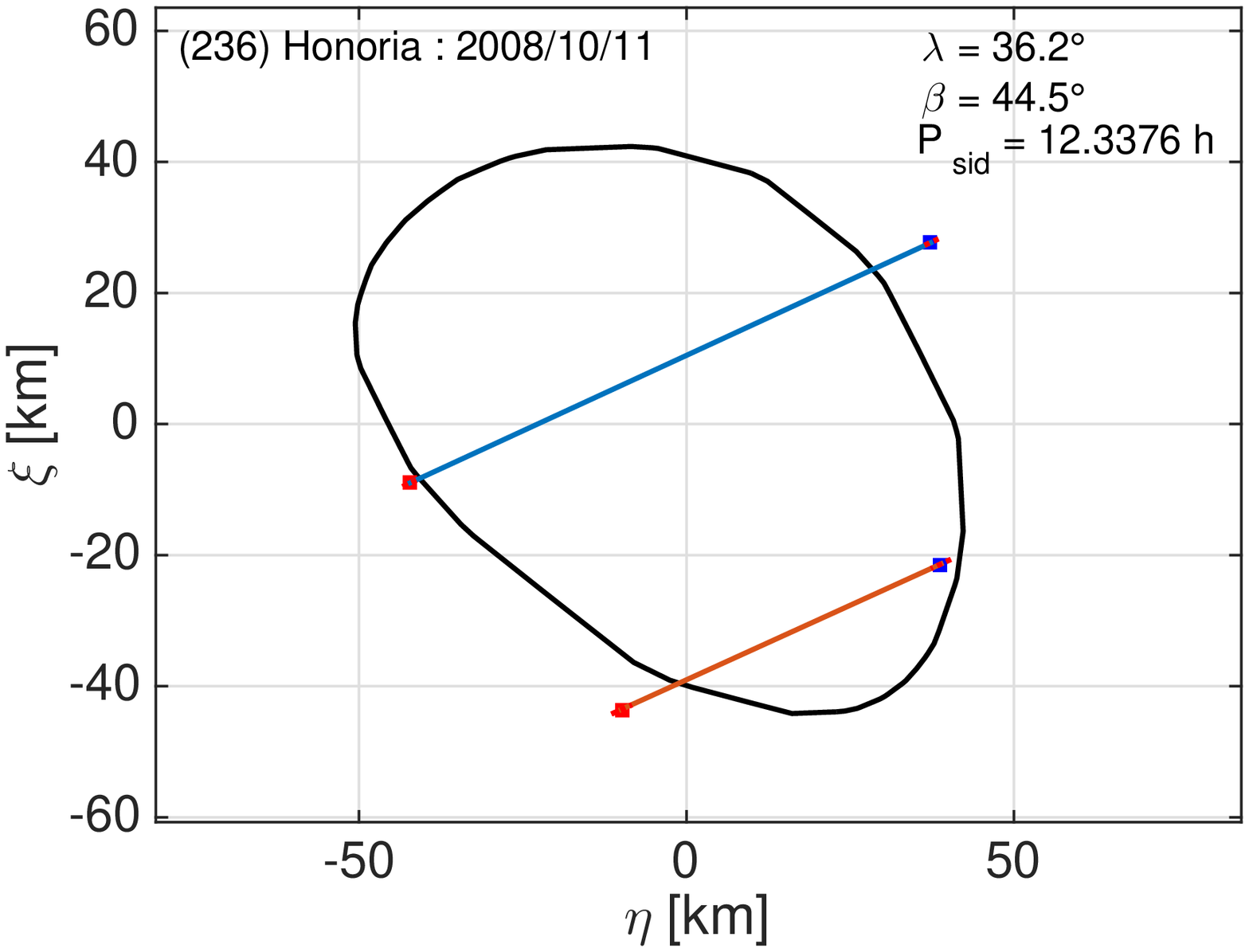}}
     {}
     &
     \subf{\includegraphics[width=8cm]{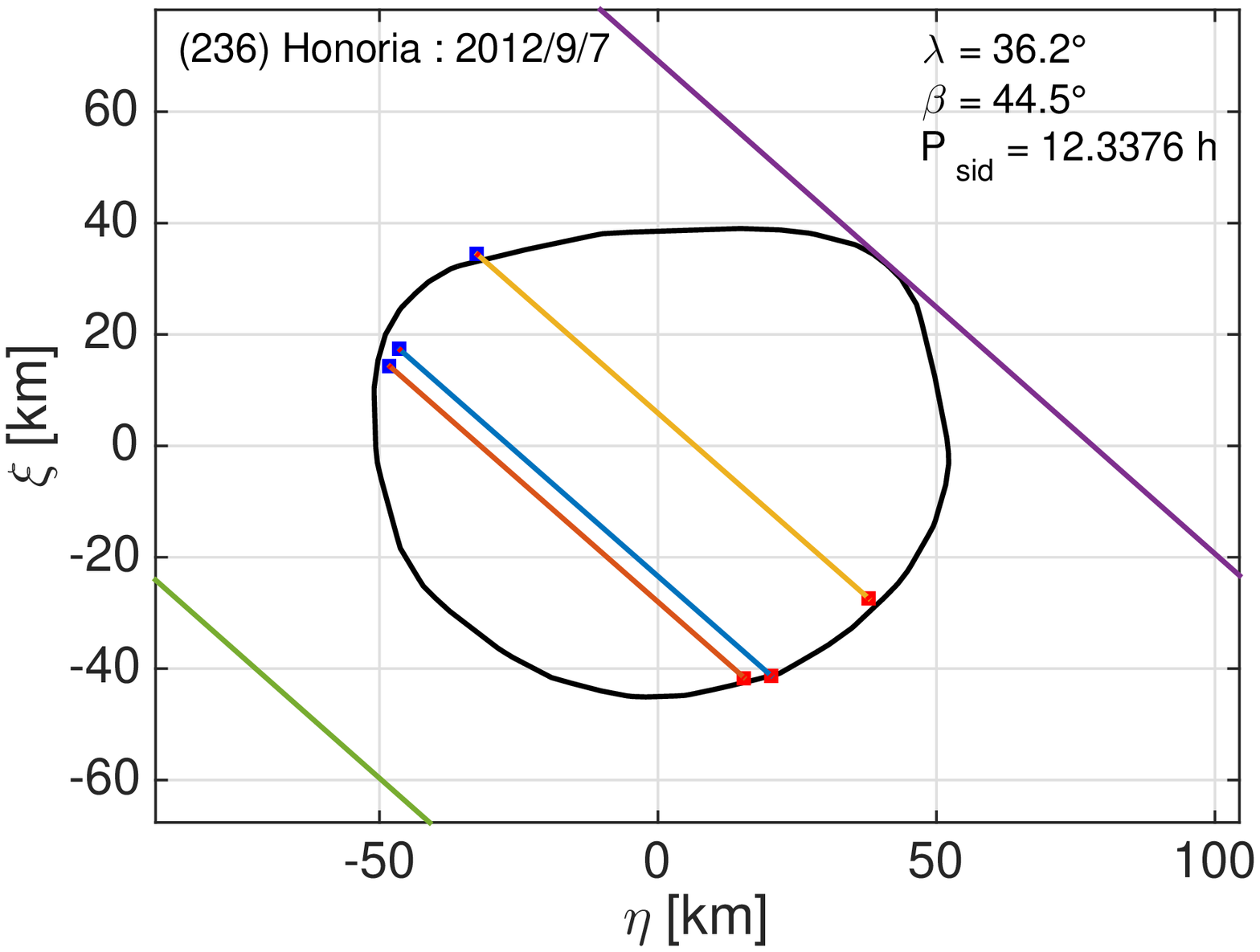}}
     {}
\\
\hline
\end{tabular}
\caption{Fit of the derived shape models for (236)~Honoria on the observed occultation chords from the 11 October 2011 and the 7 September 2012 events. The left panel represents the 2008 event with the top and bottom panels corresponding to the first and second pole solutions, respectively. The right panel is the same as the left one, but for the 2012 event.}
\label{fig:Hon_occ}
\end{figure*}

\begin{figure}
\centering
\includegraphics[width=8cm]{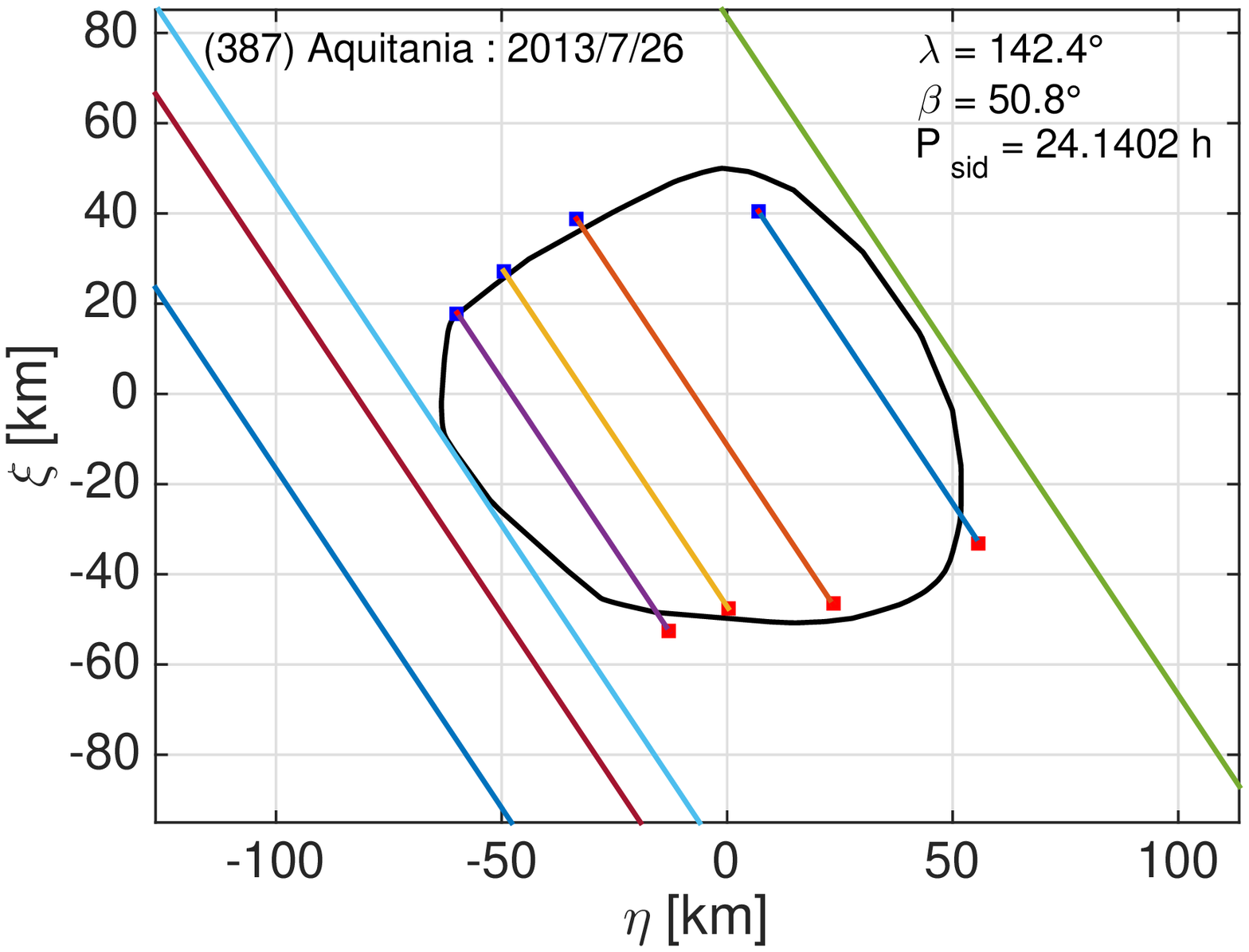}
\caption{Fit of the derived shape model for (387)~Aquitania on the observed occultation chords from the 26 July 2013 event.}
\label{fig:Aqui_occ2}
\end{figure}

\begin{figure*}
\centering
\begin{tabular}{|cc|}
\hline
\subf{\includegraphics[width=8cm]{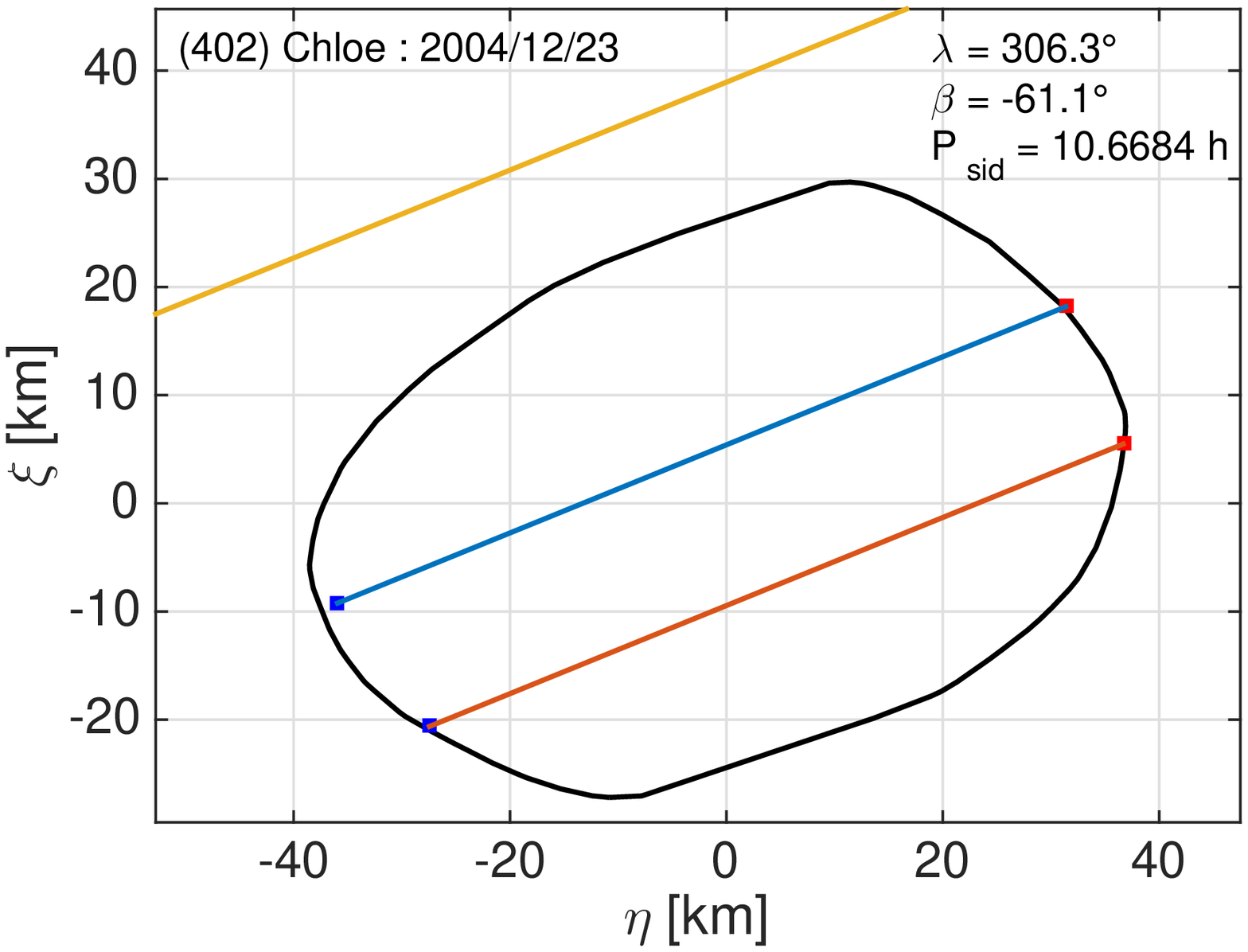}}
     {}
     &
     \subf{\includegraphics[width=8cm]{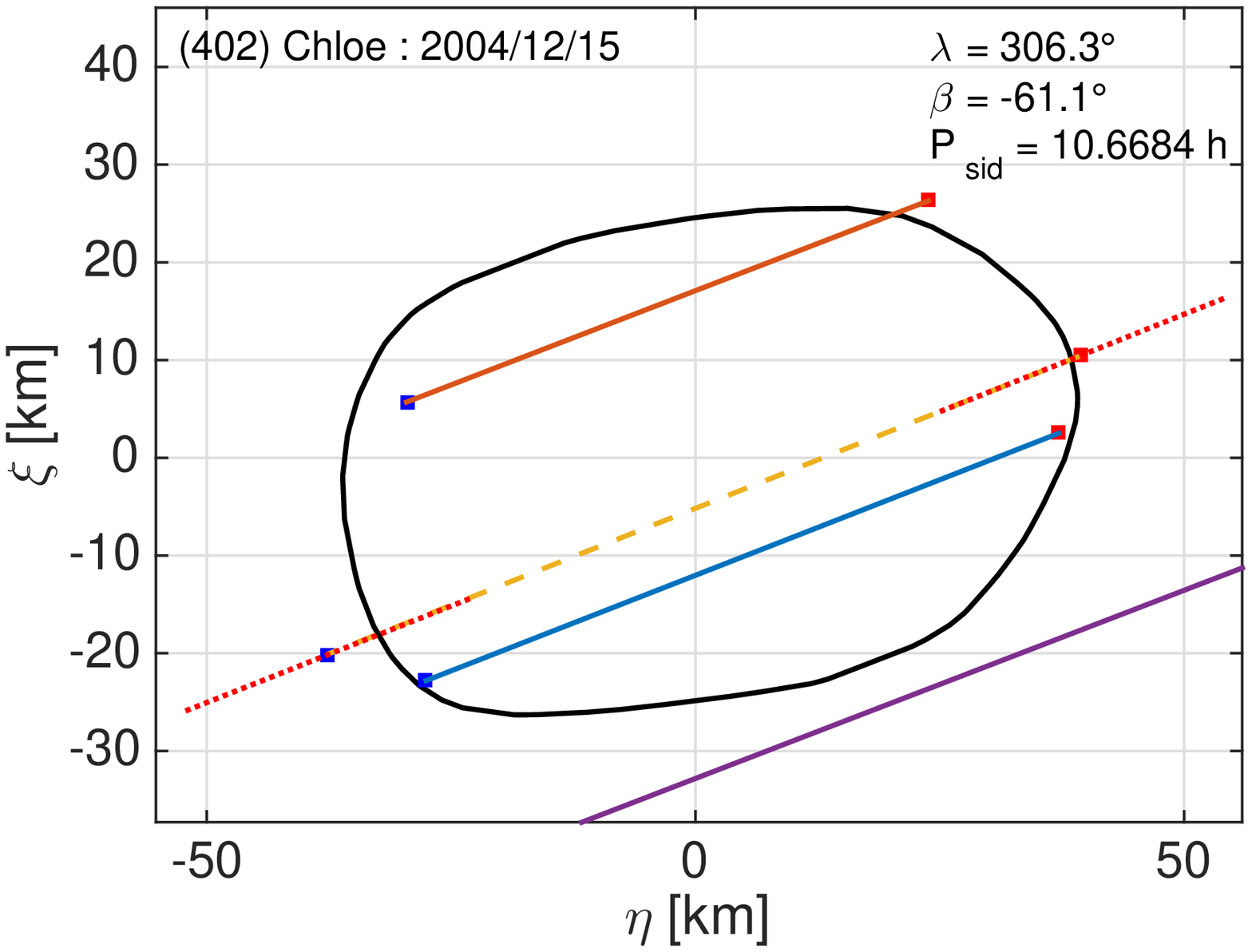}}
     {}
\\
\hline

\subf{\includegraphics[width=8cm]{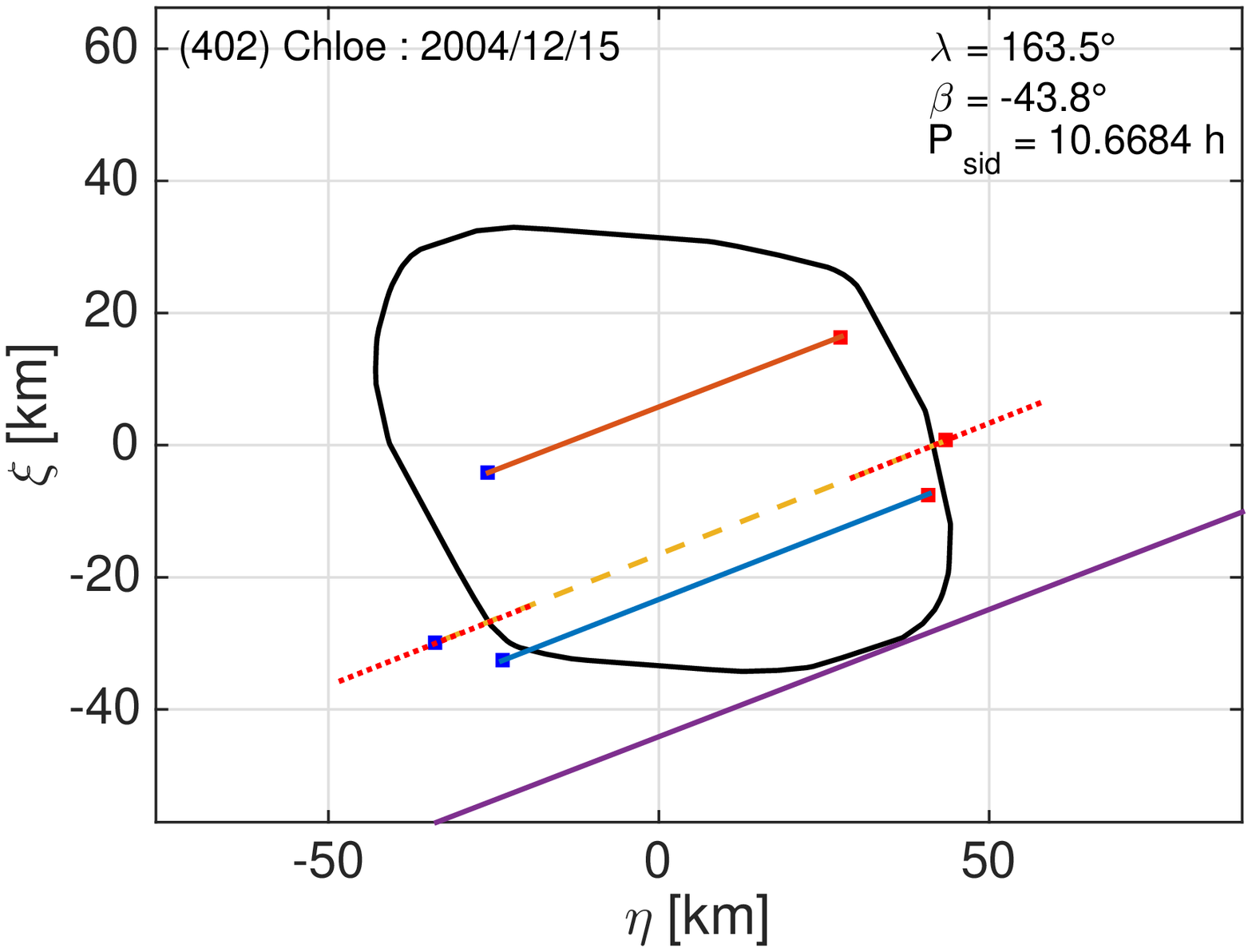}}
     {}
     &
     \subf{\includegraphics[width=8cm]{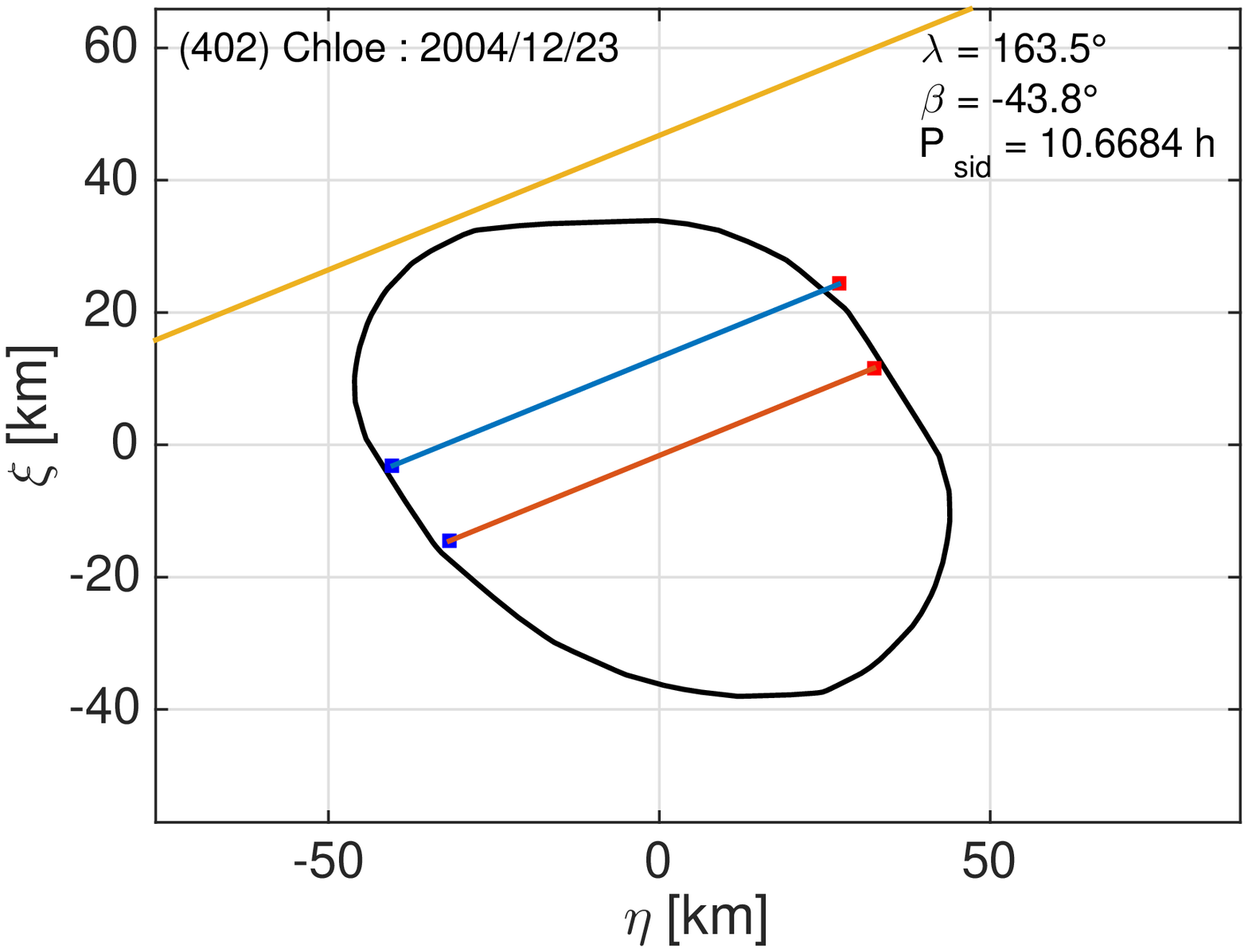}}
     {}
\\
\hline
\end{tabular}
\caption{Fit of the derived shape models for (402)~Chloe on the observed occultation chords from 15 and 23 December 2014 events. The leftmost part of the Figure represents the 15 event with the upper and lower parts corresponding to the first and second pole solutions, respectively. The rightmost part of the Figure is the same as the leftmost part, but for the 23 December event.}
\label{fig:Chloe_occdkusqhdkqs}
\end{figure*}

In Figs. \ref{fig:Occ_Baucis} to \ref{fig:Chloe_occdkusqhdkqs}, the occultations data are represented in the so-called fundamental plane (i.e. the plane passing through the centre of the Earth and perpendicular to the observer-occulted star vector) \citep{Dur_2011}. In that plane, the disappearance and reappearance of the occulted star are represented respectively by blue and red squares and the occultation chords represented by coloured continuous segments. Error bars on the disappearance and reappearance absolute timing are represented by discontinuous red lines. Negative observations (no occultation observed) are represented by a continuous coloured line. Discontinuous segments represent observations for which no absolute timing is available. For each asteroid, the corresponding volume-equivalent radius is computed (corresponding to the radius of a sphere having the same volume as the asteroid shape model). The dimensions of the ellipsoid best fitting the shape model are also given. The uncertainties on the absolute dimensions of the shape models adjusted to stellar occultations are derived by varying various parameters according to their own uncertainties. First, the different shape models obtained during the bootstrap method are used. The spin axis parameters are randomly chosen according to a normal distribution around their nominal values with standard deviation equal to the error bars given in Table~\ref{tab:Obs_Table}. Finally the extremes of each occultation chord are randomly shifted according to their timing uncertainties. The observed dispersion in the result of the scaling of the shape model on the occultations chords is then taken as the formal uncertainty on those parameters.

The list of all the observers of asteroid occultations which are used in this work can be found in Appendix D \citep{Dun_2016}. In the following we discuss some individual cases.

(172)~Baucis - There are $two$ reported occultations. The first event is a single-chord, while the other (18 December 2015) has two positive chords and two negative ones, and was used to scale the shape model (Fig. \ref{fig:Occ_Baucis}). According to the fit of the shape model to the occultation chords, both pole solutions seemed equally likely. The two solutions give similar absolute dimensions of $(41.1 \pm 2, 36.3 \pm 1.8, 31.8 \pm 1.6)$ and $(40.8 \pm 2.4, 36.1 \pm 1.7, 33.5 \pm 1.6)$ km, respectively. The corresponding volume-equivalent radii are $R_{\rm eq} = 36.2 \pm 1.8$ km and $R_{\rm eq} = 36.7 \pm 1.8$ km, to be compared to the NEOWISE \citep{Mai_2016}, AKARI \citep{Usui_2012} and IRAS \citep{Ted_1989} radii: $35.3 \pm 0.4$, $33.5 \pm 0.4$ and $31.2 \pm 0.6$ km, respectively. Even though the occultation observations cannot provide information about the best spin axis solution, the bootstrap curve indicates a clear preference for the one with retrograde rotation.

(236)~Honoria - There are eight observed occultations. However, over these eight events, only two have $\geq$  two positive chords (in 2008 with two positive chords and 2012 with three positive chords). These were used to constrain the pole orientation and scale the shape model. The result shows that the first pole solution  ($\lambda_1$ and $\beta_1$ from Table \ref{tab:Obs_Table}) is the most plausible. By simultaneously fitting the two occultations, the obtained dimensions are $(48.8 \pm 1.4, 48.3 \pm 1.3, 33.7 \pm 1.0)$ km, and the volume-equivalent radius $R_{\rm eq} = 43.0 \pm 2.1$ km. In the case of the second pole solution, the dimensions of the semi-axes are $(52.3 \pm 2.7, 52.1 \pm 2.6, 37.1 \pm 1.9)$ km and the equivalent radius is $R_{\rm eq} = 45.6 \pm 2.3$ km. These solutions are compatible with the NEOWISE, AKARI, and IRAS measurements which are respectively $38.9 \pm 0.6$, $40.6 \pm 0.5$ and $43.1 \pm 1.8$ km. The fit of the shape model for the two pole solutions is shown in Fig. \ref{fig:Hon_occ}. 

(387)~Aquitania - For this asteroid, the result of the light-curve inversion process gives different solutions for the rotation period when sparse data are included. However, the noise on the sparse data is higher than the amplitude of the light-curves themselves. For this reason we decided to discard them. As a consequence, we did not apply the bootstrap method to this asteroid.

There is one well sampled occultation of (387)~Aquitania providing four positive and four negative chords. The adjustment of the unique solution of the shape model on the occultation chords shows a good agreement. The scaling leads to an equivalent radius $R_{\rm eq} = 50.4 \pm 2.5$ km. The absolute dimensions are $(54.2 \pm 2.7, 50.6 \pm 2.5, 46.6 \pm 2.3)$ km. This is consistent with the WISE \citep{Mas_2011}, AKARI, and IRAS radii, which are respectively $48.7 \pm 1.7$, $52.5 \pm 0.7,$ and $50.3 \pm 1.5$ km. The fit of the shape model on the occultation chords is shown in Fig. \ref{fig:Aqui_occ2}.

\citet{Han_2017} found a relatively similar model, spin axis solution ($P_{\rm sid} = 24.14012$ hours, $\lambda = 123^{\circ} \pm 5^{\circ}$, and $\beta =  46^{\circ} \pm 5^{\circ}$), and size ($48.5 \pm 2$ km) using an independent approach.

(402)~Chloe - There are six reported occultations, but only two can be used to adjust the shape models on the occultation chords. These two occultations occurred on the 15 and 23 December 2004. The result clearly shows that the first pole solution is the best one. The other one would require a deformation of the shape that would not be compatible with the constraints imposed by the negative chords. The derived dimensions of (402)~Chloe, for the first spin axis solution, are $(38.3 \pm 1.9, 33.7 \pm 1.7, 24.7 \pm 1.2)$ km and the equivalent radius is $R_{\rm eq} = 31.3 \pm 1.6$ km. The second spin solution leads to the dimensions $(46.9 \pm 2.4, 41.3 \pm 2.1, 29.0 \pm 1.5 )$ km and $R_{\rm eq} = 38.3 \pm 1.9$ km. The NEOWISE radius is $27.7 \pm 0.8$ km while the AKARI and IRAS radii are $30.2 \pm 1.5$ and $27.1 \pm 1.4$ km, respectively, which agree with the first pole solution, but not with the second one. 
The fit of the shape model on the occultation chords is shown in Fig. \ref{fig:Chloe_occdkusqhdkqs}.  

(824)~Anastasia - The light-curve inversion technique provided three possible pole solutions. However, two of them present a rotation axis which strongly deviates from the principal axis of inertia. This is probably related to the extreme elongation of the shape model. The light-curve inversion technique is known to have difficulty in constraining the relative length along the axis of rotation, in particular. In the case of highly elongated bodies, the principal axis of inertia is poorly constrained.

Two occultations were reported for this asteroid. The models cannot fit the two occultations simultaneously using a single scale factor. We decided therefore not to present the shape model and spin solutions in this paper as new observations are required to better constrain the pole solution. Because of the very large rotation period of this asteroid, the presence of a possible tumbling state should also be tested.

(2085)~Henan - This asteroid possesses a relatively large rotation period. Based on our photometric observation, two synodic periods can be considered ($110 \pm 1$ hours and $94.3 \pm 1$ hours). However, the light-curve inversion technique seems to support the first one based on our observations and sparse data.

\subsection{Presence of concavities}

We have applied the flat detection method (FSDT) described in \citet{Dev_2015} to all the shape models presented in this paper, thus (Sec. \ref{Bootstrap}) providing a quantitative estimate for the presence of concavities and for the departure of an asteroid shape from a smooth surface with only gentle changes of curvature. 

\begin{figure}
\centering
\includegraphics[width=9cm]{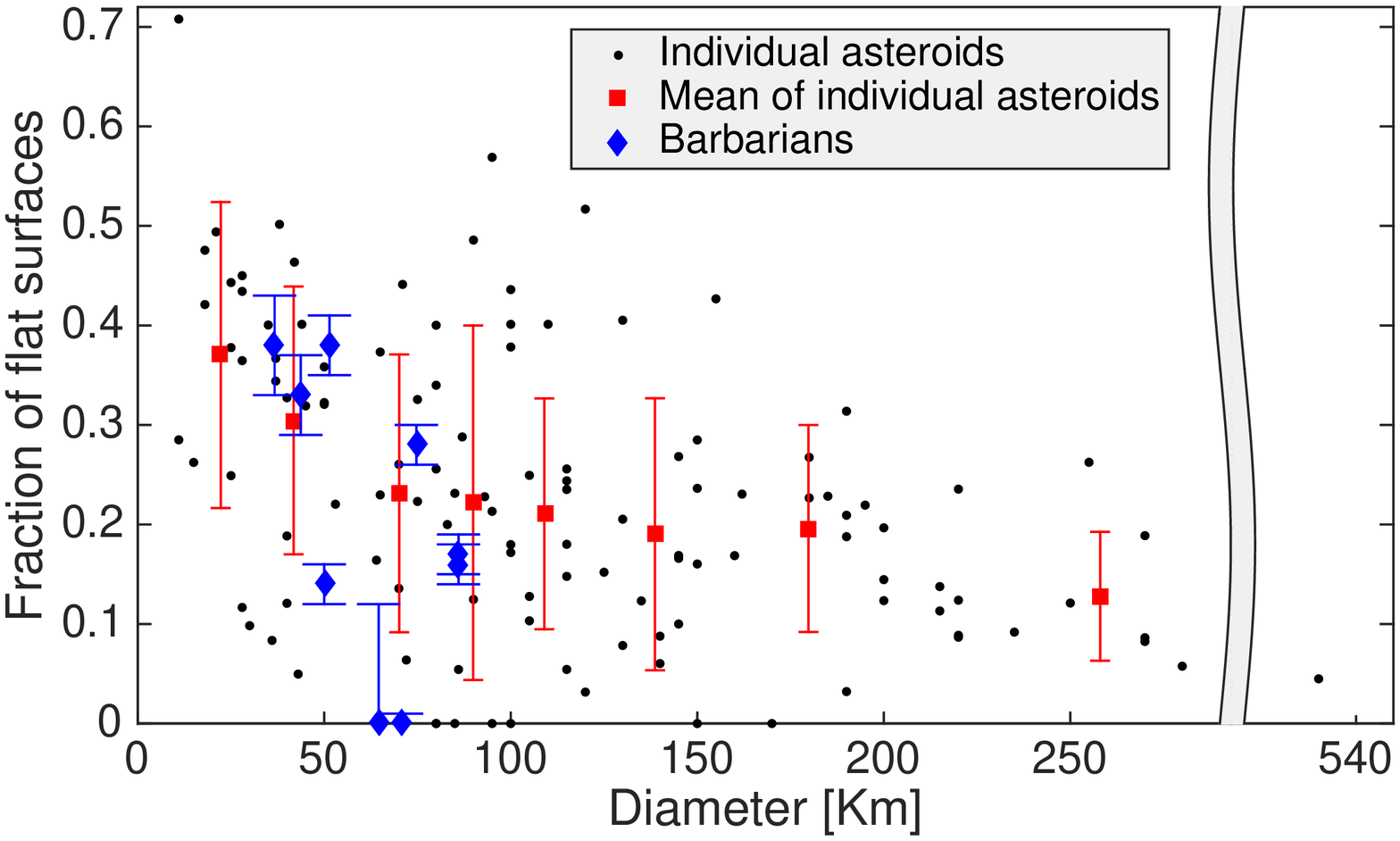}
\caption{Values of $\eta_a$ as a  function of the asteroids diameters. The black dots correspond to individual asteroids. The red squares correspond to the mean values of $\eta_a$ and diameter for bins of 15 asteroids. The blue diamonds correspond to Barbarian asteroids.}
\label{fig:FlatVSFlat}
\end{figure}

Fig. \ref{fig:FlatVSFlat} presents the result of the FSDT for the shape models presented in this work (blue diamonds, see Tables \ref{tab:Obs_Table} and \ref{tab:Obs_Tab2}). These results can be compared to the average value of $\eta_a$ computed by bins of 15 asteroids (red squares) using the asteroid population described in Sec. \ref{Sec:flatVSDiam} (black dots).

From this distribution, a correlation between diameter and fraction of flat surfaces seems to be apparent. Such correlation is expected since small asteroids are more likely to possess irregular shapes than larger ones. However, the quality of asteroid light-curves is directly correlated to the asteroids diameter. For small asteroids (< 50 km), this effect might also increase the mean value of $\eta_a$.
Fig. \ref{fig:FlatVSFlat} does not show clear differences between Barbarians and other asteroids, as Barbarians seem to populate the same interval of values. As expected, some asteroids like Barbara show a large amount of flat surfaces, but this is not a general rule for Barbarians. Of course the statistics are still not so high, but our initial hypothesis that large concave topological features could be more abundant on objects having a peculiar polarisation does not appear to be valid in view of this first attempt of verification.

\subsection{Are Barbarian asteroids slow rotators ?}

In this Section, we compare the rotation periods of Barbarians and Barbarian candidates with those of non-Barbarian type. At first sight the rotation period of confirmed Barbarian asteroids seems to indicate a tendency to have long rotation periods, as the number of objects exceeding 12h is rather large. For our analysis we use the sample of objects as in Tables \ref{tab:Obs_Table} and \ref{tab:Obs_Tab2}. By considering only asteroids with a diameter above 40 km, we are relatively sure that our sample of L-type asteroids is both almost complete, and not affected by the Yarkovsky-O’Keefe-Radzievskii-Paddack effect \citep{b10}. However, as only 13 asteroids remain in the sample, we also consider a second cut at 30 km, increasing it to 18. As shown below, our findings do not change as a function of this choice.

We proceed by comparing the rotation period of the Barbarians with that of a population of asteroids that possess a similar size distribution. Since the distribution of rotation periods is dependent on the size of the asteroids, we select a sample of asteroids with the same sizes, using the following procedure. For each Barbarian, one asteroid is picked out randomly in a population for which the rotation period is known, and having a similar radiometric diameter within a $\pm 5$~km range. This way we construct a population of asteroids with the same size distribution as the Barbarians. We now have two distinct populations that we can compare. By repeating this process several times, we build a large number of populations of `regular' asteroids. We can then check the probability that the distribution of rotation periods of such a reconstructed population matches the one of the Barbarians.

Considering the sample of asteroids with a diameter larger than 40 km, the sample is too small to derive reliable statistics. However, we notice that the median of the rotation period of the Barbarians is $20.1$ hours while the median of the population regular asteroid is only $12.0$ hours. 

In order to improve the statistic, we can also take into account asteroids with a diameter between 30 and 40 km. The median of the Barbarian asteroids is now $20.6$ hours while the regular asteroids population has a median of $11.4$ hours. 

\begin{figure}
\centering
\includegraphics[width=10cm]{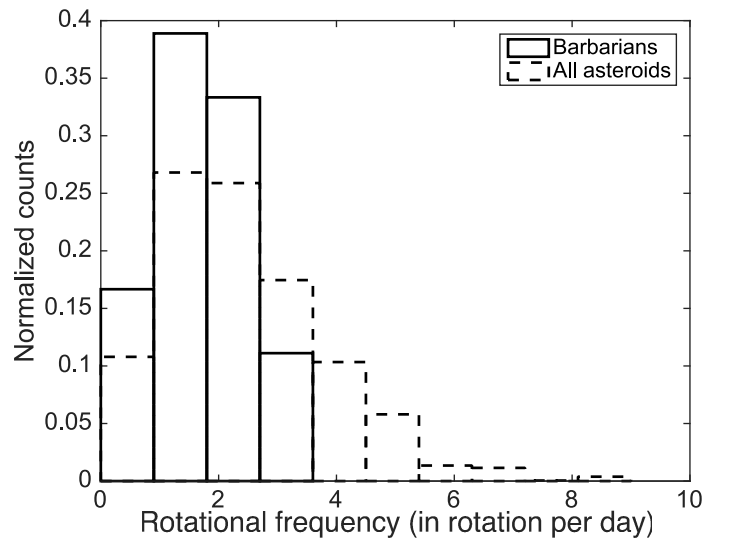}
\caption{Normalised histograms of the Barbarian compared to the histogram asteroids having a diameter between 110 and 30 km.}
\label{fig:Hist_Barb_30}
\end{figure}

Fig. \ref{fig:Hist_Barb_30} represents the histograms of the rotation periods of Barbarian asteroids (with diameter > 30 km) compared to the population of regular asteroids constructed as described at the beginning of this section. The histogram of the Barbarian asteroids shows a clear excess of long rotation periods and a lack of fast-spinning asteroids. 

We adopt the two-samples Kolmogorov-Smirnov (KS) test to compare the distributions of Barbarians and regular asteroids. This is a hypothesis test used to determine whether two populations follow the same distribution. The two-sample KS test returns the so-called asymptotic $p$-value. This value is an indication of the probability that two samples come from the same population.  

In our case, we found $p = 0.6 \%$, clearly hinting at two distinct populations of rotation period.

In conclusion our results show that there is clear evidence that the rotations of Barbarian asteroids are distinct from those of the whole population of asteroids. The abnormal dispersion of the rotation periods observed for the Barbarian asteroids is more probably the result of a true difference than a statistical bias.

\section{Conclusions}

We have presented new observations here for $15$ Barbarian or  candidate Barbarian asteroids. For some of these asteroids, the observations were secured by us during several oppositions. These observations allow us to improve the value of the rotation period. For eight of them, we were able to determine or improve the pole orientation and compute a shape model.

The shape models were analysed using a new approach based on the technique introduced by \citet{Dev_2015}. We show that this technique is capable of providing an indication about the completeness of a light-curve data set indicating whether or not further observations are required to better define a shape solution of the photometric inversion method. The extrapolation of the trend towards a large number of light-curves gives a more precise determination of $\eta_s$. This new method was applied to a large variety of shape models in the DAMIT database. This allowed us to construct a reference to which the shape models determined in this work were compared. Our results show that there is no evidence that our targets have more concavities or are more irregular than a typical asteroid. This tends to infirm the hypothesis that large-scale concavities may be the cause of the unusual polarimetric response of the Barbarians.

The improvement and new determination of rotation periods has increased the number of asteroids for which a reliable rotation period is known. This allows us to have an improved statistic over the distribution of rotation periods of the Barbarian asteroids. We show in this work possible evidence that the Barbarian asteroids belong to a population of rotation periods that contains an excess of slow rotators, and lacks fast spinning asteroids. The relation between the polarimetric response and the unusual distribution of rotation periods is still unknown.

\label{Conclusions}
 
\section*{Acknowledgements}

The authors thank the referee for his comments, which improved the manuscript.

MD and JS acknowledge the support of the Universit{\'e} de Li{\`e}ge.

MD and PT acknowledge the support of the French ``Programme Nationale de Plan\'etologie".

Part of the photometric data in this work were obtained at the C2PU facility (Calern Observatory, O.C.A.).

NP acknowledges funding from the Portuguese FCT - Foundation for Science and Technology. CITEUC is funded by National Funds through FCT - Foundation for Science and Technology (project: UID/ Multi/00611/2013) and FEDER - European Regional Development Fund through COMPETE 2020 - Operational Programme Competitiveness and Internationalisation (project: POCI-01-0145-FEDER-006922).

SARA observations were obtained under the Chilean Telescope Allocation Committee program CNTAC 2015B-4.

PH acknowledges financial support from the Natural Sciences and Engineering Research Council of Canada, and thanks the staff of Cerro Tololo Inter-American Observatory for technical support.

The work of AM was supported by grant no. 2014/13/D/ST9/01818 from the National Science Centre, Poland.

The research of VK is supported by the APVV-15-0458 grant and the VVGS-2016-72608 internal grant of the Faculty of Science, P.J. Safarik University in Kosice.

MK and OE acknowledge TUBITAK National Observatory for a partial support in using T100 telescope with project number 14BT100-648.

\begin{appendix}
\section{Summary of the new observations used in this work}

\begin{table*}
\begin{tabular}{lllll}
\hline
Asteroid & Date of observations & $N_{\rm lc}$ &Observers & Observatory \\
\hline
(122)~Gerda  & 2015 Jun 09 - Jul 06                    & 6 & M. Devog\`ele & Calern observatory, France  \\
(172)~Baucis & 2004 Nov 20 - 2005 Jan 29 & 2 & F. Manzini &Sozzago, Italy   \\
                                         & 2005 Feb 22 - Mar 1                    & 3 & P. Antonini & B\'edoin, France  \\
                                     & 2013 Mar 1 - Apr 25                        & 10 & A. Marciniak, M. Bronikowska  & Borowiec, Poland  \\
                                         &                                                         &    & R. Hirsch, T. Santana,  \\
                                         &                                                                                         &    & F. Berski, J. Nadolny,                  \\      
                                         &                                                                                       &     & K. Sobkowiak                             \\
                                         & 2013 Mar 4 - Apr 17                   & 3 & Ph.Bendjoya, M. Devog\`ele  & Calern observatory, France  \\
                                         & 2013 Mar 5 - Apr 16               & 15 &  P. Hickson  & UBC Southern Observatory,  \\
                                         &                                                              &                                       &                                                       &La Serena, Chile \\
                                         &2013 Mar 5 - 8 & 3 &M. Todd and D. Coward & Zadko Telescope, Australia   \\  
                                         &2013 Mar 29    & 1 &F. Pilcher & Organ Mesa Observatory, Las Cruces, NM  \\
                                         &2014 Nov 13             & 1   & M. Devog\`ele & Calern observatory, France  \\
                                         &2014 Nov 18 - 26 & 5 &        A. U. Tomatic, K. Kami{\'n}ski & Winer Observatory (RBT), USA  \\
(236)~Honoria    & 2006 Apr 29 -  May 3                                                 & 3       & R. Poncy  & Le Cr\`es, France  \\
                                                 & 2007 Jul 31 - Aug 2                  & 3       &       J. Coloma, H. Pallares & Sabadell, Barcelona, Spain      \\
                                                & 2007 Jun 21 - Aug 21                  &       2 &        E. Forne,  &  Osservatorio L'Ampolla, Tarragona, Spain  \\
                                                & 2012 Sep 19 - Dec 06 & 9 &  A. Carbognani, S. Caminiti& OAVdA, Italy  \\
                                                & 2012 Aug 20 - Oct 28                          &       6       &       M. Bronikowska, J. Nadolny & Borowiec, Poland  \\          
                                                &                                                                                               &               & T. Santana, K. Sobkowiak                &                                               \\                              
                                                & 2013 Oct 23 - 2014 03 20 & 8 & D. Oszkiewicz, R. Hirsch                  & Borowiec, Poland  \\
                                                &                                                                                       &       &       A. Marciniak, P. Trela,                            &                                        \\
                                                &                                                                                       &       & I. Konstanciak, J. Horbowicz            &                                        \\
                                                & 2014 Jan 03 - Feb 19          & 7& F. Pilcher & Organ Mesa Observatory, Las Cruces, NM   \\

                                             &2014 Feb 07 - Feb 17 & 2 &M. Devog\`ele & Calern observatory, France  \\     
                                                & 2014  Mar 09 - 20                             &       2       &P. Kankiewicz &            Kielce, Poland  \\                           
                                             & 2015 Mar 21 - 22                 & 2 &  W. Og{\l}oza, A. Marciniak,                                &Mt. Suhora, Poland  \\
                                             &                                                                       &    & V. Kudak                                                                             &                \\              
                                             &2015 Mar 22 - Mai 07      & 3& F. Pilcher                                        & Organ Mesa Observatory, Las Cruces, NM   \\
(387)~Aquitania & 2012 Mar 22 - May 23          & 4     & A. Marciniak, J. Nadolny, & Borowiec, Poland  \\ 
                                                &                                                                                       &               & A. Kruszewski, T. Santana, &                                                            \\
                                                & 2013 Jun 26 - Jul 25                  & 2       & M. Devog\`ele  & Calern observatory, France   \\                      
                                                & 2014 Nov 19 - Dec 29          &  4      &       A. U. Tomatic, K. Kami{\'n}ski & Winer Observatory (RBT), USA  \\                                                                                                                 &                                                                                       &               & K. Kami{\'n}ski                                                 &                                                       \\
                                                &2014 Nov 19 - Dec 18 & 2 & M. Todd and D. Coward & Zadko Telescope, Australia   \\  
                                                & 2014 Dec 17                           & 1 & F. Char                                             & SARA, La Serena, Chile  \\
                                                & 2014 Dec 18                           & 1 &                     M. Devog\`ele  & Calern observatory, France   \\         
                                                & 2015 Jan 10 - 11              & 2 & Toni Santana-Ros, Rene Duffard      & IAA, Sierra Nevada Observatory                 \\                                                
                                                  &2015 Nov 27 - 2016 Feb 16 & 2 & M. Devog\`ele  & Calern observatory, France   \\ 
(402)~Chloe             &       2015 Jul 01 - 14                        & 2 & M.C. Qui{\~n}ones           & EABA, Argentina  \\   
(458)~Hercynia & 2013 May 20 - Jun 05 & 4 & M. Devog\`ele  & Calern observatory, France   \\ 
                                                & 2014 Nov 27 - 28 & 2 &         M.-J. Kim, Y.-J. Choi           & SOAO, South Korea                                      \\
                                                & 2014 Nov 27           & 1 &             A. Marciniak                    &Borowiec, Poland        \\
                                                & 2014 Nov 28   - 29 & 2 &       A.U. Tomatic, K. Kami{\'n}ski                            & Winer Observatory (RBT), USA  \\ 
                                                & 2014 Nov 29                   & 1 &     M. Devog\`ele  & Calern observatory, France   \\ 
                                                & 2014 Nov 30                   & 1 &     N. Morales    & La Hita, Spain                                                          \\
(729)~Watsonia  & 2013 Apr 04 - Jun 06 & 16 & M. Devog\`ele  & Calern observatory, France   \\
                                                & 2014 May 26 - Jun 18 & 16 & R. D. Stephens &   Santana Observatory              \\                     
                                                & 2015 Sep 25 & 1 & O. Erece, M. Kaplan,              & TUBITAK, Turkey                \\ 
                                                &                                               &       & M.-J. Kim                                                       &                                                                       \\
                                                & 2015 Oct 05 & 1 &     M.-J. Kim, Y.-J. Choi & SOAO, South Korea                      \\
                                                & 2015 Oct 23 - Nov 30 & 5 & M. Devog\`ele  & Calern observatory, France   \\
(824)~Anastasia& 2015 Jul 05 - 25 & 7 & M. Devog\`ele  & Calern observatory, France   \\ 
 \end{tabular}
\caption{New observations used for period and shape model determination which were presented in this work and observations that are not included in the UAPC.}
\label{tab:SumarryObs1}
\end{table*}

\begin{table*}
\addtocounter{table}{-1}
\begin{tabular}{lllll}
\hline
Asteroid & Date of observations & $N_{\rm lc}$ &Observers & Observatory\\
\hline
(980)~Anacostia &2005 Mar 15 - Mar 16 & 2 & J.G. Bosch                           &Collonges, France  \\    
                                                 &2009 Feb 14 - Mar 21 & 6 & M. Audejean  & Chinon, France  \\  
                                                 &2012 Aug 18 - Nov 02 & 7 & K. Sobkowiak, M. Bronikowska & Borowiec  \\
                                                 &                                                        &      & M. Murawiecka, F. Berski                              &                                        \\
                                                 &2013 Feb 22 - Feb 24 & 3 & M. Devog\`ele  & Calern observatory, France   \\
                                                 &2013 Mar 27 - Apr 15  & 7 & A. Carbognani & OAVdA, Italy  \\ 
                                                 &2013 Dec 19 - Feb 23 & 11 & A.U. Tomatic & OAdM, Spain  \\
                                                 &2013 Dec 18 - Feb 07 & 2 & R. Hirsch, P. Trela & Borowiec, Poland   \\
                                                 &2014 Feb 22 - Mar 27& 4 & M. Devog\`ele  & Calern observatory, France   \\
                                                 &2014 Feb 25 - Mar 7   & 3 & J. Horbowicz, A. Marciniak & Borowiec, Poland  \\
                                                 &2014 Feb 27 & 1 & A.U. Tomatic & OAdM, Spain  \\
                                                 &2015 Mar 11 - May 28 & 5 & F. Char &  SARA, La Serena, Chile  \\
                                                 &2015 Jun 09 & 1 & M.C. Qui{\~n}ones & EABA, Argentina  \\
(1332)~Marconia & 2015 Apr 12 - May 18 & 8 & M. Devog\`ele  & Calern observatory, France   \\
                                                & 2015 May 20 & 1 &     A.U. Tomatic, K. Kami{\'n}ski& Winer Observatory (RBT), USA  \\ 
                                                & 2015 Jun 03 & 1 &             R.A. Artola                   & EABA, Argentina  \\  
(1372)~Haremari &2009 Nov 03 - Dec 08 & 8 & R. Durkee & Shed of Science Observatory,\\
                                                        &                                                                       &       &                                       & Minneapolis, MN, USA                                             \\
                                                        & 2014 Dec 11 - Feb 11 & 7 & M. Devog\`ele  & Calern observatory, France   \\
                                                & 2015 Jan 10 - 11              & 2 & T. Santana-Ros, R. Duffard & IAA, Sierra Nevada Observatory, Spain                                                           \\      
(1702)~Kalahari   & 2015 Apr 29 - May 18 & 3 & M. Devog\`ele  & Calern observatory, France   \\
                                                        & 2015 May 19                           & 1 & A. U. Tomatic, K. Kami{\'n}ski      & Winer Observatory (RBT), USA  \\ 
                                                        & 2015 May 20                           & 1 &M.-J. Kim, Y.-J. Choi & SOAO, South Korea             \\
                                                        &2015 Jun 04 - Jul 08 & 8 &  C.A. Alfredo          & EABA, Argentina  \\           
(2085)~Henan & 2015 Jan 09 - 10 & 2 & A. U. Tomatic & OAdM, Spain  \\
                                                & 2015 Jan 11 & 1 &T. Santana-Ros, R. Duffard & IAA, Sierra Nevada Observatory, Spain               \\     
                                                & 2015 Jan 14 - 15 & 2 &M. Todd and D. Coward & Zadko Telescope, Australia   \\  
                                                &2015 Jan 14 - Feb 11 & 7 & M. Devog\`ele  & Calern observatory, France   \\
                                                & 2015 Mar 11                           & 1 & F. Char & SARA, La Serena, Chile  \\        
(3844)~Lujiaxi & 2014 Nov 12 - Dec 23 & 5 & M. Devog\`ele  & Calern observatory, France   \\     
                                                & 2014 Nov 20 - Dec 18 & 2 &M. Todd and D. Coward&Zadko Telescope, Australia \\
                                                & 2014 Nov 20 & 1 &     A.U. Tomatic, K. Kami{\'n}ski        & Winer Observatory (RBT), USA  \\ 
                                                & 2014 Dec 25 - 26 & 2 & A.U. Tomatic & OAdM, Spain  \\
(15552)~& 2014 Oct 29 - Nov 25 & 7 & A. U. Tomatic, K. Kami{\'n}ski     & Winer Observatory (RBT), USA  \\ 
Sandashounkan                            & 2014 Oct 30 - Nov 13 & 2 & M. Devog\`ele  & Calern obervatory, France   \\    
                                                                        & 2014 Nov 20 & 1 &  A. U. Tomatic & OAdM, Spain  \\
                                                 \hline   
 \end{tabular}
\caption{Continued}
\label{tab:SumarryObs2}
\end{table*}

\section{Summary of previously published observations used in this work}

\begin{table*}
\begin{tabular}{llll}
\hline
Asteroid & Date of observations & $N_{\rm lc}$  & Reference\\
\hline
(122)~Gerda  & 1987 Jul 28 - Aug 01                & 2 &  \citet{b33} \\
                                         & 1981 Mar 13 - 25                               & 3 &  \citet{b36} \\
                                         & 1981 Mar 18 - 19                              & 2 &  \citet{Gil_1993}\\
                                         & 2003 Mai 02 - 09                                & 3 &  \citet{b33} \\
                                         & 2005 Sep 11 - Oct 04            & 2 & \citet{b33} \\
                                         & 2005 Sep 31 - Nov 28                    & 3 & \citet{b34} \\
                                         & 2009 Apr 01 - 03                                & 3 & \citet{b35} \\
(172)~Baucis & 1984 May 11                                         & 1 &  \citet{Wei_1990}        \\     
                                         & 1989 Nov 21                                             & 1 & \citet{Wei_1990}         \\             
(236)~Honoria & 1979 Jul 30 - Aug 22                                    & 7 &Harris and Young (1989) \\
                                                & 1980 Dec 30 - 1981 Feb 1                       & 7     & Harris and Young (1989) \\
(387)~Aquitania & 1979 Aug 27 - 29                              & 3     & \citet{Sch_1979} \\ 
                                                &       1981 May 30                                     & 1       &\citet{b39}                            \\
                                                & 1981 May 30 - Jul 24                  & 8       & \citet{b40} \\
(402)~Chloe     & 1997 Jan 19 - Mar 02  & 5 & \citet{Den_2000} \\
                                    &   2009 Feb 07 - 17                        & 4 &             \citet{War_2009}                         \\
                                        & 2014 May 15 - 16                      & 2 &     \citet{Ste_2014}                         \\             
(458)~Hercynia & 1987 Feb 18              & 1 & \citet{binzel_1987} \\
(729)~Watsonia& 2013 Jan  21 - Feb 14 & 8 & \citet{Ste_2013}  \\ 
(980)~Anacostia & 1980 Jul 20 - Aug 18 & 5 &   \citet{b19} \\ 
(1332)~Marconia & 2012 Aug 27 - Sep 11 & 6  & \citet{Ste_2013b} \\
(1702)~Kalahari   & 2011 Jul 28 - Aug 26 & 10   &       \citet{Oey_2012}                                                                                 \\

 \end{tabular}
\caption{All observations used for period and shape modelling which have already been published.}
\label{tab:SumarryObs3}
\end{table*}

\section{Light-curves obtained for this work \label{App:LC}}

\begin{figure*}[!h]
\centering
\begin{tabular}{|c|c|}
\hline
\includegraphics[width=8.4cm]{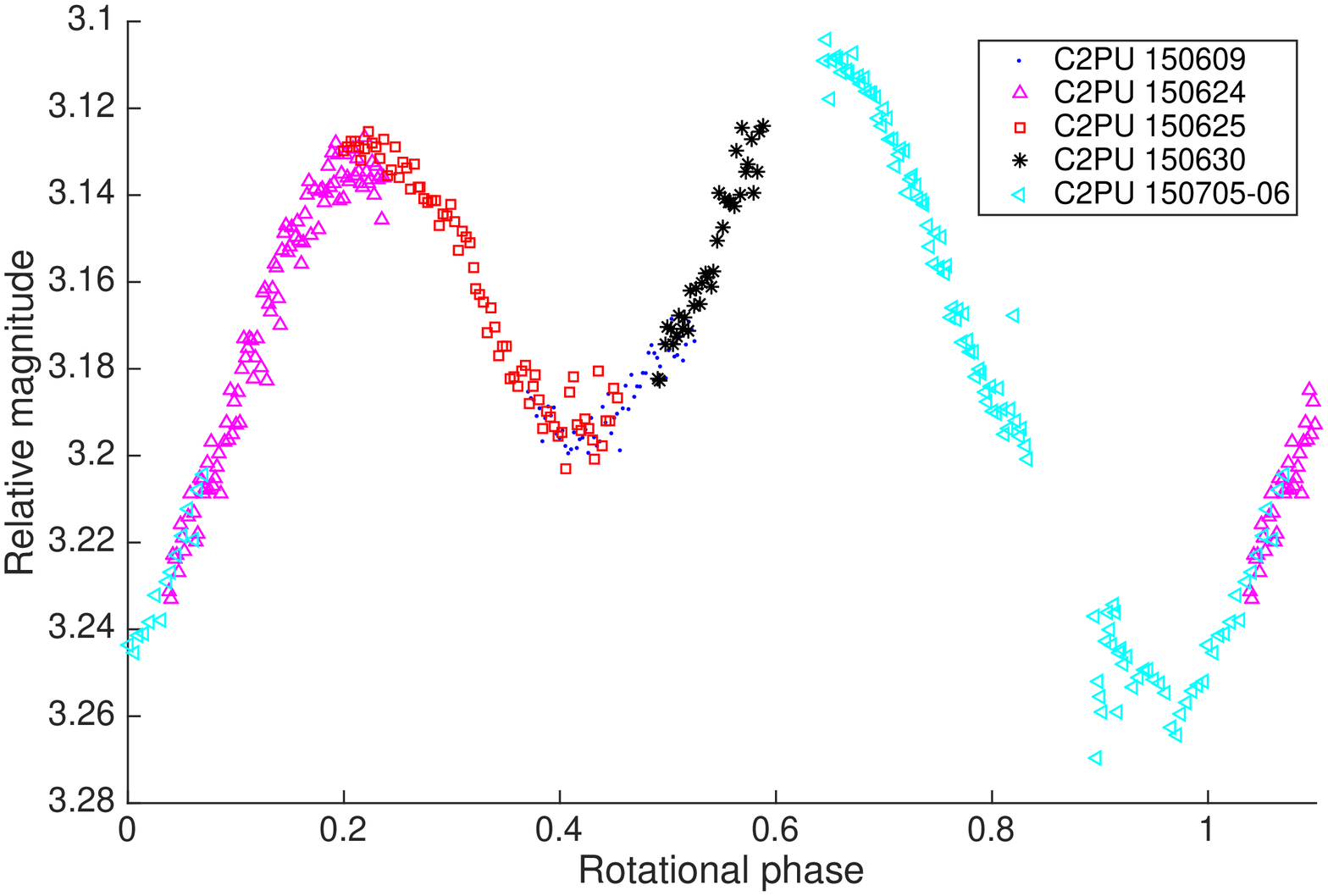}
&
\includegraphics[width=8.4cm]{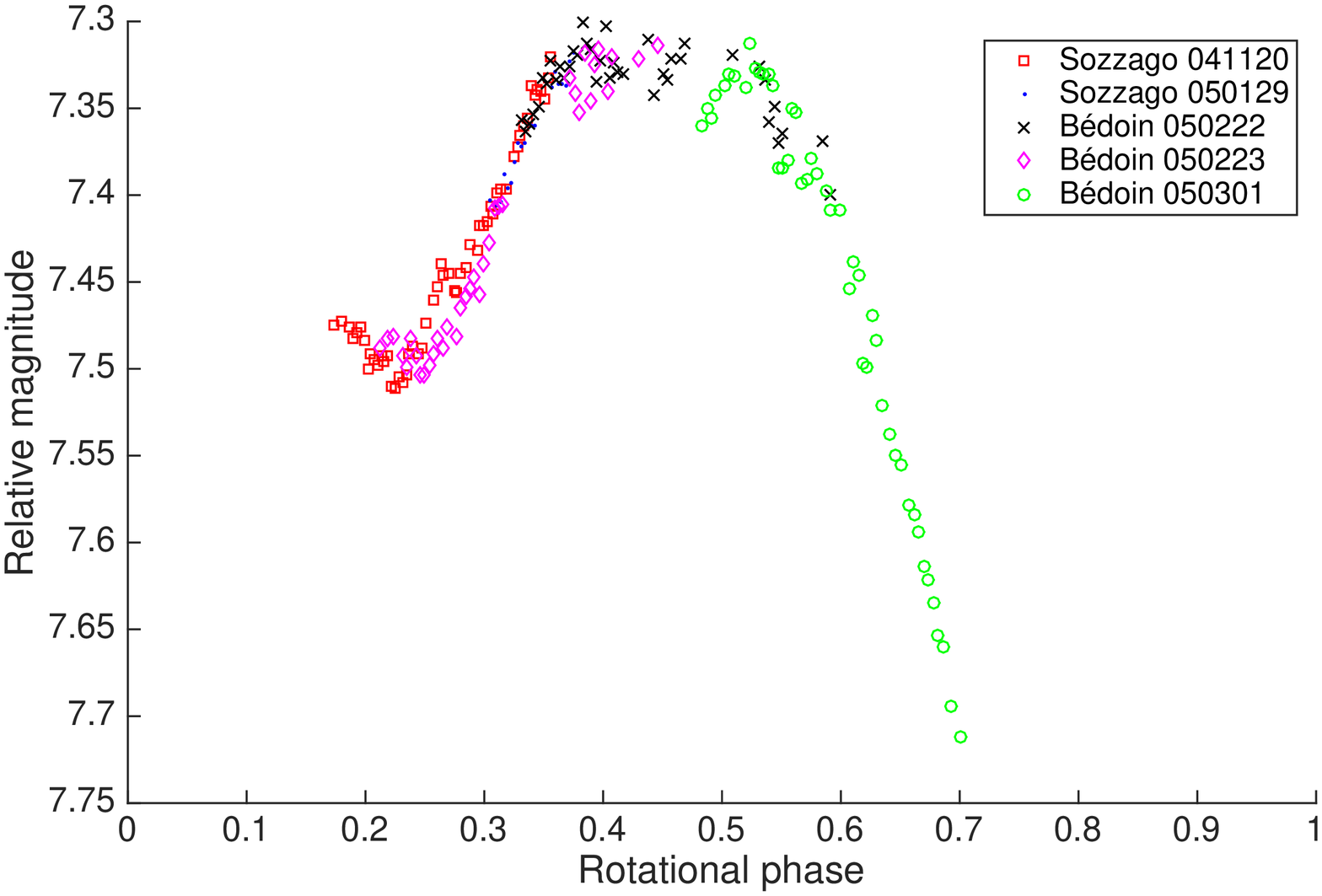}
\\
{(122)~Gerda: 2015 opposition; $P_{\rm syn} = 10.6872 \pm 0.0001 $~h}
&
     {(172)~Baucis: 2004-2005 opposition; $P_{\rm syn} = 27.417 \pm 0.005$ h}
\\
\hline     
\includegraphics[width=8.4cm]{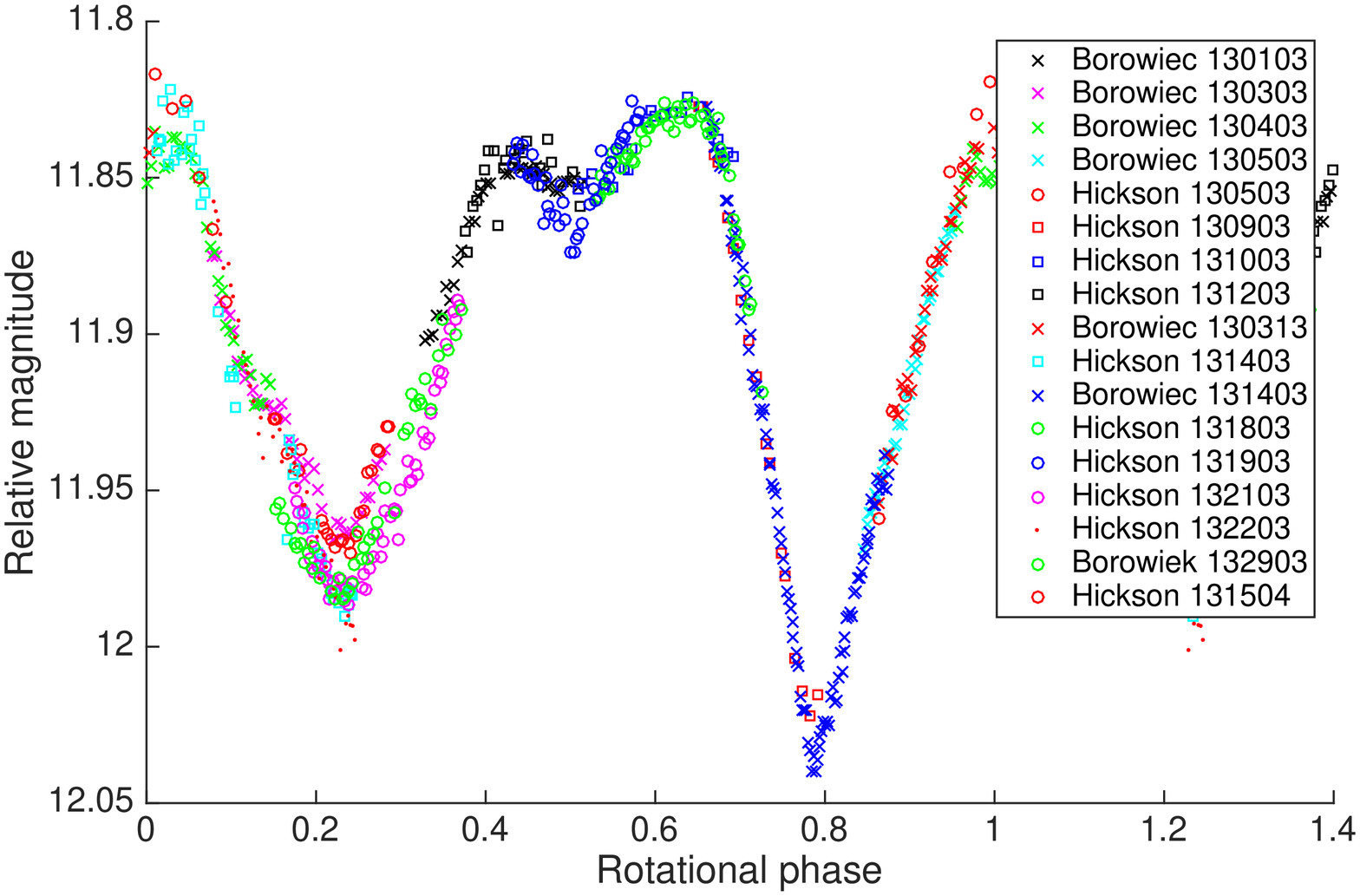}     
&
\includegraphics[width=8.4cm]{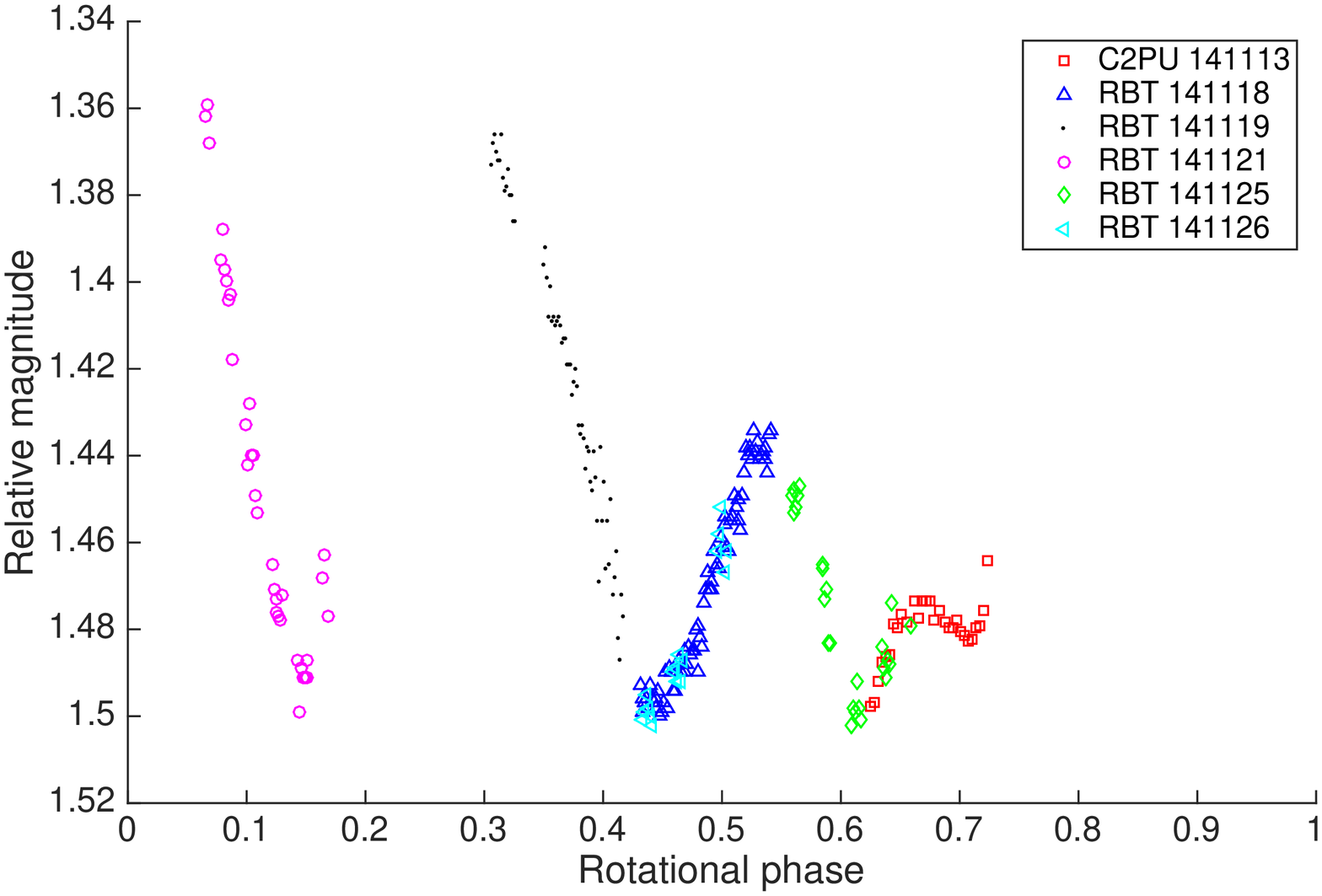}
\\
     {(172)~Baucis: 2013 opposition; $P_{\rm syn} = 27.4096 \pm 0.0004$~h}
&
     {(172)~Baucis: 2014 opposition; $P_{\rm syn} = 27.42 \pm 0.01$ h}
\\
\hline

\subf{\includegraphics[width=8cm]{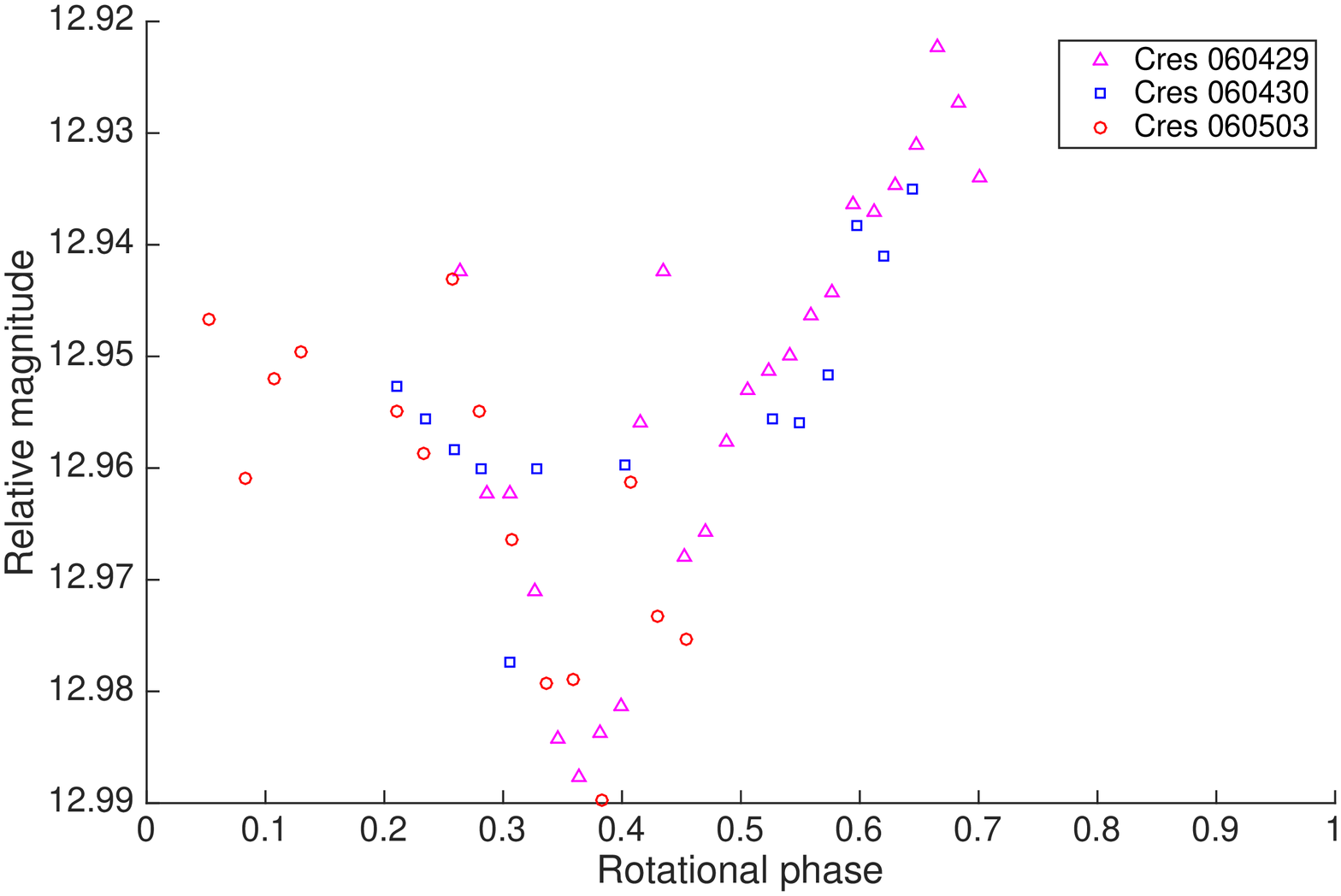}}
     {(236)~Honoria: 2006 opposition; $P_{\rm syn} = 12.35 \pm 0.10$ h}
&
\subf{\includegraphics[width=8cm]{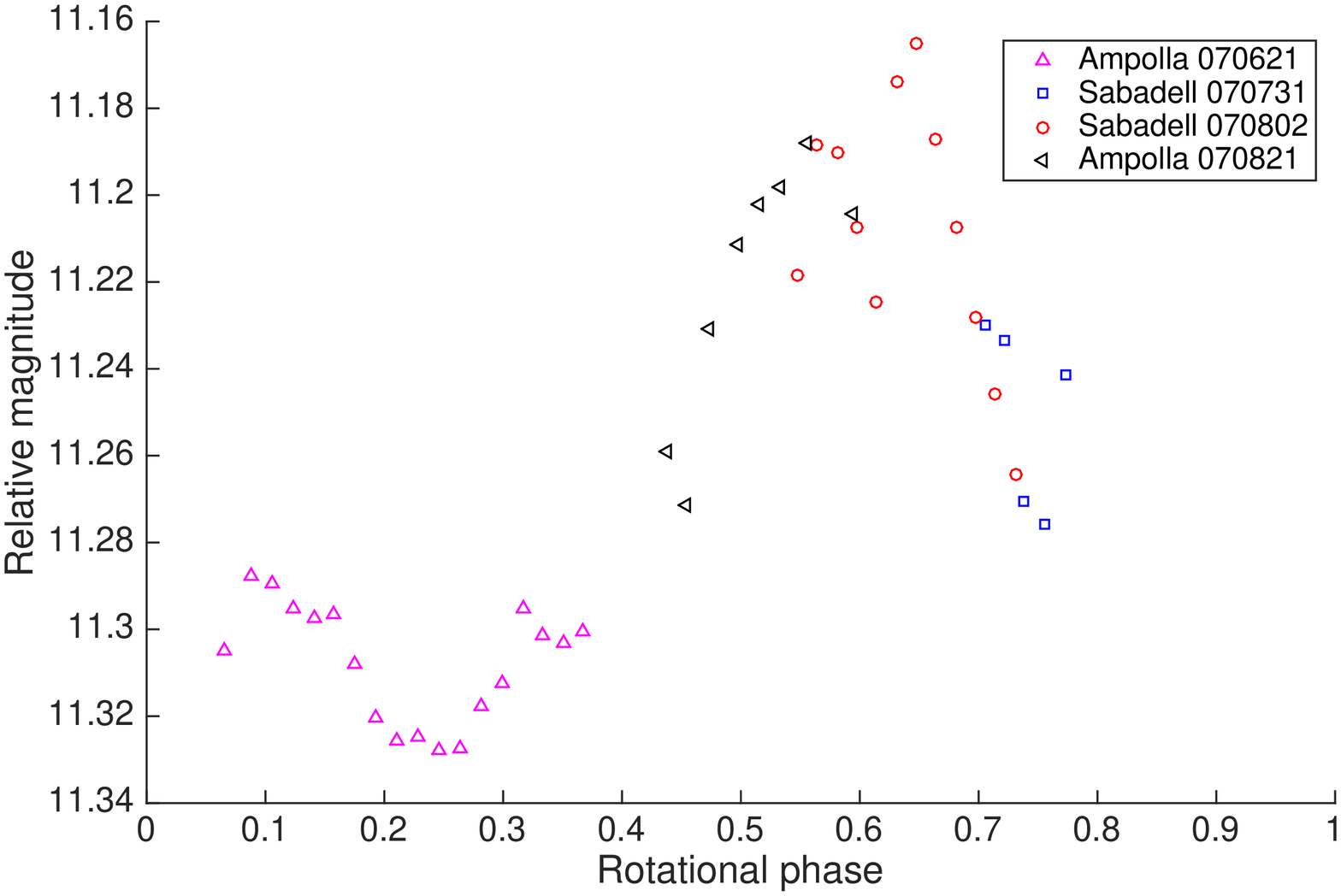}}
     {(236)~Honoria: 2007 opposition; $P_{\rm syn} = 12.34 \pm 0.02$ h}

\\
\hline
\subf{\includegraphics[width=8cm]{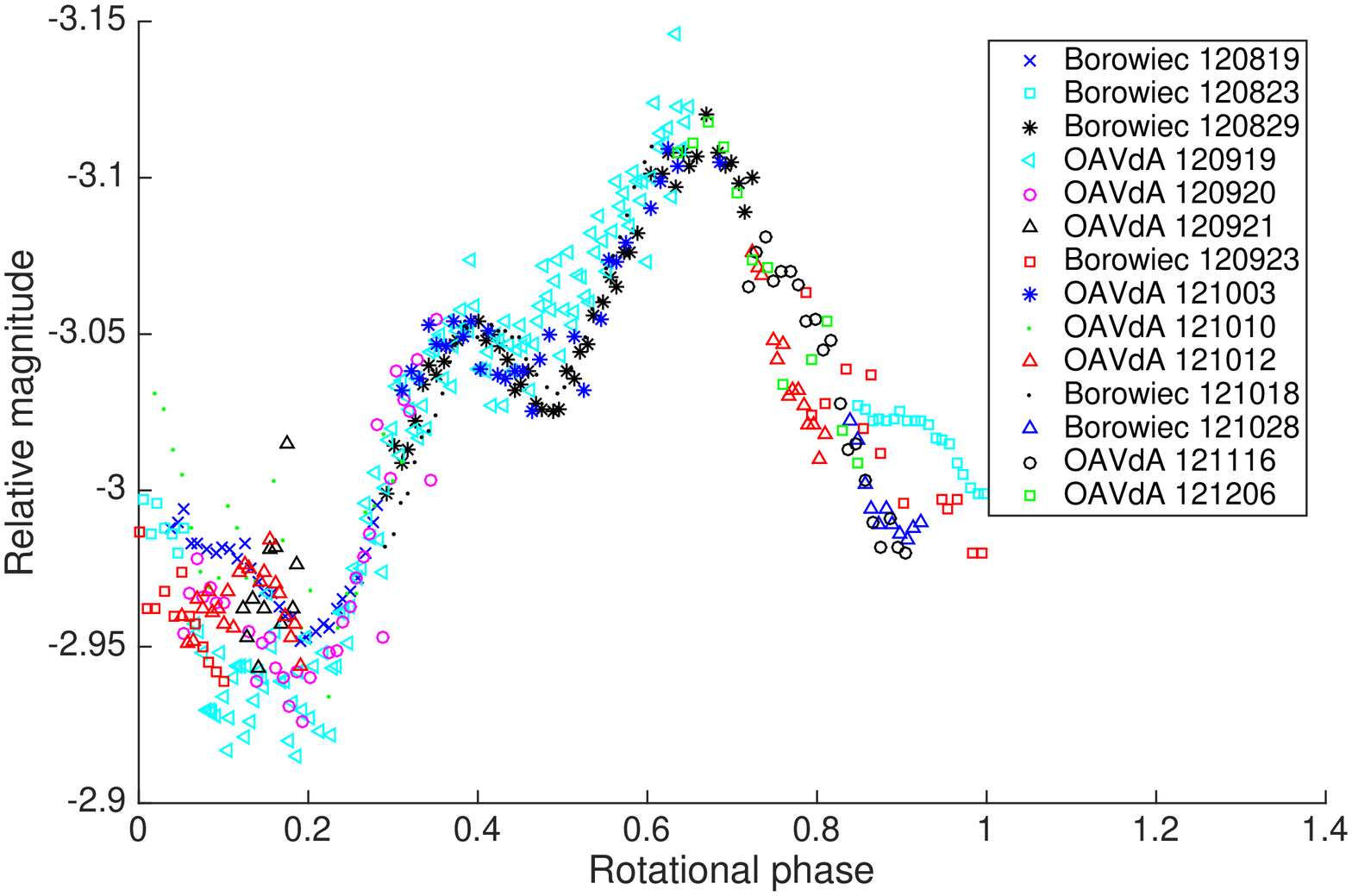}}
     {(236)~Honoria: 2012 opposition; $P_{\rm syn} = 12.337 \pm 0.001$ h}
&
\subf{\includegraphics[width=8cm]{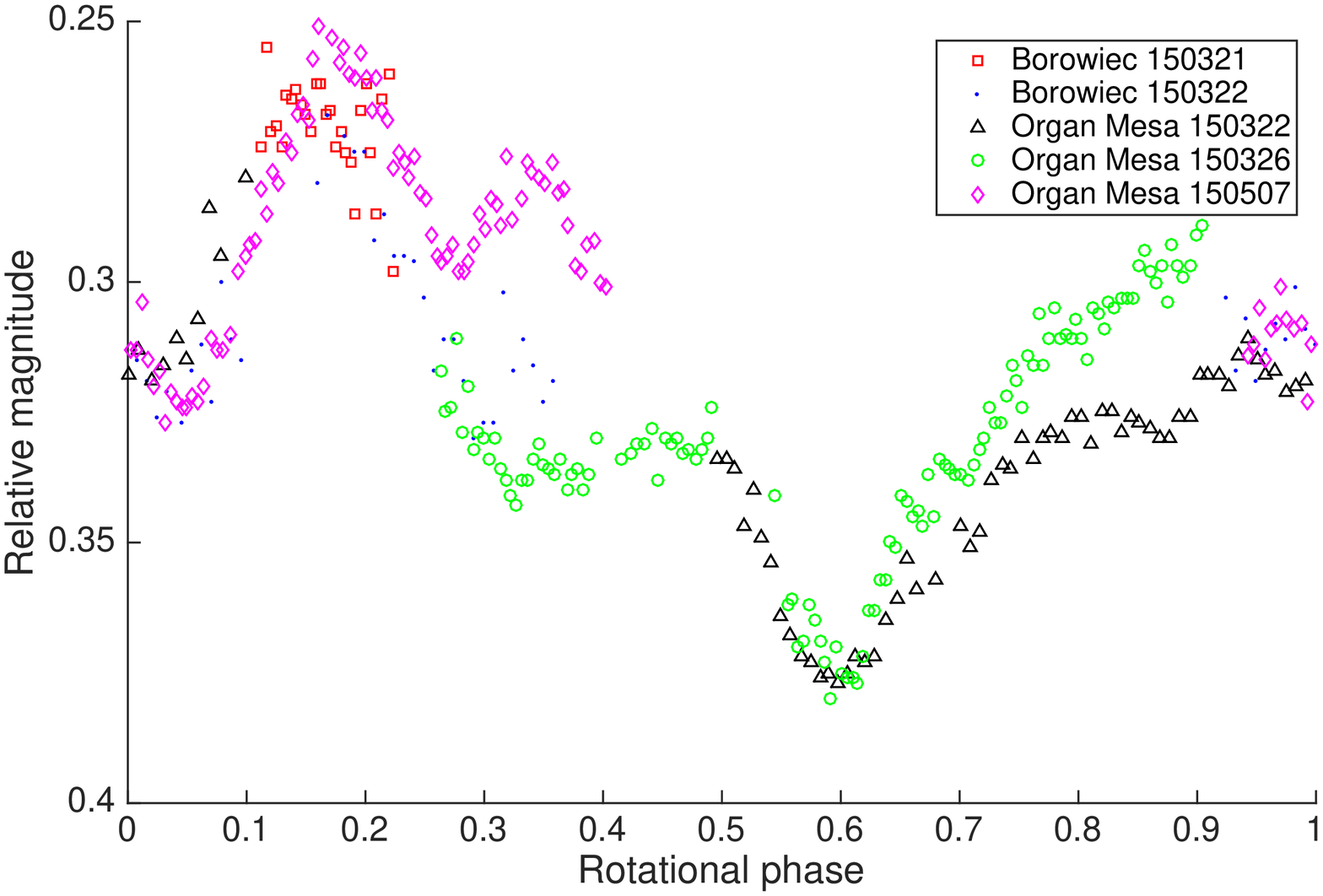}}
     {(236)~Honoria: 2015 opposition; $P_{\rm syn} = 12.33 \pm 0.01$ h}

\\
\hline

\end{tabular}
\caption{Composite light-curves of asteroids studied in this works.}
\label{LC1}
\end{figure*}

\begin{figure*}[!h]
\addtocounter{figure}{-1}
\centering
\begin{tabular}{|c|c|}
\hline

\subf{\includegraphics[width=8.4cm]{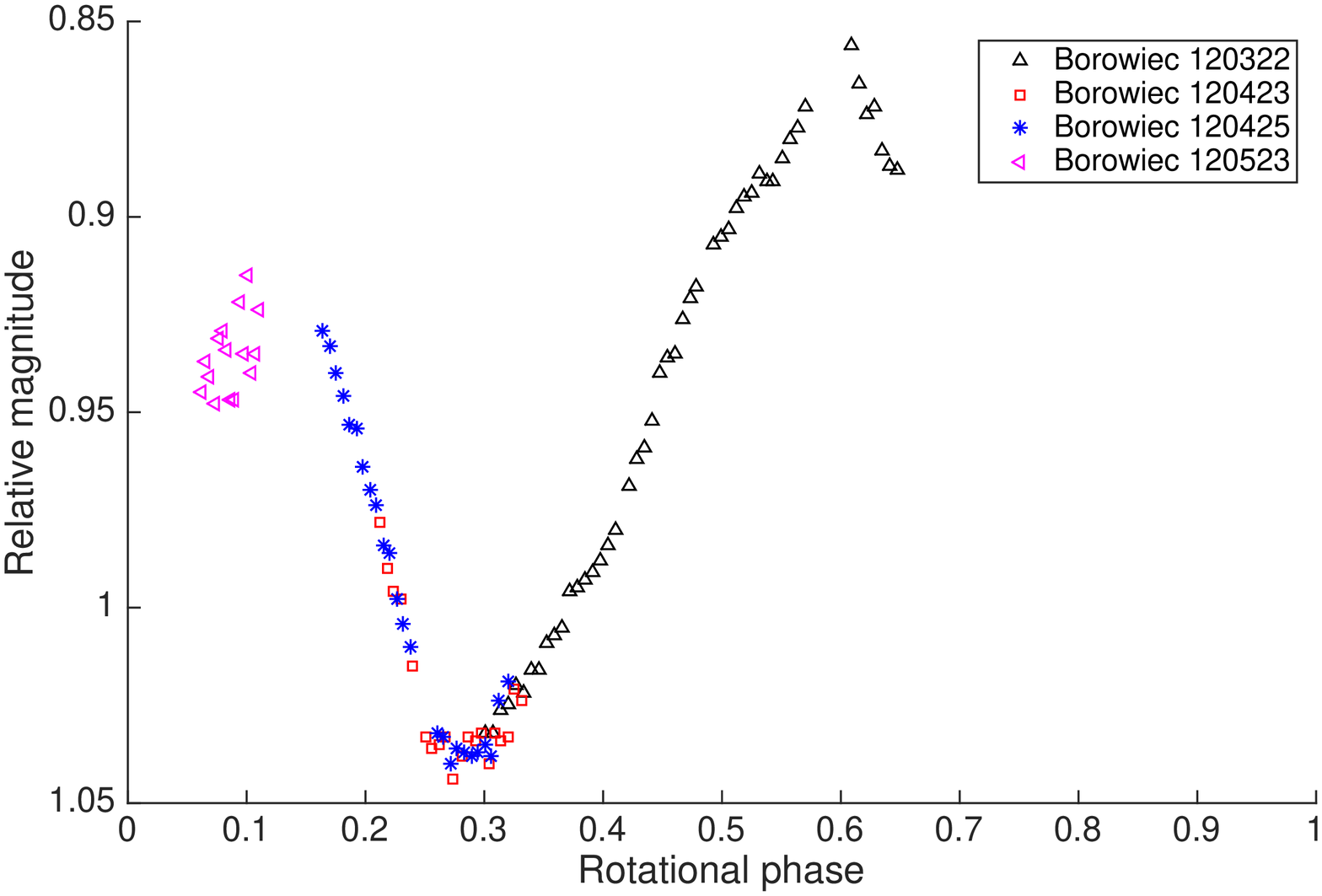}}
     {(387)~Aquitania: 2012 opposition; $P_{\rm syn} = 24.15 \pm 0.02$ h}
&
\subf{\includegraphics[width=8.4cm]{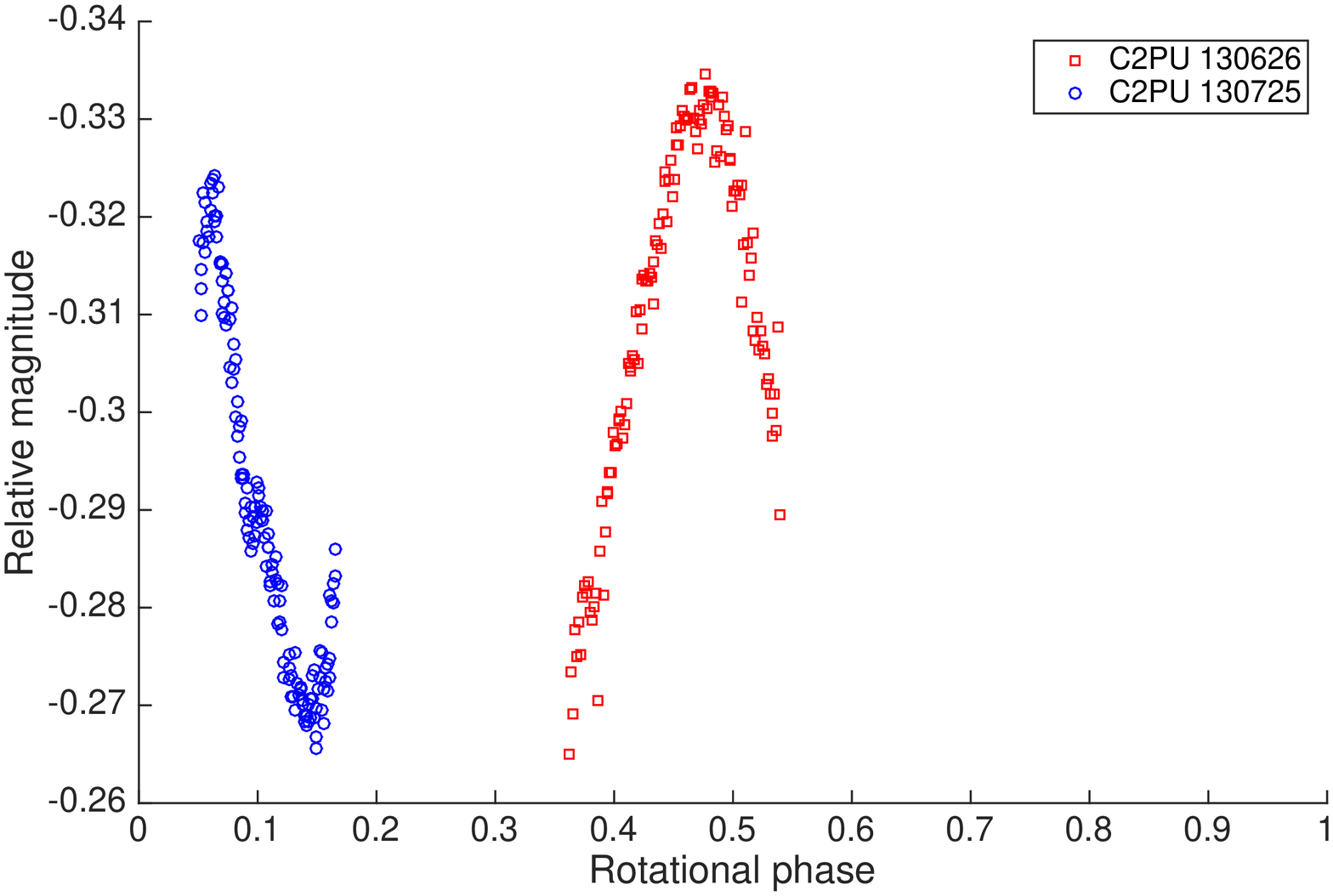}}
     {(387)~Aquitania: 2013 opposition; $P_{\rm syn} = 24.15 \pm 0.15$ h}

\\
\hline

\includegraphics[width=8.4cm]{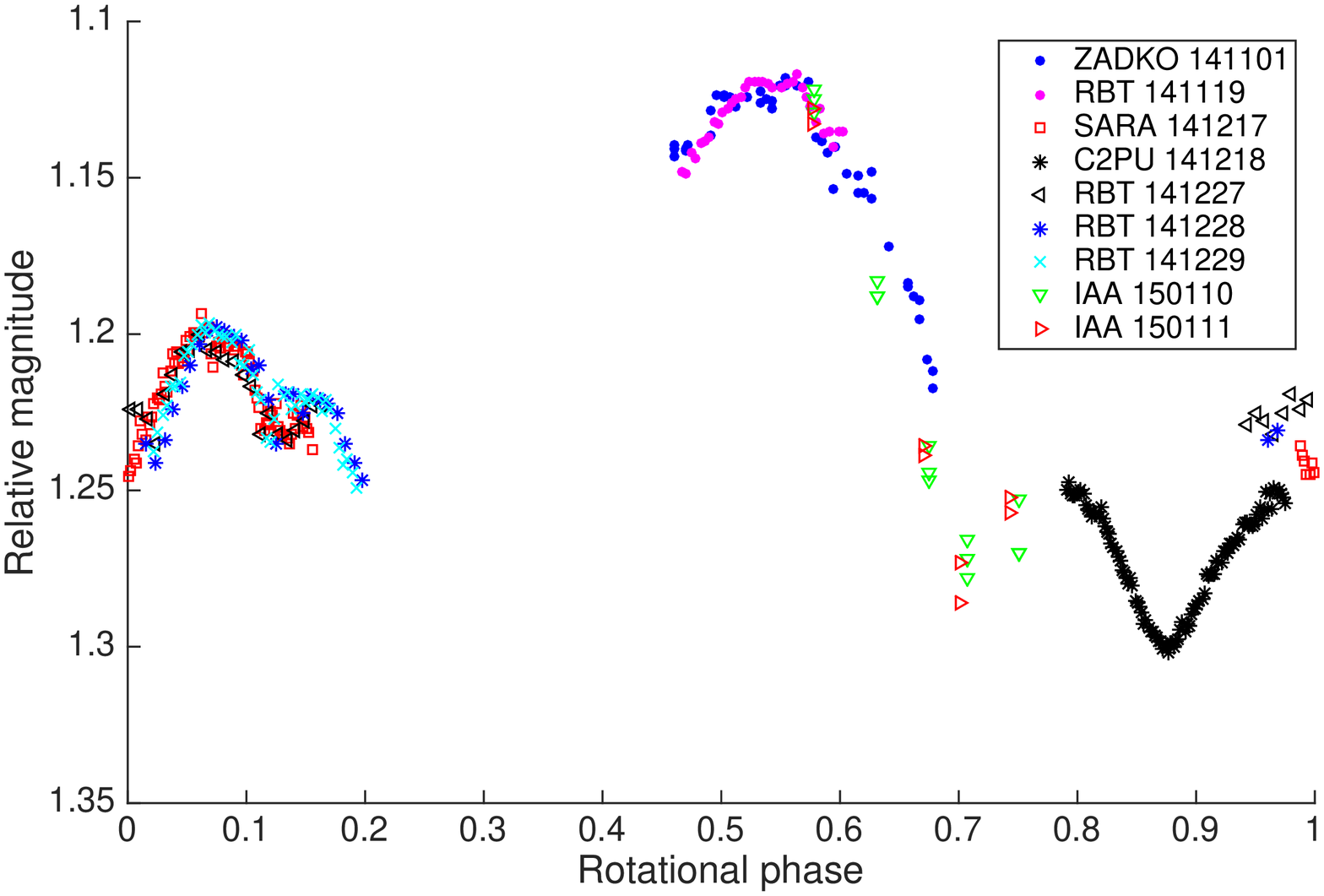}
&

\includegraphics[width=8.4cm]{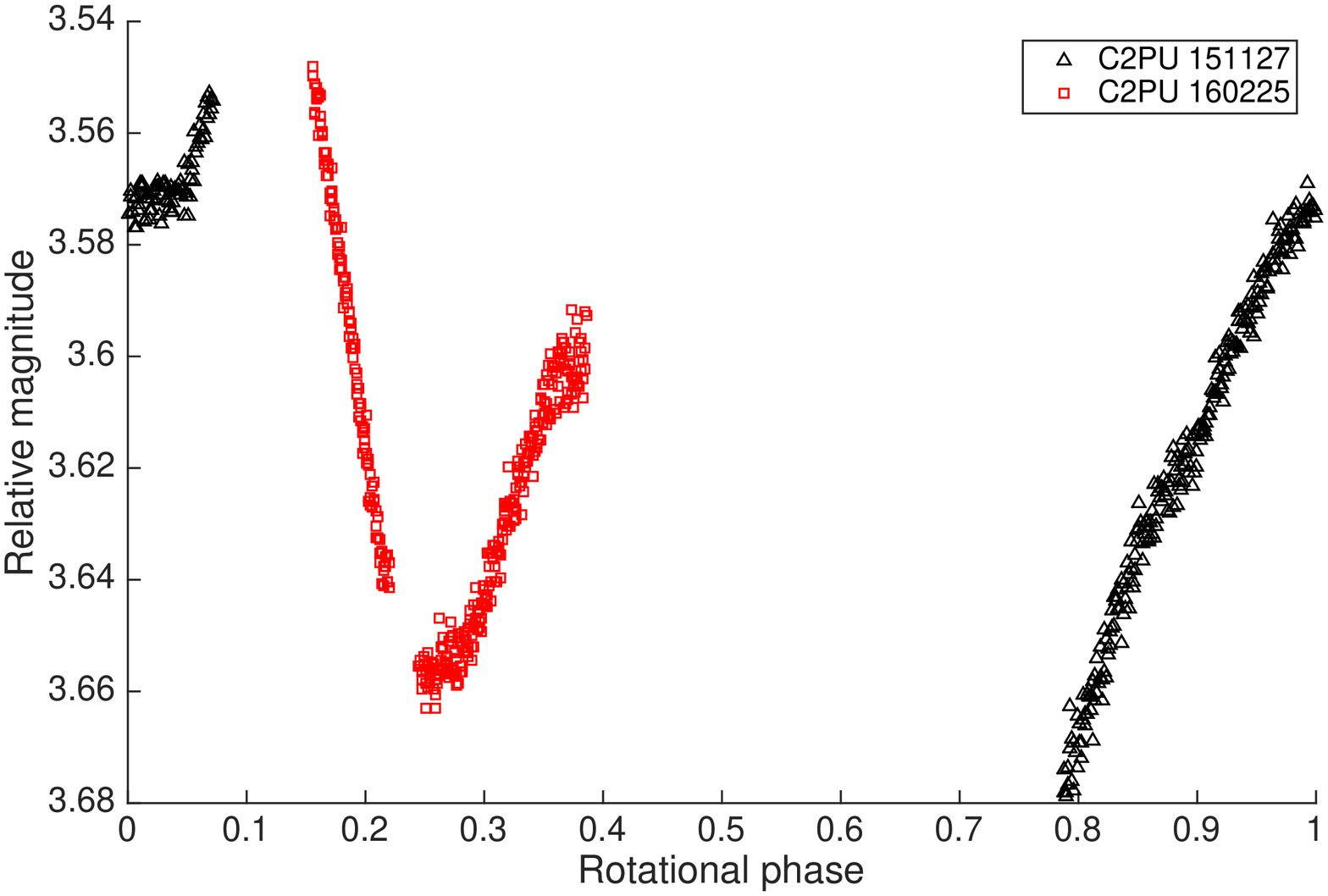}
\\
     {(387)~Aquitania: 2014-2015 opposition; $P_{\rm syn} = 24.14 \pm 0.02$~h}
&
     {(387)~Aquitania: 2015-2016 opposition; $P_{\rm syn} = 24.147$ h}

\\
\hline

\includegraphics[width=8.4cm]{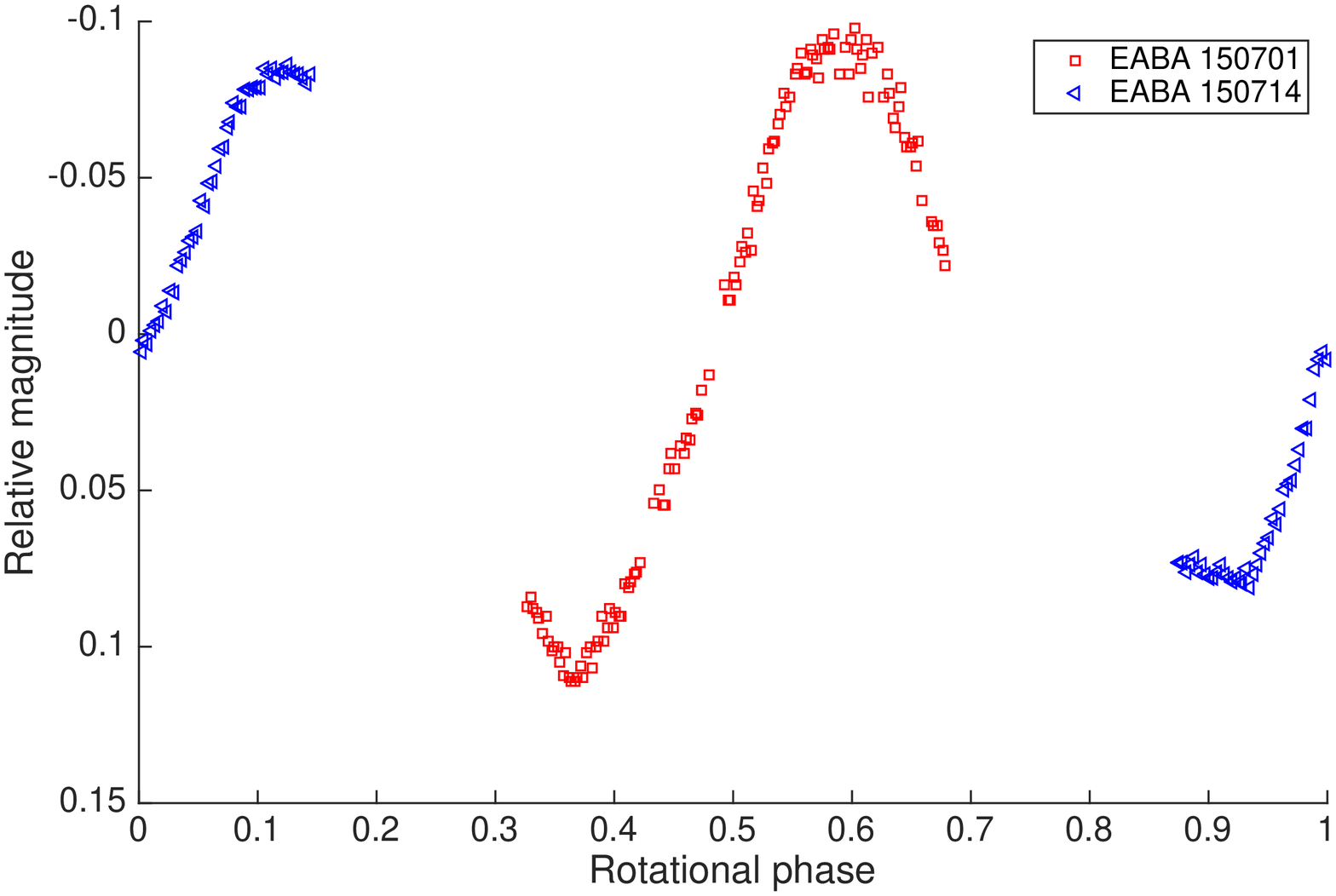}

&
\includegraphics[width=8.4cm]{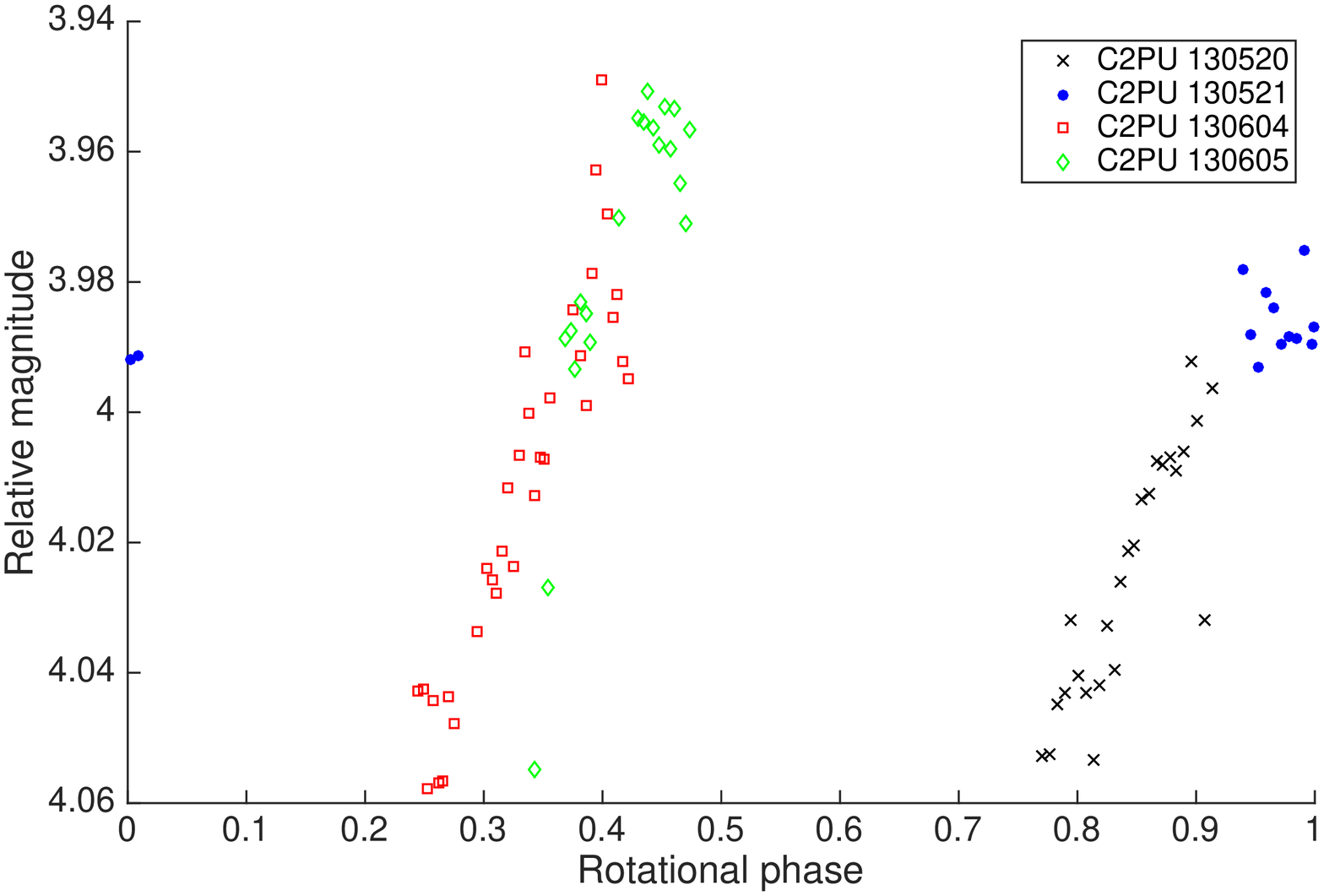}
\\
    {(402)~Chloe: 2015 opposition;  $P_{\rm syn} = 10.70 \pm 0.05$~h}

     &
          {(458)~Hercynia: 2013 opposition; $P_{\rm syn} = 21.8 \pm 0.1 $ h}  

\\
\hline

\includegraphics[width=8.4cm]{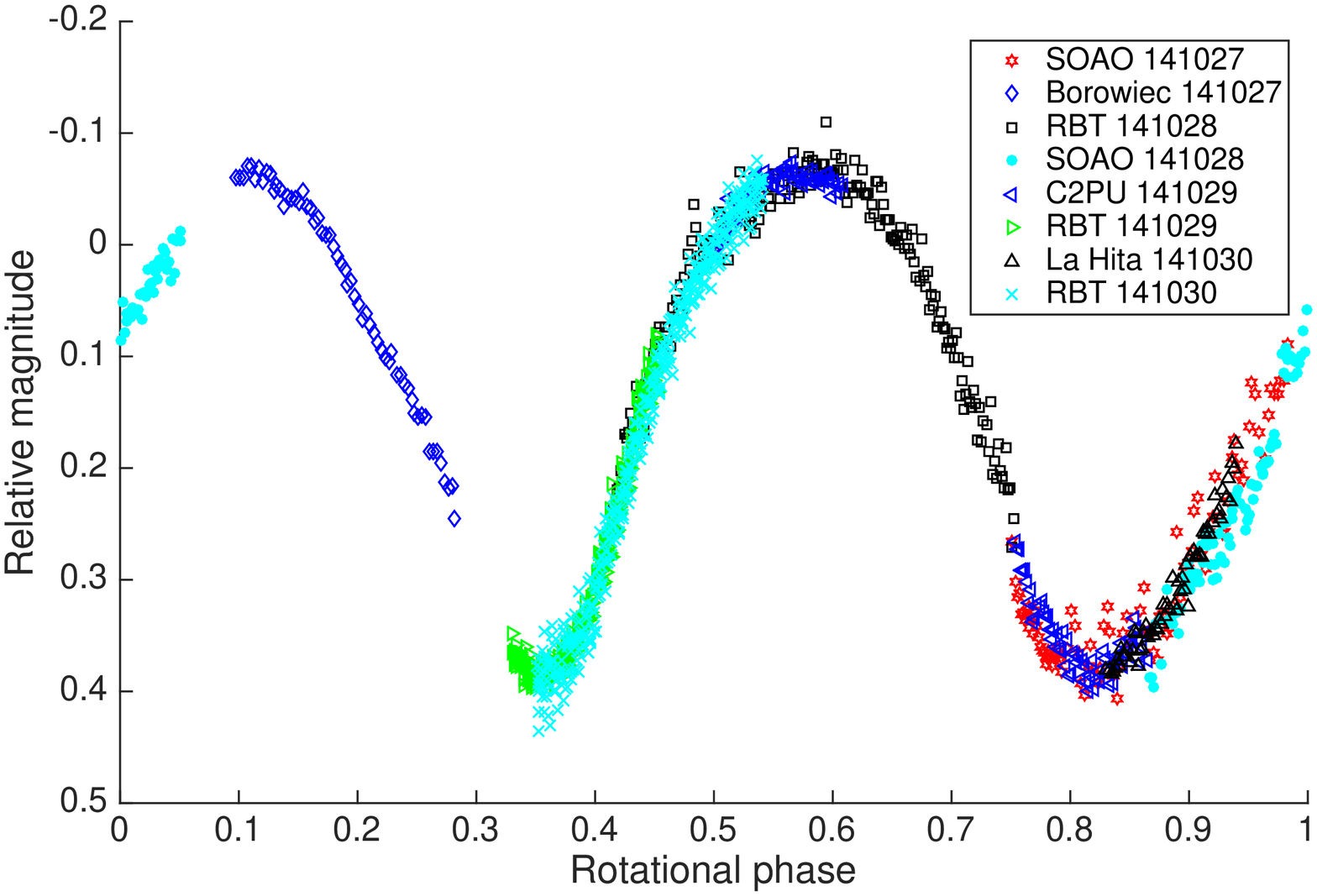}
&
\includegraphics[width=8.4cm]{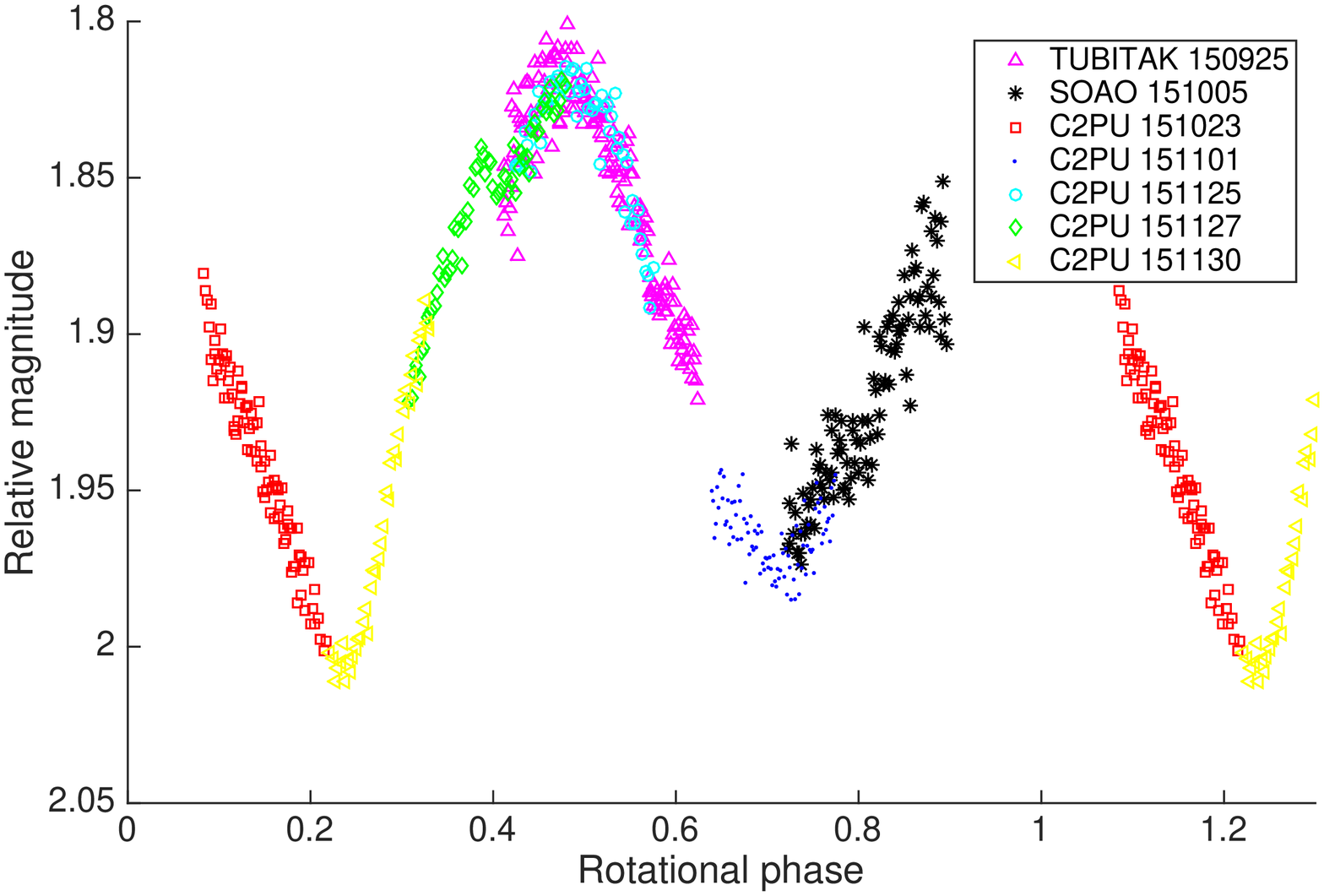}
\\
{(458)~Hercynia: 2014 opposition; $P_{\rm syn} = 21.81\pm0.03$~h}
&
     {(729)~Watsonia: 2015 opposition; $P_{\rm syn} = 25.190 \pm 0.003$ h} 
     
\\
\hline

\end{tabular}
\caption{Continued}
\end{figure*}

\begin{figure*}
\centering
\begin{tabular}{|c|c|}
\hline

\includegraphics[width=8.4cm]{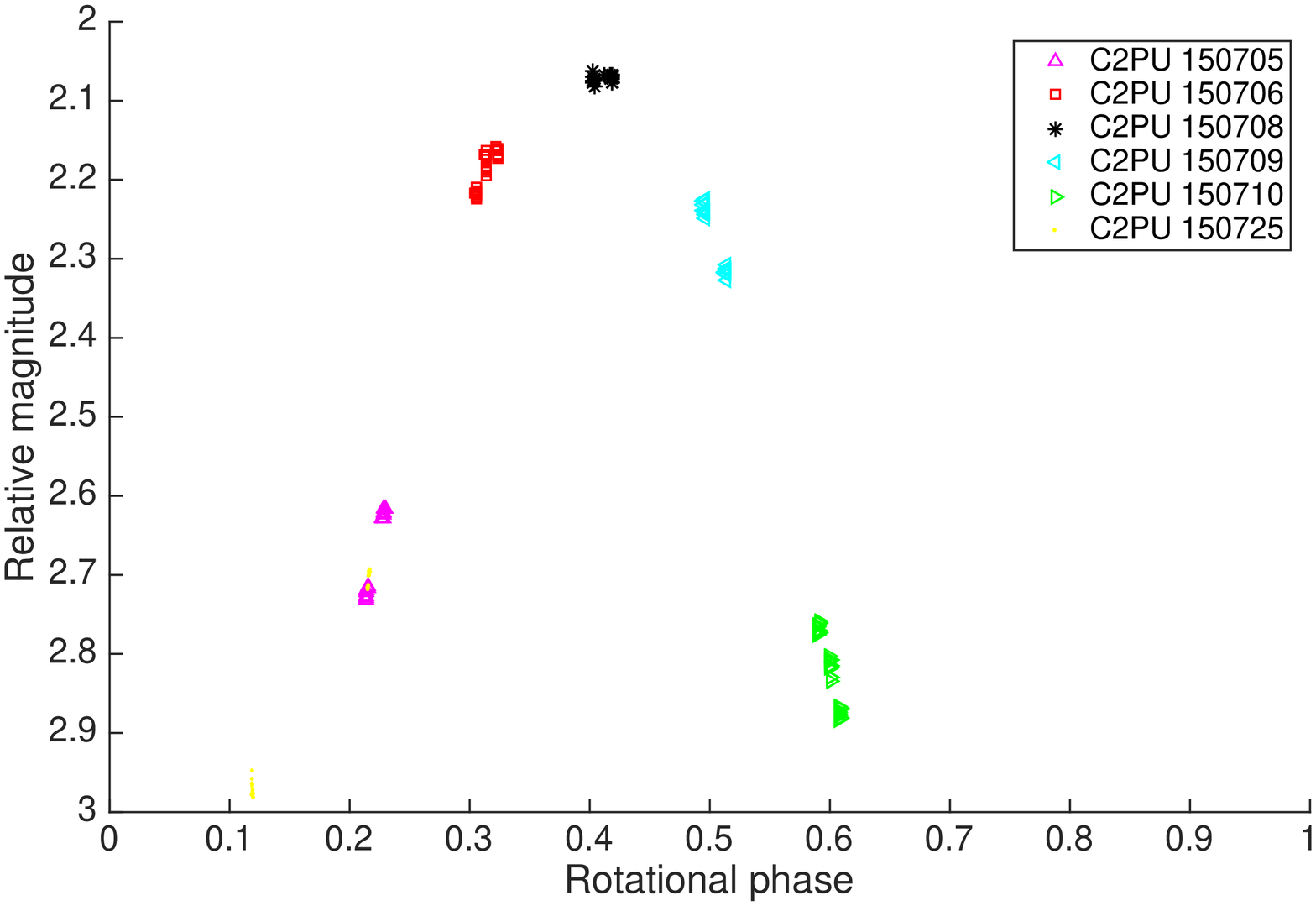}
   
&
\includegraphics[width=8.4cm]{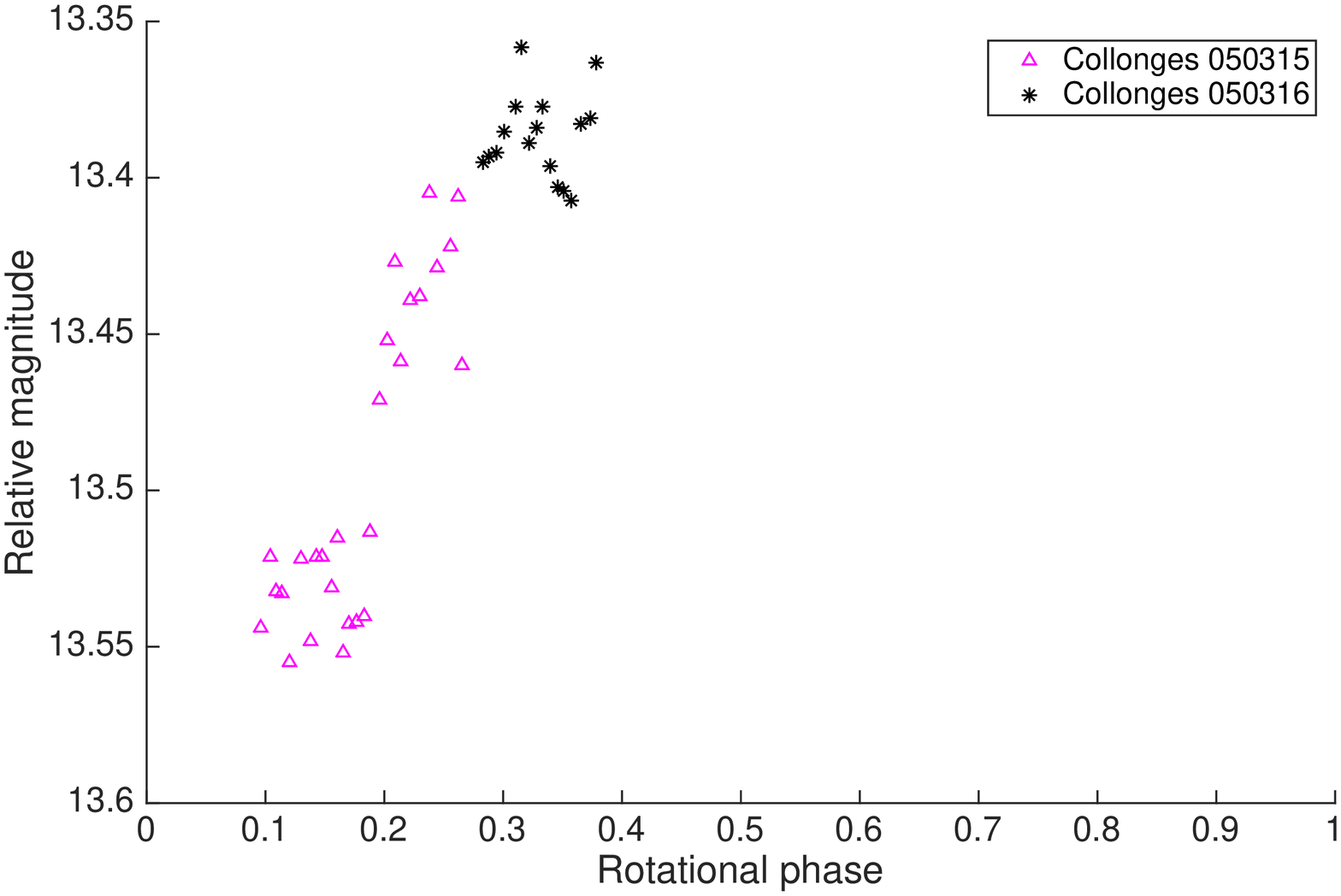}
\\

     {(824)~Anastasia: 2015 opposition; $P_{\rm syn} = 252$~h}
     &
          {(980)~Anacostia: 2005 opposition; $P_{\rm syn} = 20.1 \pm 0.1$ h}  

\\
\hline

\subf{\includegraphics[width=8.4cm]{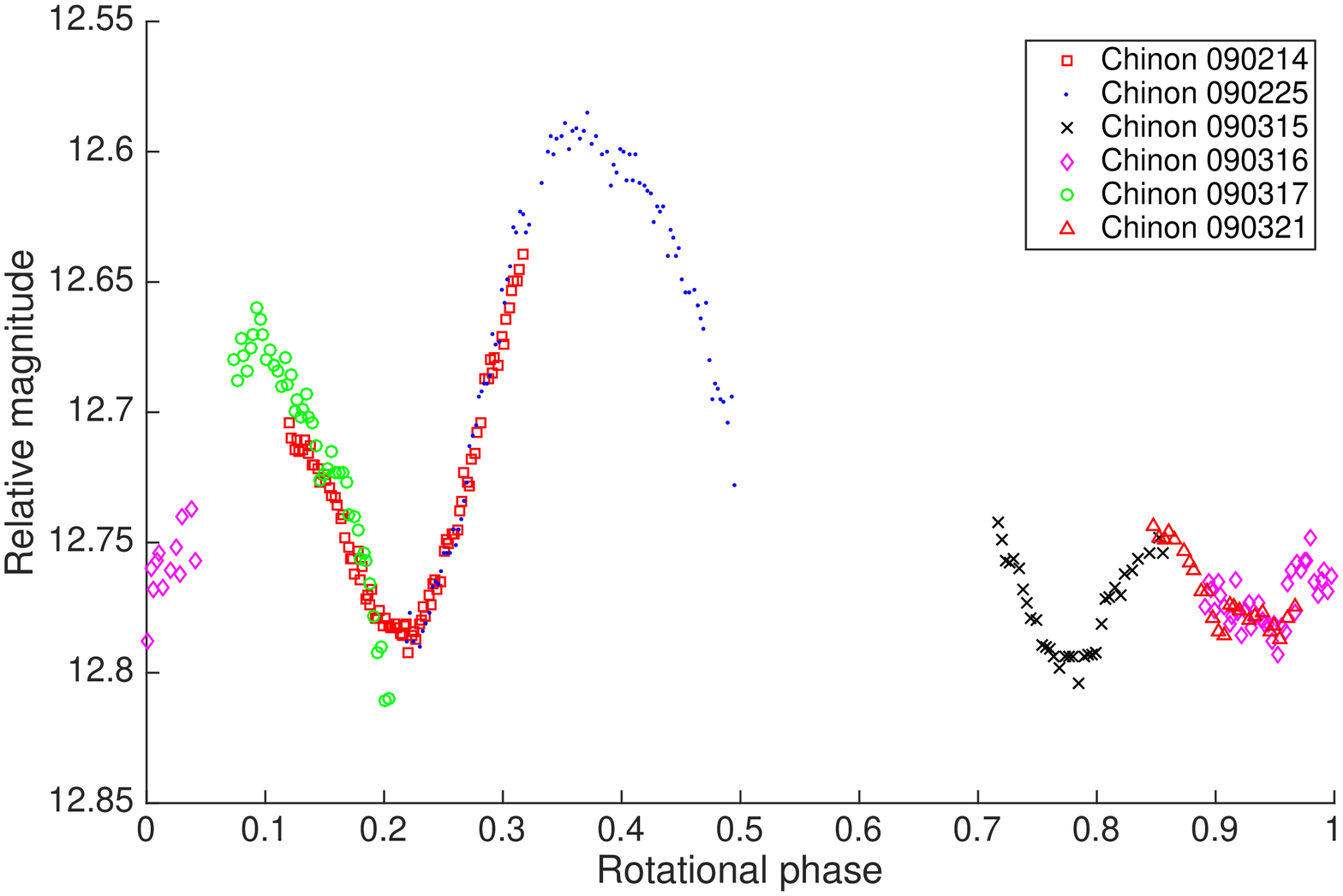}}
     {(980)~Anacostia: 2009 opposition; $P_{\rm syn} = 20.124 \pm 0.001$ h} 
   
&
\subf{\includegraphics[width=8.4cm]{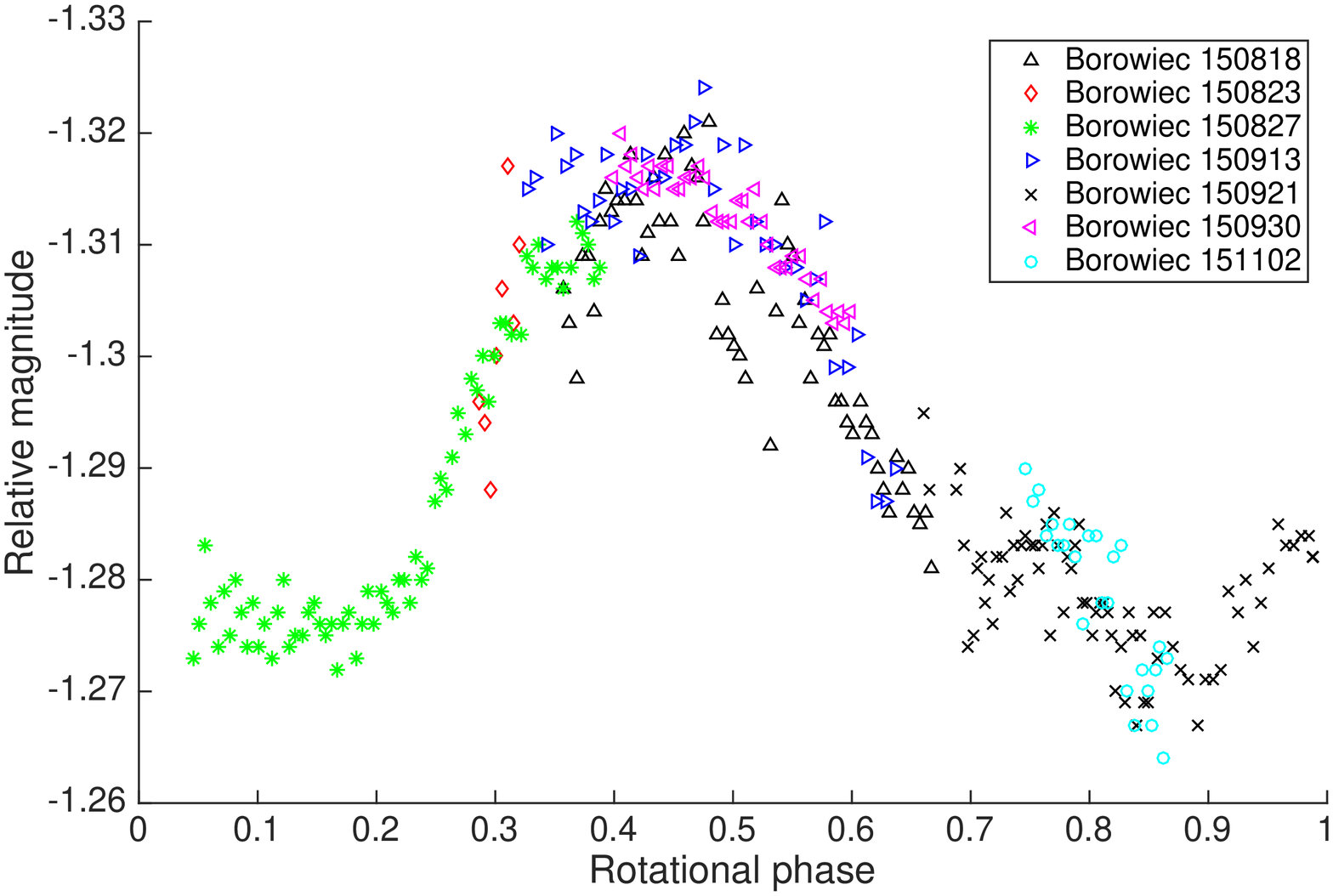}}
     {(980)~Anacostia: 2012 opposition; $P_{\rm syn} = 20.17 \pm 0.01$ h}

\\

\hline

\subf{\includegraphics[width=8.4cm]{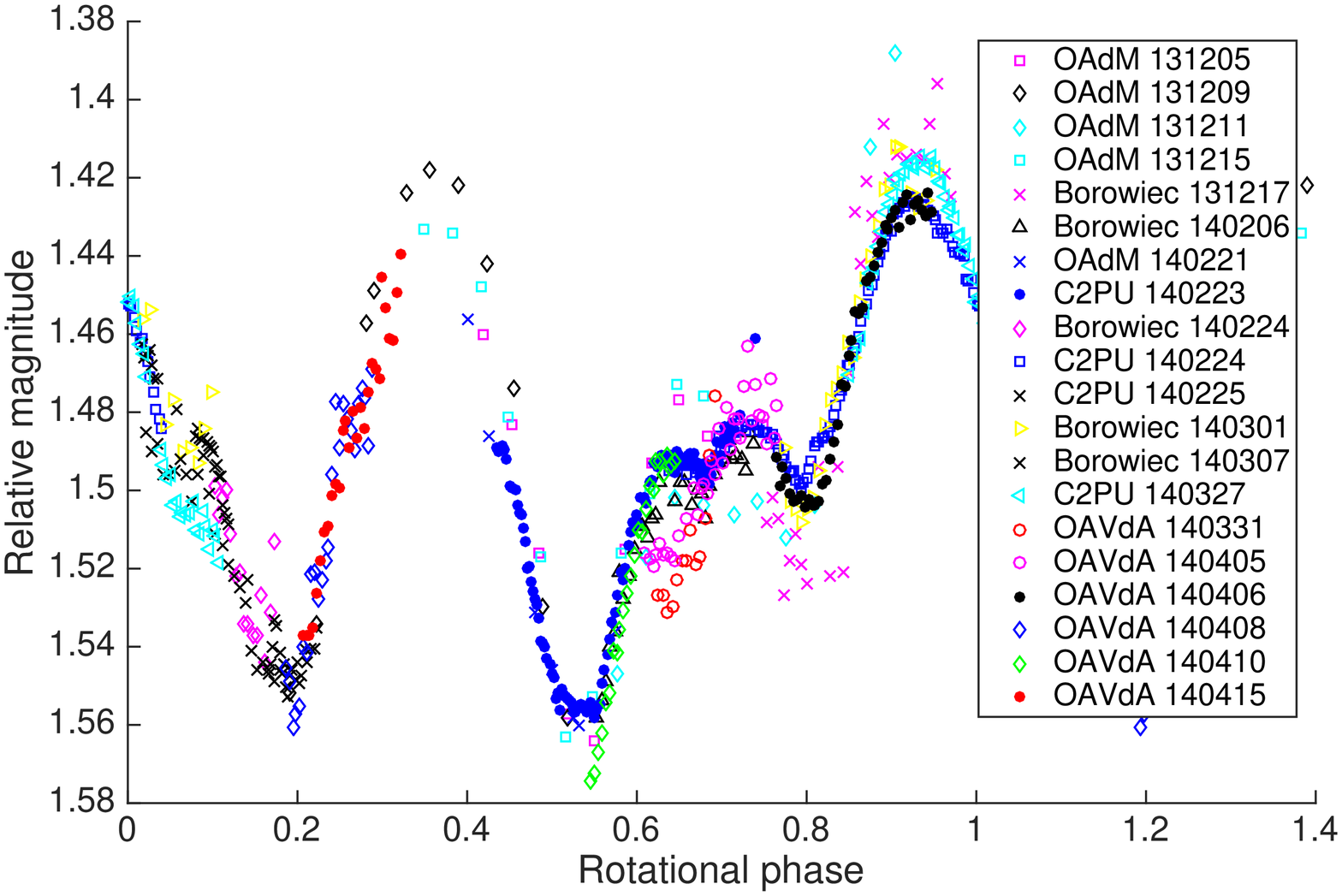}}
     {(980)~Anacostia: 2013-2014 opposition; $P_{\rm syn} = 20.1137 \pm 0.0002$~h}

&

\subf{\includegraphics[width=8.4cm]{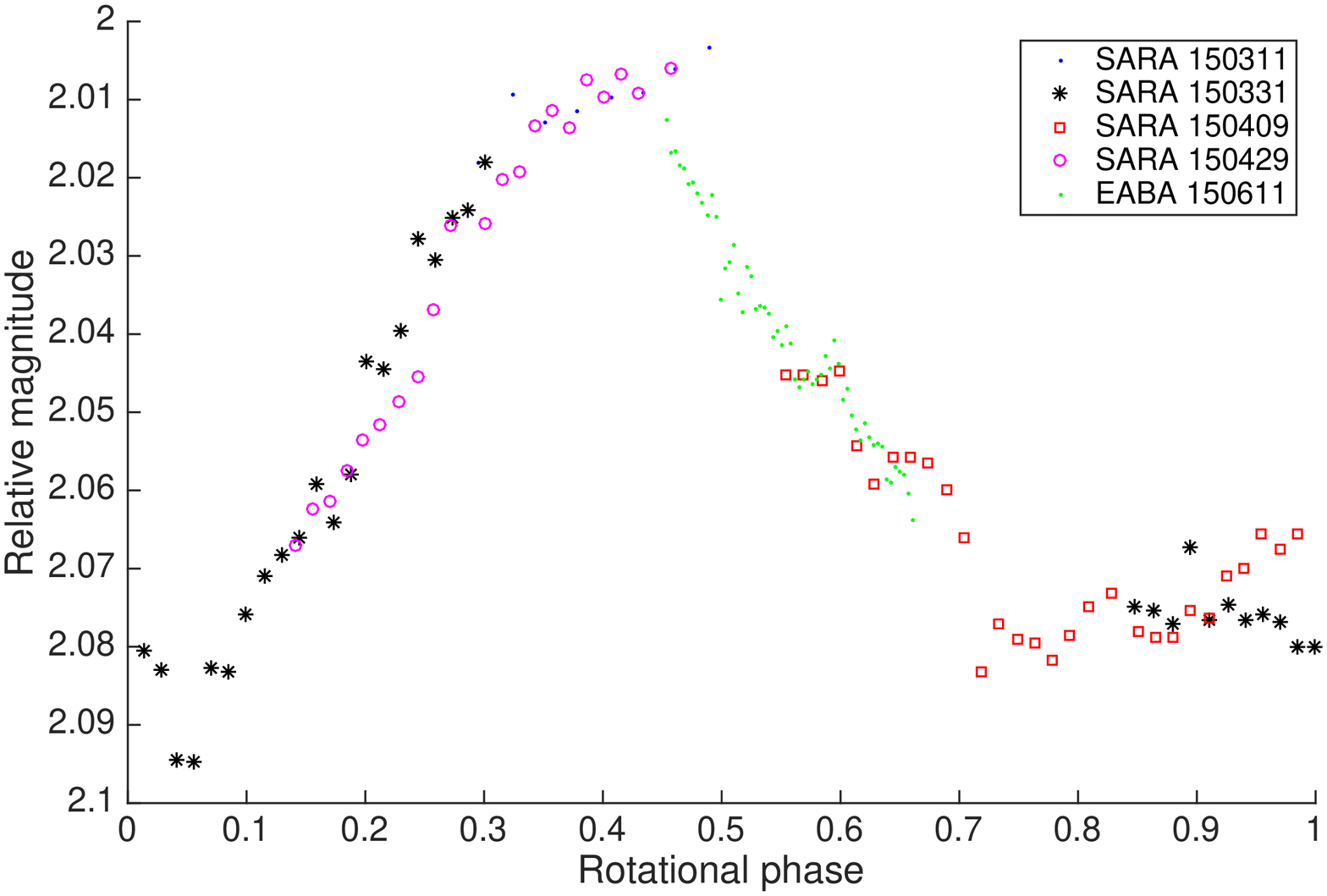}}
     {(980)~Anacostia: 2015 opposition; $P_{\rm syn} = 20.16 \pm 0.01$ h}

\\
\hline

\subf{\includegraphics[width=8.4cm]{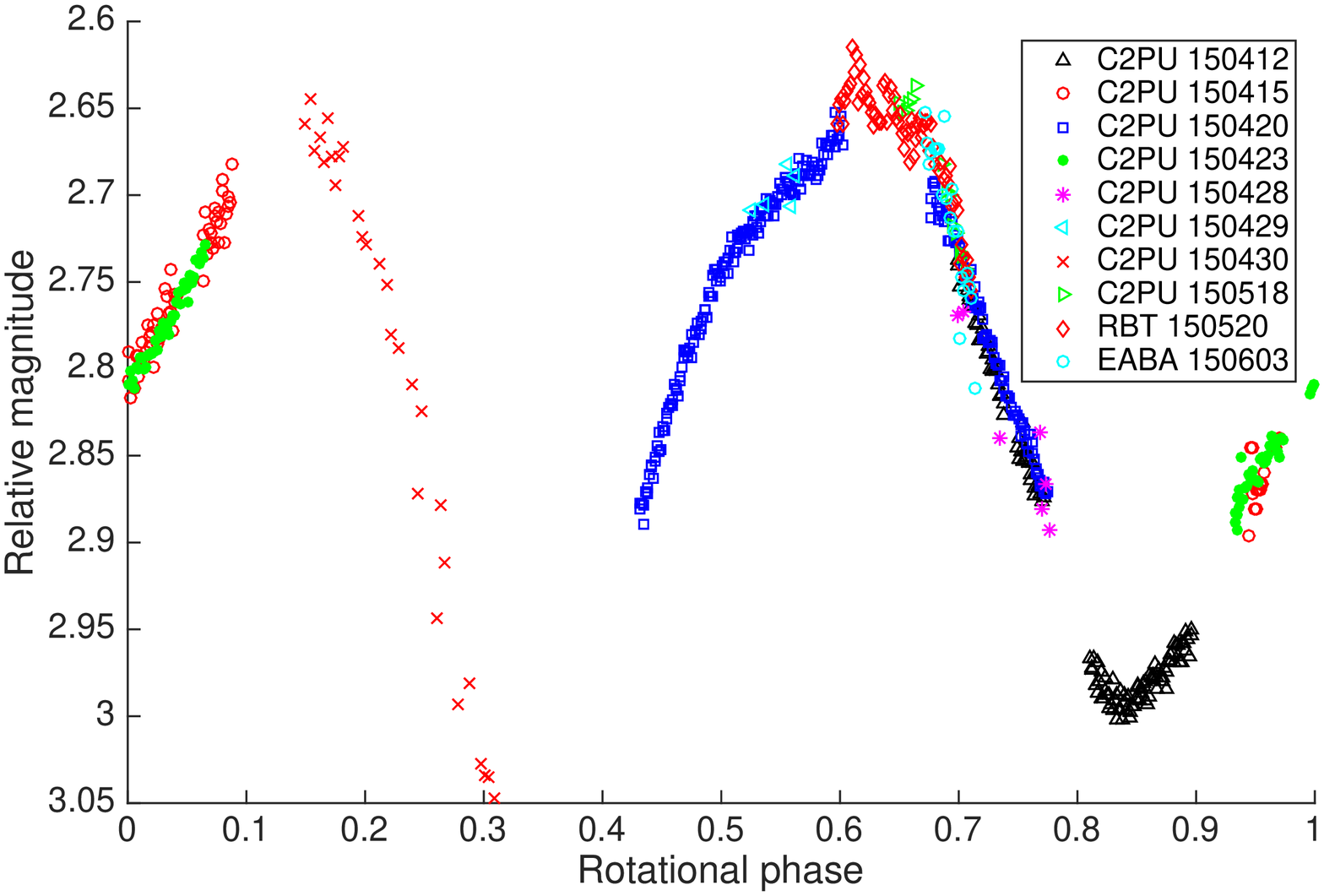}}
     {(1332)~Marconia, $P_{\rm syn} = 32.1201 \pm 0.0005$~h}

&
     \subf{\includegraphics[width=8.4cm]{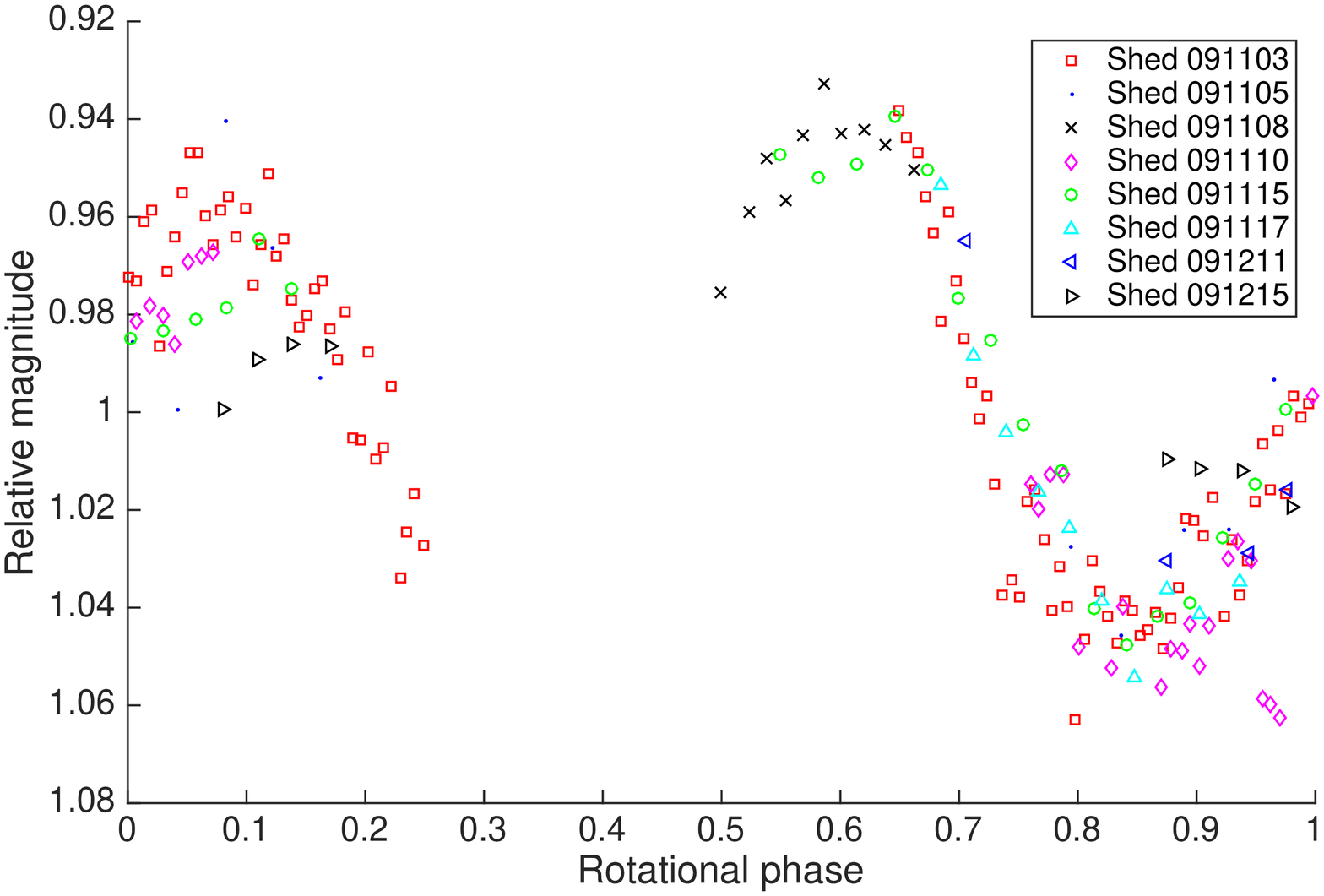}}
     {(1372)~Haremari: 2009 opposition; $P_{\rm syn} = 15.22 \pm 0.01$ h}

\\
\hline

\end{tabular}
\caption{Continued}
\label{fig:App_LC}
\end{figure*}

\begin{figure*}
\addtocounter{figure}{-1}
\centering
\begin{tabular}{|c|c|}
\hline

\subf{\includegraphics[width=8.5cm]{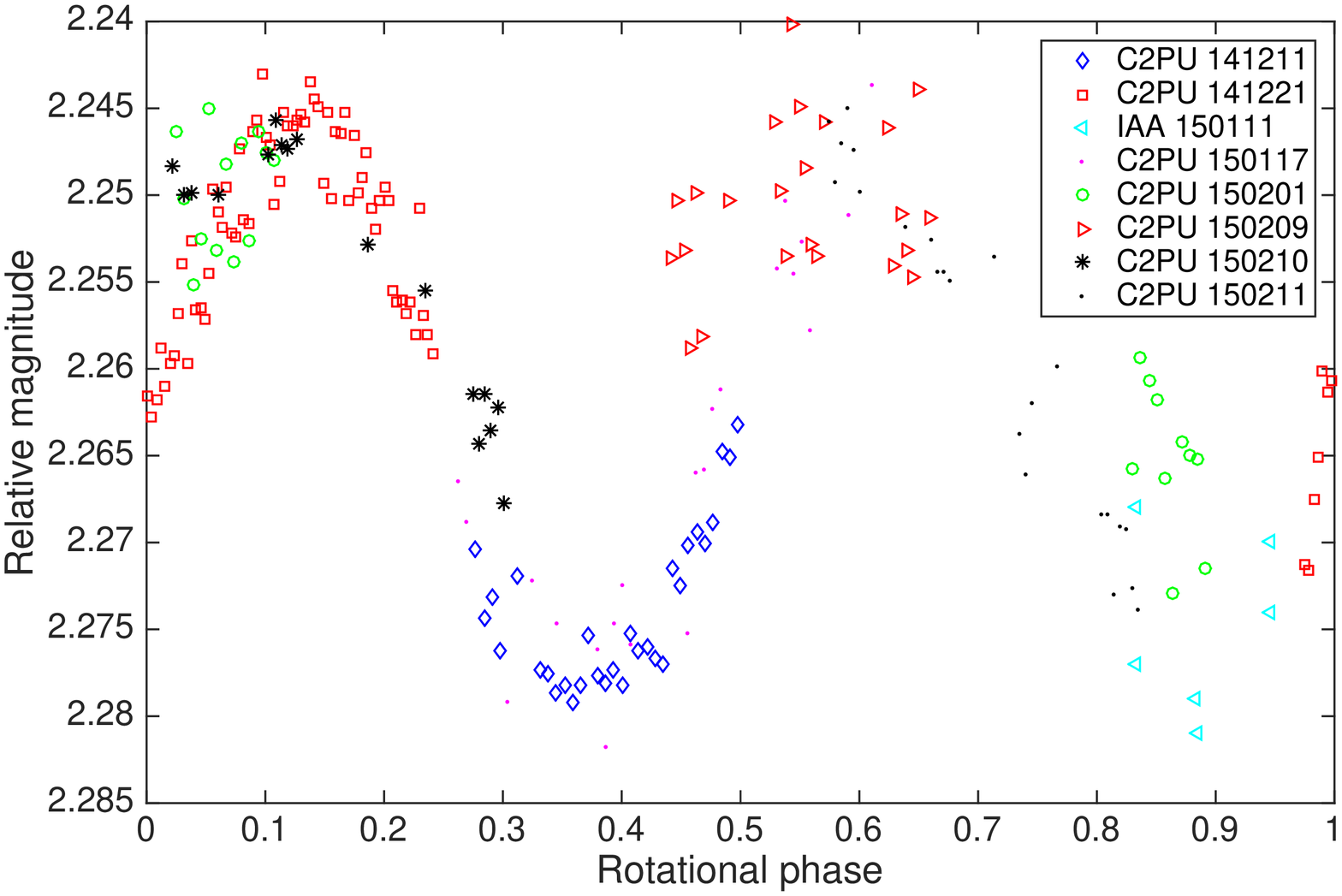}}
     {(1372)~Haremari, $P_{\rm syn} = 15.24 \pm 0.03$~h}

&
\subf{\includegraphics[width=8.5cm]{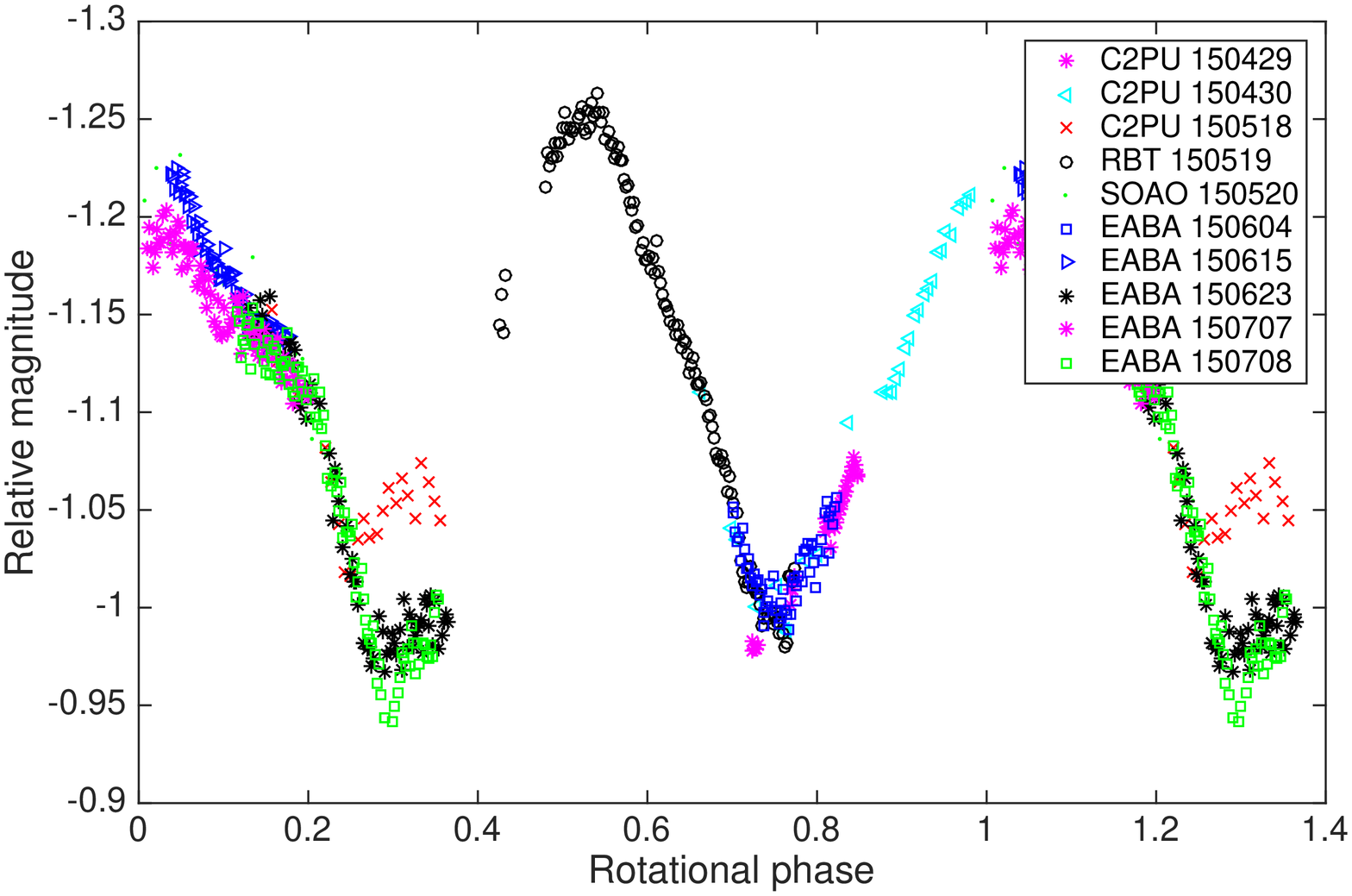}}
     {(1702)~Kalahari, $P_{\rm syn} = 21.15 \pm 0.03$~h}

\\
\hline

\subf{\includegraphics[width=8.5cm]{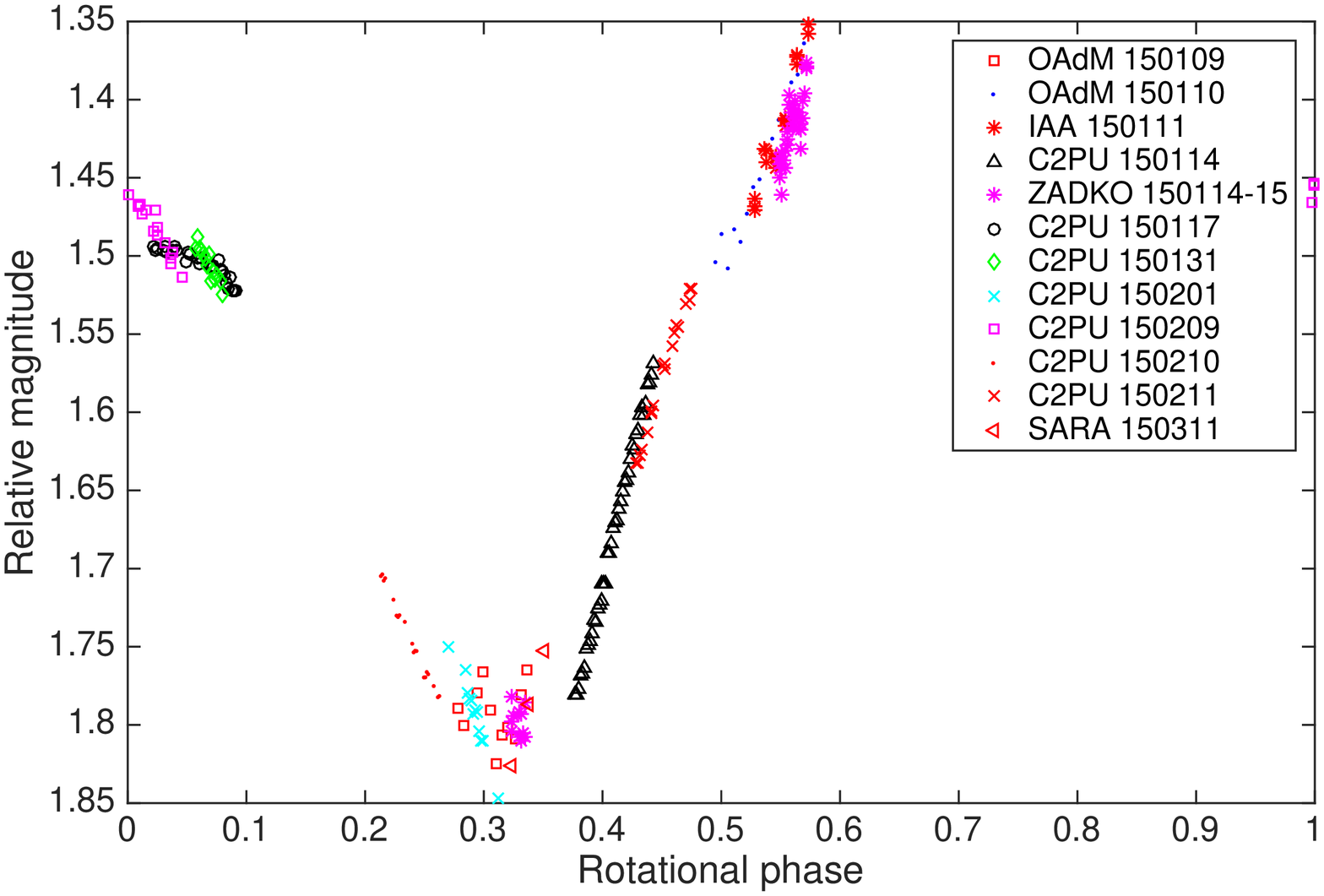}}
     {(2085)~Henan, $P_{\rm syn} = 110.5 \pm 1$~h ($94.3 \pm 1$~h)}

&
\subf{\includegraphics[width =8.5cm]{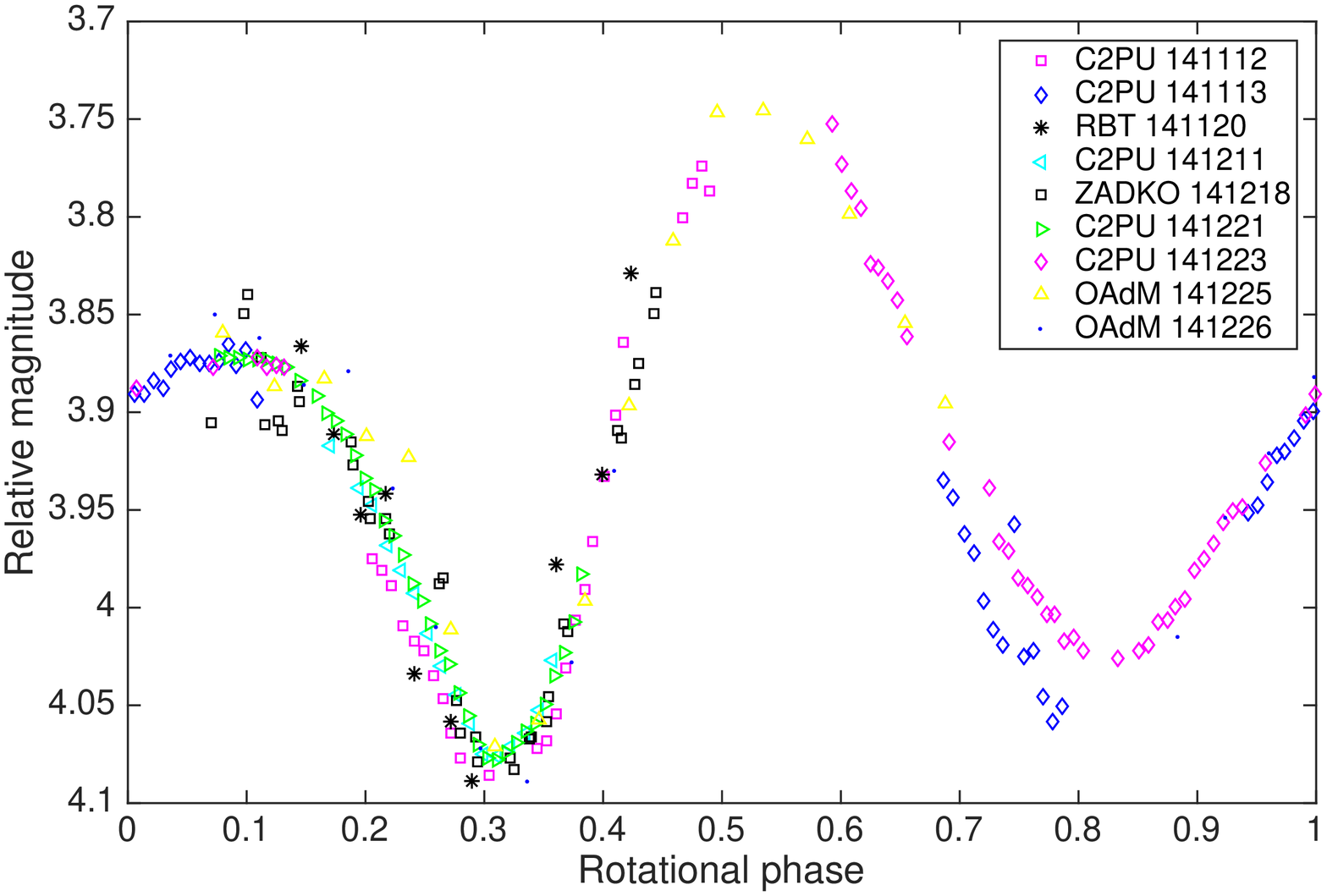}}
     {(3844)~Lujiaxi, $P_{\rm syn} = 13.33 \pm 0.01$~h}

\\
\hline
\subf{\includegraphics[width=8.5cm]{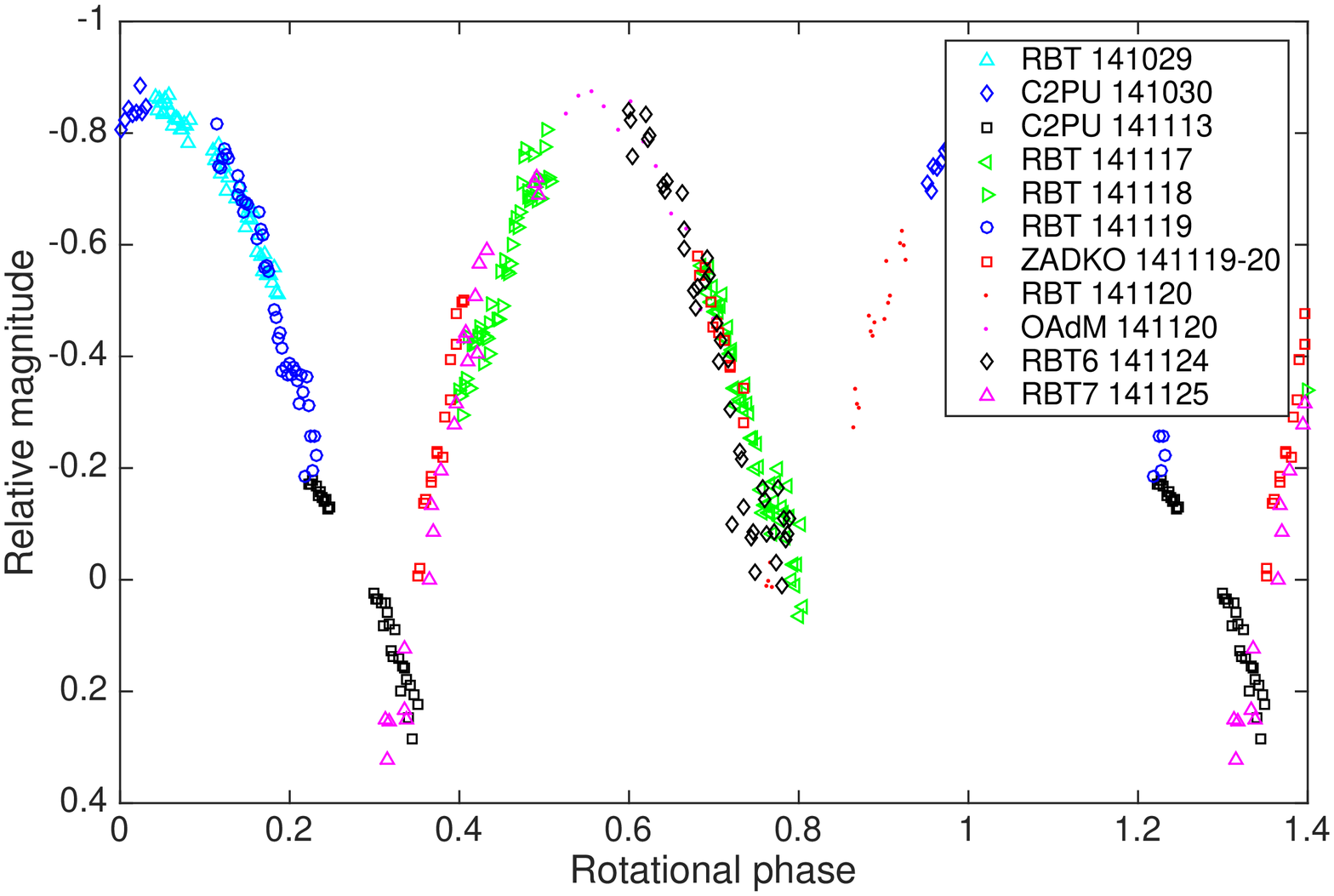}}
{(15552)~Sandashounkan, $P_{\rm syn} = 33.62 \pm 0.3$~h}

&

\subf{\includegraphics[width=8.5cm]{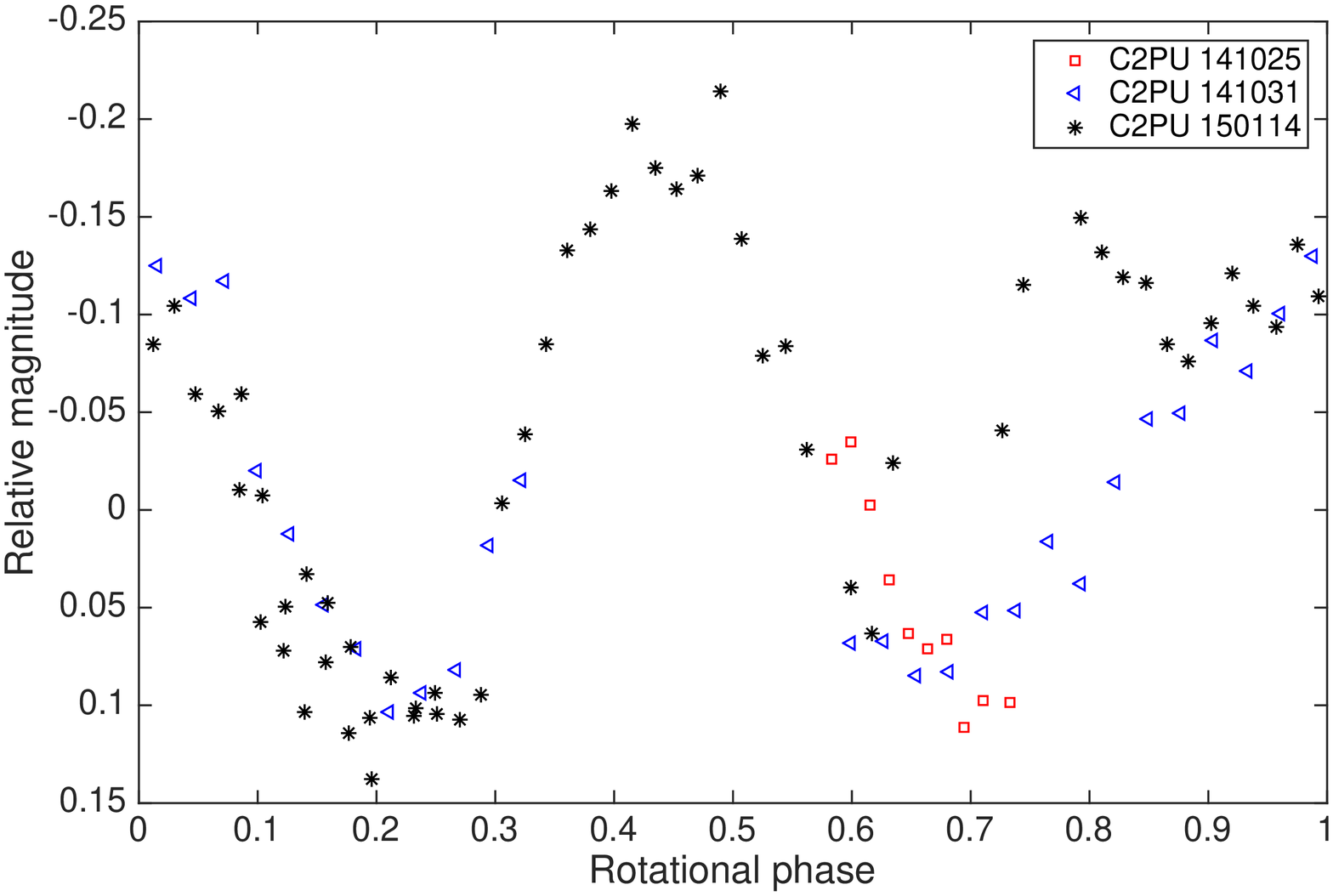}}{(67255)~2000ET109, $P_{\rm syn} = 3.70 \pm 0.01$~h}
\\
\hline
\end{tabular}
\caption{Continued}
\end{figure*}

\section{Shape models derived in this work}
\label{App:Shape}

\begin{figure*}
\centering
\begin{tabular}{|c|}
\hline

\subf{\includegraphics[width=17cm]{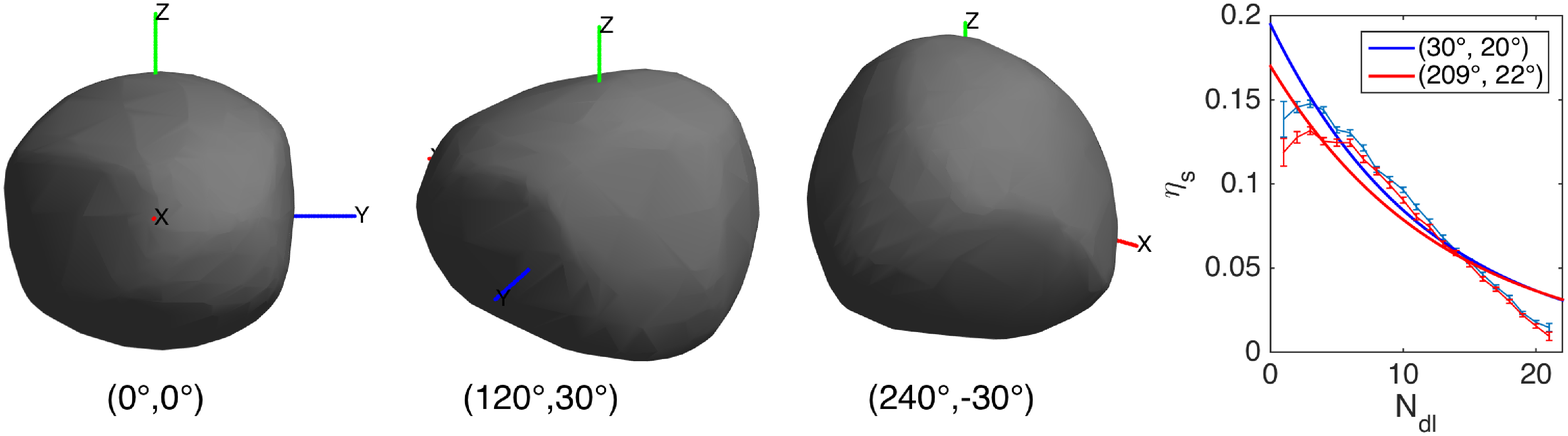}}
     {(172)~Baucis, pole solution ($209^{\circ}$, $22^{\circ}$).}

\\
\hline
\subf{\includegraphics[width=17cm]{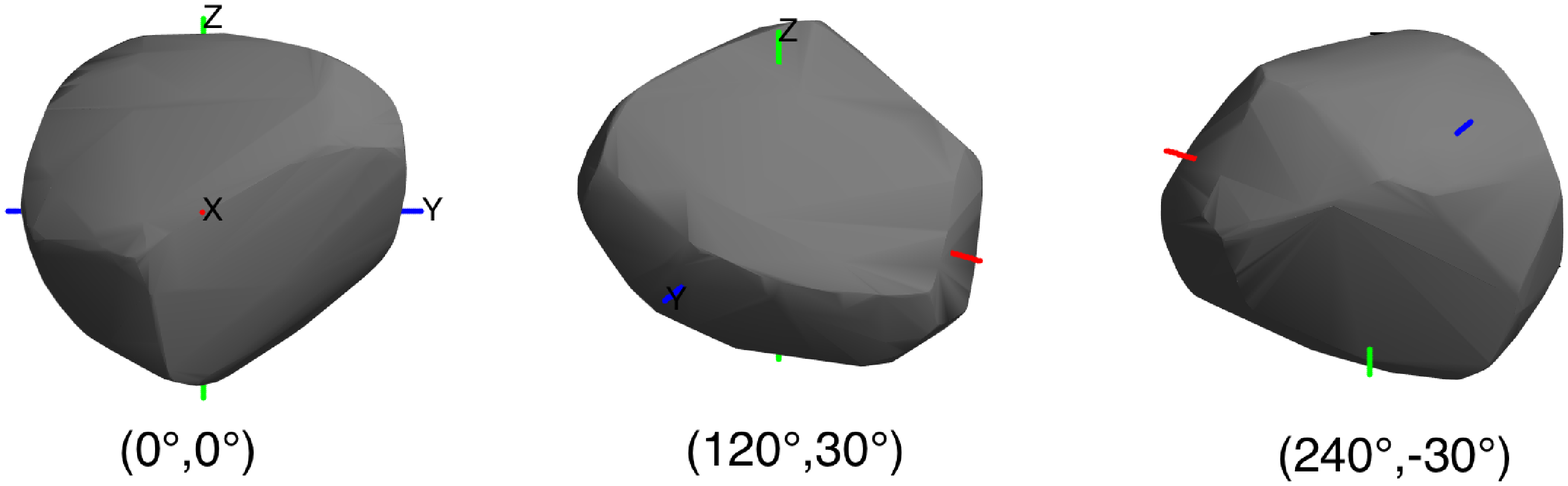}}
     {(387)~Aquitania, pole solution ($142^{\circ}$, $51^{\circ}$).}

\\
\hline
\subf{\includegraphics[width=16cm]{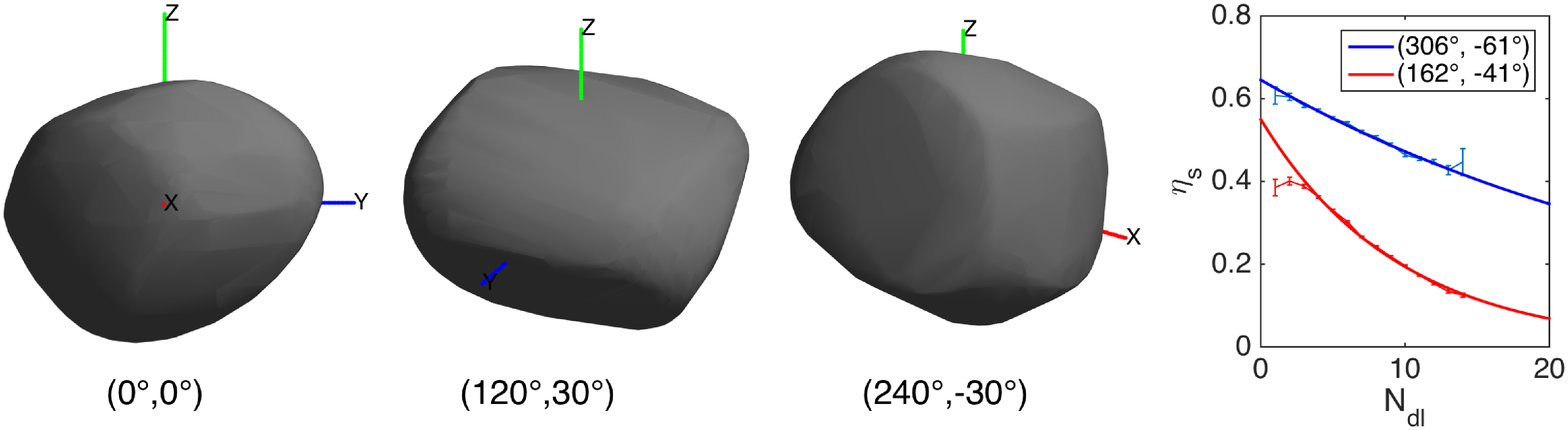}}
     {(402)~Chloe, pole solution ($306^{\circ}$, $-61^{\circ}$).}

\\
\hline
\end{tabular}
\caption{Asteroid shape models derived in this work. For each shape model, the reference system in which the shape is described by the inversion procedure, is also displayed. The $z$ axis corresponds to the rotation axis. The $y$ axis is oriented to correspond to the longest direction of the shape model on the plane perpendicular to $z$. Each shape is projected along three different viewing geometries to provide an overall view. The first one (left-most part of the figures) corresponds to a viewing geometry of $0^{\circ}$ and $0^{\circ}$ for the longitude and latitude respectively (the $x$ axis is facing the observer). The second orientation corresponds to ($120^{\circ}$, $30^{\circ}$) and the third one to ($240^{\circ}$, $-30^{\circ}$). The inset plot shows the result of the Bootstrap method. The $x$ axis corresponds to the number of light-curves used and the $y$ axis is $\eta_a$. Since the shape model of (387)~Aquitania was derived without sparse data, the bootstrap method was not applied.}
\label{fig:Shape}
\end{figure*}

\begin{figure*}
\addtocounter{figure}{-1}
\centering
\begin{tabular}{|c|}
     \hline
\subf{\includegraphics[width=16cm]{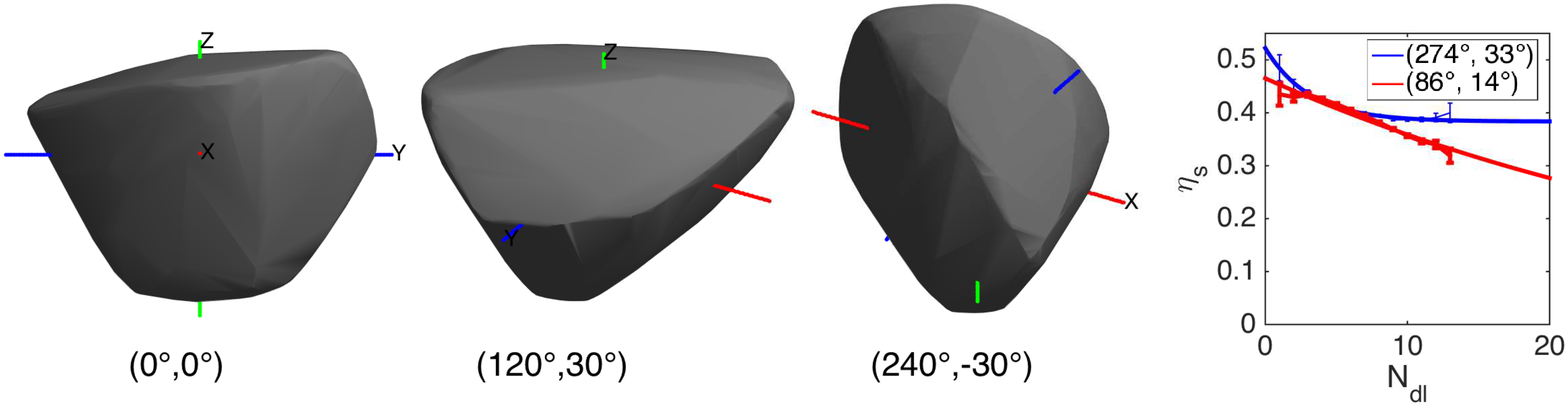}}
     {(458)~Hercynia, pole solution ($274^{\circ}$, $33^{\circ}$).}

\\
\hline
\subf{\includegraphics[width=16cm]{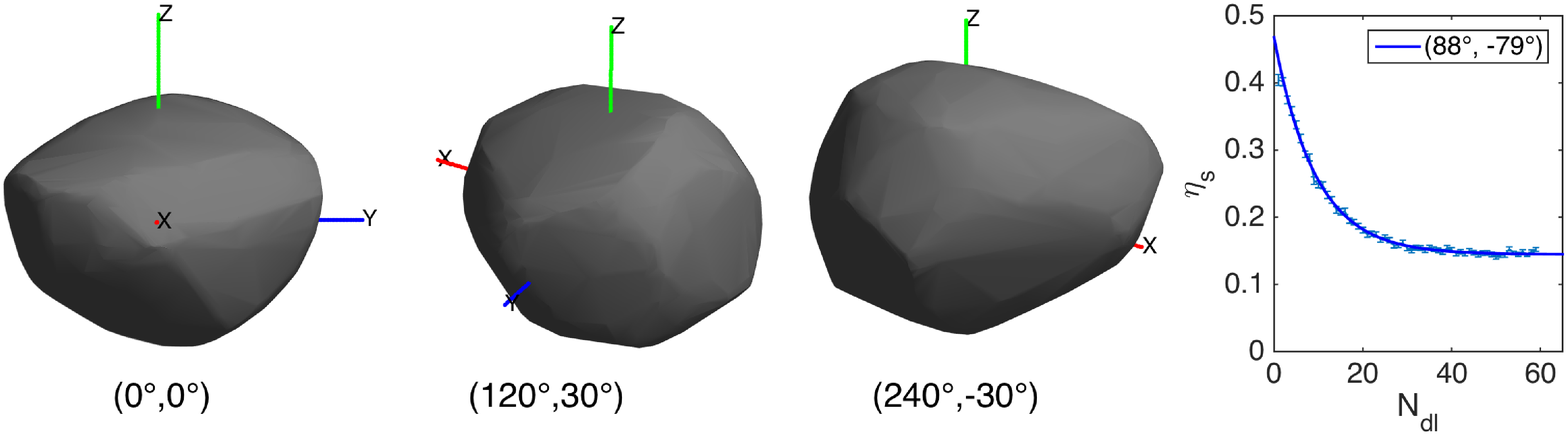}}
     {(729)~Watsonia, pole solution ($88^{\circ}$,  $-79^{\circ}$)}

\\
\hline
\subf{\includegraphics[width=16cm]{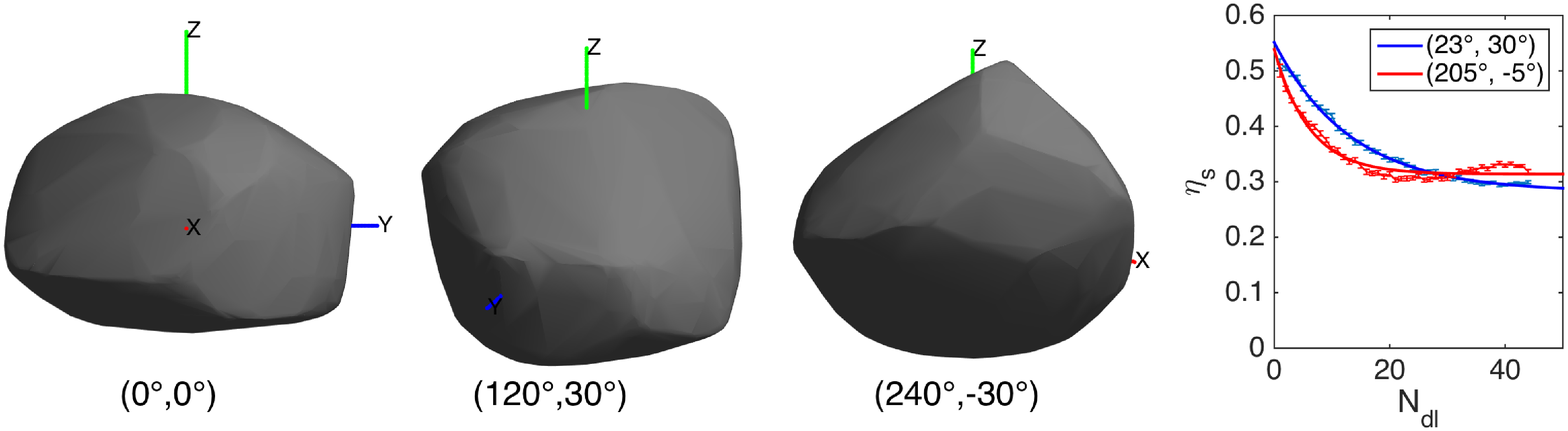}}
     {(980)~Anacostia, pole solution ($23^{\circ}$, $30^{\circ}$)}

\\
\hline
\end{tabular}
\caption{Continued}
\end{figure*}

\section{List of observers for the occultation data used in this work}

\begin{figure}
\centering
\begin{tabular}{l}
\hline
List of observers \\
\hline
\bf{(172)~Baucis} (2015-12-18) \\
J.A. de los Reyes, Sensi Pastor, SP \\ 
F. Reyes, SP \\
J.L. Ortiz, SP \\ 
F. Aceituno, SP \\
\\
\bf{(236)~Honoria} (2008-10-11) \\
S Conard, Gamber, MD, USA \\                          
R Cadmus, Grinnell, IA, USA \\
\\
\bf{(236)~Honoria} (2012-09-07) \\
C. Schnabel, SP \\
C. Perello, A. Selva, SP \\
U. Quadri, IT \\
P. Baruffetti, IT \\ 
J. Rovira, SP \\      
\\   
\bf{(387)~Aquitania} (2013-07-26) \\
T. Blank, West of Uvalde, TX, USA \\                           
T. Blank, Uvalde, TX , USA \\        
T. Blank, D'Hanis, TX, USA \\    
T. Blank, Devine, TX, USA \\                                     
S. Degenhardt, Moore, TX, USA \\            
M. McCants, TX, USA \\
S. Degenhardt, Bigfoot, TX, USA \\              
S. Degenhardt, Poteet, TX, USA \\                 
S. Degenhardt, Pleasanton, TX, USA \\
\\
\bf{(402)~Chloe} (2004-12-15) \\
R. Nugent, Buffalo, TX, USA \\                            
R. Venable, Bunnell, FL, USA \\                               
R. Venable, DeLand, FL, USA \\                            
J. Stamm, Oro Valley, AZ, USA \\                         
R. Peterson, Scottsdale, AZ, USA  \\
B. Cudnik, Houston, TX, USA \\
\\
\bf{(402)~Chloe} (2004-12-23) \\
S. Preston, Ocean Park, WA, USA \\                  
J. Preston, Ilwaco, WA, USA  \\          
T. George,  Moro, OR, USA \\
\hline
\end{tabular}
\caption{List of observers of stellar occultation by an asteroid.}
\label{Tab:Occ_Obs}
\end{figure}

\end{appendix}

\end{document}